\documentclass{ws-ijmpa}
\usepackage[super,compress]{cite}
\usepackage{graphicx}
\usepackage{color}

\newcommand{\be}{\begin{equation}}
\newcommand{\ee}{\end{equation}}
\newcommand{\bea}{\begin{eqnarray}}
\newcommand{\eea}{\end{eqnarray}}
\newcommand{\beq}{\begin{equation}}
\newcommand{\eeq}{\end{equation}}
\newcommand{\nn}{\nonumber}
\newcommand{\ctg}{\rm ctg}
\newcommand{\tg}{\rm tg}

\begin{document}
\title{THE GROUP THEORETICAL DESCRIPTION
OF THE THREE-BODY PROBLEM}

\author{J. NYIRI}
\address{Institute for Particle and Nuclear Physics, Wigner
RCP, Budapest 1121, Hungary }

\author{V.A. NIKONOV}
\address{National Research Centre "Kurchatov Institute",
Petersburg Nuclear Physics Institute, Gatchina 188300, Russia}

\maketitle

\begin{history}
\received{Day Month Year}
\revised{Day Month Year}
\end{history}

\begin{abstract}
The group theoretical description of the three-particle problem provides
successful techniques for the solution of different questions. We present here
a review of this approach.
\end{abstract}

\ccode{PACS numbers: 11.10.Ef, 11.30.-j, 02.20.-a}

\section{Introduction}

The three-body problem in quantum mechanics in general, and in
topics like molecular, nuclear and particle physics in particular,
provides a variety of interesting fields for investigations. The
developed technics were useful for the solution of different,
sometimes quite unexpected problems, and because of that the
three-body problem was and remains a rather attractive topic.

In the thirties and forties the Hartree-Fock, Thomas-Fermi, and
variational methods were suggested, and many different approaches
were considered in the past decades \cite{1,2,3}. Developments and
improvements have never stopped; see, for example,
Refs.~\refcite{4,5,6}.

In classical mechanics the three-body problem appeared long ago when
the motion of planets became the subject of investigations. A
solution for newtonian interactions was given by Euler \cite{euler}
around 1760; a little later Lagrange \cite{lagrange} solved a
generalized problem with additional linear forces. The story of
further developments  can be found, {\ e.g.} in
Refs.~\refcite{whit,murr}.

In quantum mechanics the three-particle problem was taken into
consideration from the very beginning, in the end of the twenties.
In the thirties the problem of falling on the center was formulated
\cite{thoma}. In the fifties and the sixties a set of specific
topics were investigated. First, that was the formulation of the
equation for calculations of energy levels in three-nucleon systems,
He$^3$/H$^3$, using point-like forces for two-particle interactions,
the Skornyakov -- Ter-Martirosyan equation \cite{sk-tm}. Its
analysis and the determination of the neutron-deuteron scattering
length was given by Danilov \cite{danil}. The general solution of
the equation was formulated by Minlos and Faddeev \cite{mi-fa}.
Namely, there are two sets of levels below and above the value of
the basic one (which is not determined by the equation, it is a
parameter of the model). The lower set corresponds to the three-body
collapse \cite{thoma}, while the second set corresponds to the
concentration of levels towards zero binding energy. The effect of
concentration of the three-body spectra at zero total energy with
increasing two-particle scattering lengths was emphasized in
\cite{efimov}. The three-body collapse branch of levels was
eliminated in the Efimov solution \cite{efimov} by a cutting
procedure at large relative momenta -- this method of regularization
implicitly introduces short range three-body forces.

At present the short-range approach (Skornyakov--Ter-Martirosyan
equation) with a set of levels concentrating at zero energy is
widely used in molecular physics, see {\ e.g.}
Refs.~\refcite{BHK,penkov,motovilov,ef-review} and references therein.


The Faddeev equation \cite{faddeev} for non-relativistic
three-particle systems and that for four particles \cite{yakubovsky}
are logical consequences of the investigation of many particle
systems in the framework of quantum mechanics \cite{fm}.

In the sixties the dispersion relation technics was used for the
investigation of three-particle systems: first, within the
non-relativistic approach, to expand the amplitudes near the
thresholds \cite{aag,aa}, then to describe certain relativistic
processes with resonances in the intermediate state
\cite{ad-tri,ak-box}.

The relativistic three-particle dispersion relation equation was
written in Refs.~\refcite{a-eta,a-3mesons}, see also
Ref.~\refcite{book-4}.
Relativistic three-body equations were seriously discussed before
\cite{kh-tr} but at that time the analytical continuation of the
amplitude situated it on the second ({\it i.e.}, unphysical) sheets
of complex variables.

As a rule, the relativistic description of a three particle system
is also a description of the surrounding states which are related to
each other by transitions. The problem of coupled states is
discussed in Ref.~\refcite{book-4}, and this line of research requires
significant efforts.

A further way for the development of methods for investigating three
particle systems is the elaboration of techniques for the expansion
of wave functions (or amplitudes) over a convenient set of states.
It is a problem both for the non-relativistic and relativistic
approaches. Different sets of expansion are appropriate for
different problems depending on the type of interactions in the
system.

In this review paper we present the group theoretical description of
the three-particle problem suggested in the middle of the sixties
\cite{1n,2n,3n,4n,5n,6n}; the last papers on the subject were
written in the eighties. It is just a reminder of
this part of the three-body story. It can be considered as an
addition to the book \cite{book-4}.

From a group theoretical point of view the most interesting
questions are related to the fifth quantum number $\Omega$. This has
to be introduced because the quantum numbers describing rotations
and permutations are not sufficient to characterize the states in
the three-body system. In the considered papers a complete set of
basis functions for the quantum mechanical three-body system is
chosen in the form of hyperspherical functions, characterized by
quantum numbers corresponding to the chain $O(6) \supset SU(3)
\supset O(3)$. Equations are derived to obtain the basis functions
in an explicit form.

The problem of constructing a basis for a system of three free
particles, making use of representations of the three-dimensional
rotation group and of the permutation group, is quite simple in
principle. Nevertheless, problems appear in implementing a
straightforward way for the construction of a general solution for
the set of equations which determine the eigenfunctions. As it turns
out, the eigenvalue equations can be simplified considerably, then
the solution is derived in a closed form, the coefficients are
calculated in different ways, numerical results are obtained.

The developed technics can be applied in a variety of cases. First
of all, as soon as the quantum mechanical problem which we have
considered has the same symmetry properties as the classical one, it
is possible to investigate the classical problem. The equations of
motion are obtained very easily for both the case of free particles
and that of different potentials.

The classification of a three-body system can be used also for the
analysis of three-particle decay processes. For example, dealing
with a Dalitz plot for decay processes, it turned out to be useful
to expand the point density inside the physical region into a series
of orthonormal functions. (Such an expansion is similar to the usual
phase analysis for two-particle decays. It was helpful in analyzing
experimental data, for the calculation of different correlation
functions etc.) The set of basis functions chosen as K-harmonics was
especially suitable for the description of correlations between the
momenta of particles. Also, from a practical point of view it was
essential to develop a method to calculate matrix elements of
two-particle interactions introducing different potentials and to
obtain a proper approximation for bound states as well.

%
%

Let us note that investigation of the question for a special case
was carried out by Badalyan and Simonov \cite{7n,8n}. The basis for
expansion was formed by the so-called K-polynomials, which are
harmonic functions corresponding to the Laplace operator on the
six-dimensional sphere. Further, a complete set of solutions was
considered with five commuting operators. It was shown how in
principle one can construct the polynomials being eigenfunctions of
these operators. But the calculations were done in a rather
complicated way, so only the lowest polynomials were obtained, which
can be characterized by four quantum numbers.

Another possibility of constructing a basis was demonstrated by
Zickendraht \cite{9n}, however the used method is also too
complicated and does not give a sufficiently general result.

In the paper of L\'{e}vy-Leblond and L\'{e}vy-Nahas \cite{10n} the
connection between the basis and the representations of SU(3)
is pointed out. The authors have used a proper parametrization and
obtained the Laplace operator expressed in terms of angular
variables. Yet, they did not discuss a general solution either.

If one intends to construct harmonic functions for the
three-particle system analogous to the spherical functions forming
the basis in case of two particles, it is natural to use angular
variables on the six-dimensional sphere or on the three-dimensional
complex sphere and build up the required functions in terms of these
coordinates. In this survey we demonstrate a way to carry out this
program. Let us note that the full group of motion on the
six-dimensional sphere is too large for our aim. The problem is just
to find the suitable subgroup.

Introducing angular variables, we have to separate similarity
transformations and take into consideration only those
transformations under which the sum of squares of coordinates of the
three particles is invariant, {\it i.e.} the radius of the six
dimensional sphere remains constant.

Consider now a triangle, the vertices of which are determined by
three particles. If we exclude the similarity transformations, two
possible types of transformations are left: rotations in the
ordinary three dimensional space which are described by the O(3)
group, and deformations of the triangle. It can be easily seen that
the deformation leads to SU(2).

It is obvious that different forms of a deformed, non-rotating
triangle can be considered as the projections onto its plane of all
the possible positions of a rotating rigid triangle.

Studying both types of transformations at the same time, one can say
that all the transformations of a triangle besides the similarity
transformations are described by the projections onto the
three-dimensional space of a rigid triangle which rotate in the
four-dimensional space. This means that an arbitrary motion of three
particles is equivalent to the rotation of a rigid triangle and the
similarity transformations.

Let us turn our attention to a formal analogy with the Kepler
problem. The planetary motion along the elliptic trajectory can be
described as the projection of motion along the great circle on the
four dimensional sphere onto its equatorial section. Ellipses with
equal major axes are corresponding to different great circles. After
carrying out the transformation of time, we can show that the Kepler
motion will be described by the free motion of a point on the
four-dimensional sphere, famous Fock-symmetry \cite{focko4}. This
way we arrive at the local O(4) symmetry which will also be
considered.

Both the representation of the group of motions on the
six-dimensional sphere and its reduction to SU(3) or O(4) involve
the representation of the permutation group P$_2$(3). That is why
this description is extraordinary convenient for the system of three
equivalent particles. Here we will restrict ourselves to this simple
case. In the general case of arbitrary masses some new features will
appear only when we expand the amplitudes or the wave functions of
the interacting particles over the basis functions \cite{11n}. As it
is well known, the boundary of the definition of the functions
depends on the masses.

Concerning the construction of the basis, a question arises whether
it is necessary to build up the basis with the help of K-polynomials
of the harmonic functions of O(6). Obviously, if the interaction
between the particles is weak and their motion differs only slightly
from the free one, this choice of the basis functions will be
natural. If, on the contrary, the particles are strongly bounded and
form an almost rigid triangle, a basis, which do not obey the Laplace
equation on the six dimensional sphere, turns out to be more
convenient. As an example of such a basis, so-called B-polynomials
will be constructed.

An interesting subject for discussion is given by the fifth quantum
number $\Omega$ (see Ref.~\refcite{12n}). The introduction of this
quantity becomes necessary since it is not sufficient to use the
quantum numbers coming from the reduction O(6)= O(3)x O(2), {\it i.e.}
the quantum numbers characterizing the rotations and permutations. In
fact for a system consisting of more than three particles one has to
introduce additional quantum numbers: three quantum numbers in the
case of four particles, and four in the case of five particles. In
the case of a system including six or more particles five new
quantum numbers are necessary. It is a rather remarkable fact that
for more than six particles the number of additional quantum numbers
remains constant.

It is worthwhile to study also the energy spectrum of the
three-particles system in the case when the triangle formed by three
particles is getting rigid. The transition from the spectrum of
non-interacting particles to the spectrum of the top may be
investigated with the help of the presented basis.

\section{Coordinates and Observables}

The usual way of choosing the coordinates is the following. Let
$x_i(i=1,2,3)$ be the radius vectors of the three particles, and fix
\begin{equation}
x_1+x_2+x_3 = 0  \,.
\label{I.1}
\end{equation}
The Jacobi coordinates for equal masses will be defined as
\begin{eqnarray}\label{I.2-4}
&&  \xi\ =\ -\sqrt{\frac32}\ (x_1+x_2)\,,
\qquad \eta\ =\ \sqrt{\frac12}\ (x_1-x_2)\,,
 \\
&& \xi^2+\eta^2\ =\ 2x^2_1+2x_1x_2+2x^2_2\ =\ x^2_1+x^2_2+x^2_3\ =\
\rho^2.
\nn
\end{eqnarray}
We may define similar coordinates in the momentum space as well. In
that case condition (\ref{I.1}) means that we are in the
centre-of-mass frame, and $\rho^2$ is a quantity proportional to the
energy.

The quadratic form $\rho^2$  can be understood as an invariant of the
$0(6)$ group. In fact, we are interested in the direct product
$O(3)\times O(2)$, as we have to introduce the total angular momentum
observables $L$ and $M$ (group $O(3)$), and quantum numbers of the
three-particle permutation group $O(2)$.

To characterize our three-particle system we need five quantum
numbers. Thus the $O(6)$ group is too large for our purposes and it
is convenient to deal with $SU(3)$ symmetry, in case of which we
dispose exactly of the necessary four quantum numbers.

Let us introduce the complex vector
\be
z = \xi+i\eta\,,
\qquad
z* = \xi-i\eta\,.
\label{I.5-6}
\ee
The permutation of two particles leads in terms of these coordinates
to rotations in the complex $z$-plane:
\be
P_{12}\Big({z \atop z^*}\Big)=\Big({z^* \atop z}\Big), \quad
P_{13}\Big({z \atop z^*}\Big)=\Big({e^{i\pi/3}z^* \atop e^{-i\pi/3}z}
\Big), \quad P_{23}\Big({z \atop z^*}\Big) =
 \Big({e^{-i\pi/3}z^* \atop e^{i\pi/3}z}\Big).
\label{I.7}
\ee
The condition
\begin{equation}
\xi^2+\eta^2\ =\ |z|^2\ =\ \rho^2
\label{I.8}
\end{equation}
gives the invariant of the group $SU(3)\subset O(6)$. In the following
we will take $\rho=1$.

The generators of $SU(3)$ are defined, as usual:
\begin{equation}
A_{ik}\ =\ iz_i\,\frac{\partial}{\partial
z_k}-iz^*_k\,\frac{\partial}{\partial z^*_i}\,. \label{I.9}
\end{equation}
The chain $SU(3)\supset SU(2)\supset U(1)$ familiar from the theory
of unitary symmetry of hadrons is of no use for us, because it does
not contain $O(3)$, {\it i.e.} going this way we cannot introduce the
angular momentum quantum numbers. Instead of that, we consider two
subgroups $O(6)\supset O(4)\sim SU(2)\times O(3)$ and
$O(6)\supset SU(3)$. In other words, we have to separate
from (\ref{I.8}) the
antisymmetric tensor-generator of the rotation group $O(3)$
\begin{equation}
L_{ik}\,=\,\frac12(A_{ik}-A_{ki})\,=\,\frac12\left(iz_i
\frac{\partial}{\partial z_k}-iz_k\frac{\partial}{\partial
z_i}+iz^*_i\frac{\partial}{\partial z^*_k}
-iz^*_k\frac{\partial}{\partial z^*_i}\right). \label{I.10}
\end{equation}
The remaining symmetric part
\begin{equation}
B_{ik}\ =\ \frac12(A_{ik}+A_{ki})\ =\ \frac12\left(iz_i
\frac{\partial}{\partial z_k}+iz_k\frac{\partial}{\partial
z_i}-iz^*_i\frac{\partial}{\partial z^*_k} -iz^*_k
\frac{\partial}{\partial z^*_i}\right) \label{I.11}
\end{equation}
is the generator of the group of deformations of the triangle which
turns out to be locally isomorphic with the rotation group. Finally,
we introduce a scalar operator
\begin{equation}
N\ =\ \frac1{2i}\mbox{ Sp }A\ =\ \frac12\sum_k\left(z_k
\frac{\partial}{\partial z_k}-z^*_k\frac{\partial}{\partial
z^*_k}\right). \label{I.12}
\end{equation}
For characterizing our system, we choose the following quantum numbers:

\begin{equation} \left.  \begin{array}{l}
K(K+4)~-~\mbox{eigenvalue of the Laplace operator (quadratic Casimir}\\
\mbox{operator for }\ SU(3));\\
L(L+1)~-~ \mbox{eigenvalue of the square of the angular momentum
operator}\\
 L^2\ =\ 4\sum\limits_{i>k} L^2_{ik},\\
M~-~ \mbox{eigenvalue of }\ L_3\ =\ 2L_{12}\,,\\
\nu~-~\mbox{ eigenvalue of }\ N\,.
\end{array}  \right\}.
\label{I.13}
\end{equation}
Although the generator (\ref{I.12}) is not a Casimir operator of
$SU(3)$, the representation might be characterized by means of its
eigenvalue, because, as it can be seen, the eigenvalue of the
Casimir operator of third order can be written as a combination of
$K$ and $\nu$. If the harmonic function belongs to the
representation $(p,q)$ of $SU(3)$, then it is the eigenfunction of
$\Delta$ and $N$ with eigenvalues $K(K+4)$ and $\nu$ respectively,
where $K=p+q$ and $\nu=p-q$.

The fifth quantum number is not included in any of the considered
subgroups, we have to take it from $O(6)$. We define it as the
eigenvalue of
\begin{equation}
\Omega\ =\ \sum_{ikl} L_{ik}B_{kl}L_{li}\ =\ \mbox{Sp }LBL\,.
\label{I.14}
\end{equation}
This cubic generator was first introduced by Racah \cite{19n}. Its
physical meaning will be discussed later.

\subsection{Parametrization of a complex sphere}

Dealing with a three-particle system, we have to introduce
coordinates which refer explicitly to the moving axes. One of the
possible parametrizations of the vectors $z$ and $z^*$ is the
following:
\begin{eqnarray} \label{I.15-18}
&& z\ =\ \frac1{\sqrt2}\,e^{-i\lambda/2}\Big(e^{ia/2}l_1 +ie^{-ia/2}l_2
\Big),
\\
&& z^*\ =\ \frac1{\sqrt2}\,e^{i\lambda/2}\Big(e^{-ia/2}l_1-ie^{ia/2}
l_2\Big),  \nn
\\
&&  |z|^2\ =\ 1,
\quad
 l^2_1\ =\ l^2_2\ =\ 1, \quad l_1l_2\ =\ 0.
\nn
\end{eqnarray}
In terms of these variables the (diagonal) moment of inertia has the
following components:
\be
\sin^2\Big(\frac a2 -\frac\pi4\Big), \quad
\cos^2\Big( \frac a2-\frac\pi4\Big), \quad l.
\ee

The three orthogonal unit vectors $l_1,l_2$ and $l=l_1\times l_2$
form the moving system of coordinates. Their orientation to the
fixed coordinate system can be described with the help of the Euler
angles $\varphi_1,\theta,\varphi_2$:
\begin{eqnarray} \label{I.19-21}
I_1 &=& \Big\{-\sin\varphi_1\sin\varphi_2+\cos\varphi_1\cos\varphi_2
\cos\theta \\
&&-\ \sin\varphi_1\cos\varphi_2-\cos\varphi_1\sin\varphi_2\cos\theta;
-\cos\varphi_1\sin\theta\Big\},
\nn \\
I_2 &=& \Big\{-\cos\varphi_1\sin\varphi_2-\sin\varphi_1\cos\varphi_2
\cos\theta;
\nn \\
&& -\cos\varphi_1\cos\varphi_2+\sin\varphi_1\sin\varphi_2\cos\theta;
\sin\varphi_1\sin\theta\Big\},
\nn \\
I &=& \Big\{ -\cos\varphi_2\sin\theta;\sin\varphi_2\sin\theta;
-\cos\theta\Big\}.
\end{eqnarray}

In the following it will be simpler to introduce a new angle
\be
a=\alpha - \frac{\pi}{2}
\ee
and work with the vectors:
\begin{eqnarray} \label{I.22-23}
z\ &=&\ e^{-i\lambda/2}\Big(\cos\frac a2\,l_+
+i\sin\frac a2\,i_-\Big),
 \qquad
  z^*\ =\ e^{-i\lambda/2}\Big(\cos\frac a2\,l_-
-\sin\frac a2\,l_+
\Big),\nn \\
l_+&=&\frac1{\sqrt2}\,(l_1+il_2)\,, \qquad
l_-=\frac1{\sqrt2}\,(l_1-l_2)\,.
\end{eqnarray}
Vectors $l_+$ and $l_-$ have the obvious properties
\begin{equation}
l^2_+=l^2_-=0\,, \quad l_0=(l_+\times l_-)=-il\,, \quad l_+l_-=1\,,
\quad l^*_+=l_-\,.
\label{I.25}
\end{equation}
Let us turn our attention to the fact that the components of $l_+$
and $l_-$ may be expressed in terms of the Wigner D-functions,
defined as
\begin{equation}
D^l_{mn}(\varphi_1\theta\varphi_2)\ =\ e^{-i(m\varphi_1+n\varphi_2)}
P^l_{mn}(\cos\theta)
\label{I.26}
\end{equation}
in the following way:
\begin{eqnarray}
l_+ &=& \left\{D^1_{1-1}(\varphi_1\theta\varphi_2);\
D^1_{10}(\varphi_1\theta\varphi_2);\
D^1_{11}(\varphi_1\theta\varphi_2)\right\},
\nonumber\\
l_0 &=& \left\{D^1_{0-1}(\varphi_1\theta\varphi_2);\
D^1_{00}(\varphi_1\theta\varphi_2),\
D^1_{01}(\varphi_1\theta\varphi_2)\right\},
\nonumber\\
\label{I.27}
l_- &=& \left\{D^1_{-1-1}(\varphi_1\theta\varphi_2);\
D^1_{-10}(\varphi_1\theta\varphi_2);\
D^1_{-11}(\varphi_1\theta\varphi_2)\right\},
\end{eqnarray}
These equations demonstrate the possibility to construct the Wigner
functions from the unit vectors corresponding to the moving
coordinate system, in a way similar to the construction of spherical
harmonics from the unit vectors of the fixed coordinate system.
However, we see that the traditional parametrization of the vectors
$l_i$ which we have introduced is not fortunate; it would be much
more aesthetical to go over to a parametrization in which
\be
D^1_{mn}(\varphi_1\theta\varphi_2)\ =\ l_mk_n\,, \nn
\ee
where $l_m$ and $k_n$ are unit vectors of the moving and fixed
coordinate systems, respectively. Yet, so far we will not change the
parametrization.

 The vectors $z$ and $z^*$  can be written as
\begin{eqnarray} \label{I.28-29}
&& z_M\ =\ \sum_{M'=\pm1/2}D^{1/2}_{1/2,M'}(\lambda,a,0)D^1_{2M',M}
(\varphi_1\theta\varphi_2)\,,
 \\
&&z^*_M\ =\ D^{1/2}_{-1/2,-1/2}(\lambda,a,0) D^1_{-1,M}
(\varphi_1\theta\varphi_2)-D^{1/2}_{-1/2,1/2}(\lambda,a,0)
D^1_{1,M}(\varphi_1\theta\varphi_2). \nn
\end{eqnarray}

\subsection{The Laplace operator}

We have now to write the operators, the eigenvalues of which we are
looking for. Let us first construct the Laplace operator. We could
do that by a straightforward calculation of $\Delta=|A_{ik}|^2$, but
we choose a simpler way. We calculate
\begin{equation}
dz\ =\ -\frac i2\,zd\lambda +\frac12\,e^{-i\lambda}(l\times z^*)
da-(d\omega\times z)\,.
\label{I.30}
\end{equation}
This rather simple expression is obtained by introducing the
infinitesimal rotation $d\omega$. Its projections onto the fixed
coordinate $k_1=(1,0,0)$, $k_2=(0,1,0)$, $k_3=(0,0,1)$ given in
terms of the Euler angles are well known:
\begin{eqnarray}
&& d\omega_1\ =\ \cos\varphi_2\sin\theta d\varphi_1-\sin\varphi_2d\theta
\,, \nonumber\\
&& d\omega_2\ =\ -\sin\varphi_2\sin\theta
d\varphi_1-\cos\varphi_2d\theta\,,
\nonumber\\
&& d\omega_3\ =\ \cos\theta d\varphi_1+d\varphi_2\,.
\label{I.31}
\end{eqnarray}

This provides
\begin{eqnarray}
&& \frac{\partial}{\partial\omega_1}\ =\
\cos\varphi_2\frac1{\sin\theta}
\frac{\partial}{\partial\varphi_1}-\cos\varphi_2
\frac{\partial}{\partial\varphi_2}-\sin\varphi_2\frac{\partial}{\partial\theta}\,,
\nonumber\\
&& \frac{\partial}{\partial\omega_2}\ =\
-\sin\varphi_2\frac1{\sin\theta}
\frac{\partial}{\partial\varphi_1}+\sin\varphi_2\ctg\theta
\frac{\partial}{\partial\varphi_2}
-\cos\varphi_2\frac{\partial}{\partial\theta}\,,
\nonumber\\
&& \frac{\partial}{\partial\omega_3}\ =\
\frac{\partial}{\partial\varphi_2}\,, \label{I.32}
\end{eqnarray}
and the permutation relations

\begin{equation}
\Big[\frac{\partial}{\partial\omega_1}\,,\frac{\partial}{\partial\omega_2}\Big]=
\frac{\partial}{\partial\omega_3}\,, \quad
\Big[\frac{\partial}{\partial\omega_2}\,,
\frac{\partial}{\partial\omega_3}\Big]=\frac{\partial}{\partial\omega_1}\,,
\quad
\Big[\frac{\partial}{\partial\omega_3}\,,\frac{\partial}{\partial\omega_1}\Big]=
\frac{\partial}{\partial\omega_2}\,. \label{I.33}
\end{equation}
The effect of this operator on an arbitrary vector $A$ is

\begin{equation}
\frac{\partial}{\partial\omega_i}\,A\ =\ \omega_i\times A\, ;
\label{I.34}
\end{equation}
Here ${\bf \omega}_i$ is a vector of the length $\omega_i$, directed
along the $i$ axis. The expression can be checked using the
perturbation relation.

Let us determine now the rotation around the moving axes:
\begin{equation}
d\Omega_i\ =\ l_id\omega\,.
\label{I.35}
\end{equation}
In an explicit form $\partial/\partial\Omega_i$ can be written
\begin{eqnarray}
&& \frac{\partial}{\partial\Omega_1}\ =\ \cos\varphi_1\ctg\theta
\frac{\partial}{\partial\varphi_1}-\cos\varphi_1 \frac1{\sin\theta}
\frac{\partial}{\partial\varphi_2} +\sin\varphi_1
\frac{\partial}{\partial\theta}\,,
\nonumber\\
&& \frac{\partial}{\partial\Omega_2}\ =\ -\sin\varphi_1\ctg\theta
\frac{\partial}{\partial\varphi_1} +\sin\varphi_1\frac1{\sin\theta}
\frac{\partial}{\partial\varphi_2}+ \cos\varphi_1
\frac{\partial}{\partial\theta}\,,
\nonumber\\
&& \frac{\partial}{\partial\Omega_3}\ =\
-\frac{\partial}{\partial\varphi_1}\,. \label{I.36}
\end{eqnarray}
The minus sign in the third component reflects our choice of
normalization of the $D$-function with a minus in the exponent
(\ref{I.26}).

The permutation relations for the operators
$\partial/\partial\Omega_i$ are
\begin{equation}
\Big[\frac{\partial}{\partial\Omega_1}\,,\frac{\partial}{\partial\Omega_2}\Big]=
-\frac{\partial}{\partial\Omega_3}\,, \quad
\Big[\frac{\partial}{\partial\Omega_2}\,,
\frac{\partial}{\partial\Omega_3}\Big]=-\frac{\partial}{\partial\Omega_1}\,,
\quad
\Big[\frac{\partial}{\partial\Omega_3}\,,\frac{\partial}{\partial\Omega_1}\Big]
= -\frac{\partial}{\partial\Omega_2}\,. \label{I.37}
\end{equation}
The effect on $A$ is defined, correspondingly, as
\begin{equation}
\frac{\partial}{\partial\Omega}\,A\ =\ -\Omega_i\times a\,,
\label{I.38}
\end{equation}
which differs from (\ref{I.34}) by the sign, as a consequence of the
different signs in the permutation relations (\ref{I.33}) and
(\ref{I.37}).

From (\ref{I.30}) we obtain
\bea
ds^2&=&dz\,dz^*=g_{ik}x^ix^k
\nn \\
&=&\frac14 da^2+\frac14 d\lambda^2 +\frac12
d\Omega^2_1+\frac12 d\Omega^2_2+d\Omega^2_3 -\sin a\,d\Omega_1
d\Omega_2-\cos a\,d\Omega_3d\lambda\,.\quad
\label{I.39}
 \eea
This expression determines the components of the metric tensor
$q_{ik}$, and it becomes easy to calculate the Laplace operator
\begin{eqnarray}
\Delta' & =& \frac14\,\Delta = \frac14\frac1{\sqrt g}
\frac{\partial}{\partial x^i}\,g^{ik}\sqrt g
\frac{\partial}{\partial x^k}\ =
\nonumber \\
&=& \frac{\partial^2}{\partial a^2}+2{\ctg 2}a
\frac{\partial}{{\partial}
a} +\frac{1}{{\sin^2}a}\left(\frac{\partial^2}{\partial\lambda^2}+
{\cos} a\frac{\partial^2}{\partial\lambda\,\partial\Omega_3}+\frac14
\frac{\partial^2}{\partial\Omega^2_3}\right) +
\nonumber\\
&& +\
\frac1{2\cos^2a}\left[\frac{\partial^2}{\partial\Omega^2_1}+\sin a
\Big( \frac{\partial^2}{\partial\Omega_1\partial\Omega_2}
+\frac{\partial^2}{\partial\Omega_2\partial\Omega_1}\Big)
+\frac{\partial^2}{\partial\Omega^2_2}\right]. \label{I.40}
\end{eqnarray}
If $\Phi$ is the eigenfunction of $\Delta'$, corresponding to a
definite representation of $SU(3)$, then
\begin{equation}
\Delta'\Phi\ =\ -\frac14\,K(K+4)\Phi\ =\
-\frac K2\Big(\frac K2+2\Big)\Phi
\label{I.41}
\end{equation}
and
\begin{equation}
N\Phi\ =\ \nu\Phi\,, \qquad N\ =\ i\frac\partial{\partial\lambda}\,
\label{I.42}
\end{equation}
has to be fulfilled.

Expressing (\ref{I.40}) in terms of the Euler angles, we get the
Laplace operator in the form obtained in Ref.~\refcite{10n}:
\begin{eqnarray}
\Delta' &=& \Delta_a-{\tg\,} a\frac{\partial}{{\partial}
a}+\frac1{2{\cos}^2a}
\Big(\Delta_\theta-\frac{\partial^2}{\partial\varphi^2_1}\Big)
-\frac{{\sin} a}{2{\cos}^2a}\ \times
\nonumber\\
&& \times\
\bigg[\cos2\varphi_1\Big(\frac{1+\cos^2\theta}{\sin^2\theta}
\frac{\partial}{\partial\varphi_1} -2\frac{\cos\theta}{\sin^2\theta}
\frac{\partial}{\partial\varphi_2}
-2\ctg\theta\frac{\partial^2}{\partial\varphi_1\partial\theta}
+2\frac1{2\sin\theta}
\frac{\partial^2}{\partial\varphi_2\partial\theta}\Big)
\nonumber\\
&& + \quad
\sin2\varphi_1\Big(\Delta_\theta-
\frac{\partial^2}{\partial\varphi^2_1}
-2\frac{\partial^2}{\partial\theta^2}\Big)\bigg],
\label{I.43}
\end{eqnarray}
where $\Delta_a$ and $\Delta_\theta$ are the Laplace operators
\begin{eqnarray}
&& \Delta_a\ =\ \frac{\partial^2}{{\partial} a^2}+{\ctg} a
\frac{\partial}{{\partial} a} +\frac1{{\sin}^2a}
\Big(\frac{\partial^2}{\partial\lambda^2}+{\cos} a
\frac{\partial^2}{\partial\lambda\,\partial\Omega_3} +\frac14
\frac{\partial^2}{\partial\Omega^2_3}\Big),
\nn \\
&& \Delta_\theta\ =\ \frac{\partial^2}{\partial\theta^2}+\ctg\theta
\frac{\partial}{\partial\theta}+\frac1{\sin^2\theta}
\Big(\frac{\partial^2}{\partial\varphi^2_1} -2\cos\theta
\frac{\varphi^2}{\partial\varphi_1\partial\varphi_2}
+\frac{\partial^2}{\partial\varphi^2_2}\Big) \label{I.44-45}
\end{eqnarray}
of the $O(3)$ group. The Laplace operator (\ref{I.43}) differs from
that calculated in Ref.~\refcite{10n} by the parametrization. They are
connected, however, by a unitary transformation.

\subsection{Calculation of the generators \boldmath$L_{ik}$ and
\boldmath$B_{ik}$}

To obtain the generators directly from $dz$, we have to invert a
$5\times5$ matrix in the case of a three-particle system. That
requires rather a long calculation, which is getting hopeless for a
larger number of particles. Instead of performing the
straightforward calculation, we get the wanted expressions in the
following way. Let us first consider $L_{ik}$, or rather one of its
components, {\ e.g.} $L_{12}$. We introduce a parameter $\sigma_{ik}$
which defines the motion along the particular trajectory which
corresponds to the action of the operator $L_{ik}$. Thus, formally
we can write
\begin{equation}
L_{12}\ =\ \frac12\left(iz_1\frac{\partial}{\partial z_2}
-iz_2\frac{\partial}{\partial z_1} +iz^*_1 \frac{\partial}{\partial
z^*_2} -iz^*_2\frac{\partial}{\partial z^*_1} \right) \equiv\
\frac{\partial}{\partial\sigma_{12}}\,. \label{I.46}
\end{equation}
Acting with $L_{12}$ on the vectors $z$ and $z^*$
\begin{equation}
L_{12} \left( \begin{array}{c}
z_1 \\ z_2\\ z_3 \end{array}\right) =\ \frac12
\left( \begin{array}{c} -iz_2\\ iz_1\\ 0 \end{array}\right) ,
\quad  L_{12}\left(\begin{array}{c}
z^*_1\\ z^*_2\\ z^*_3 \end{array} \right)=\ \frac12 \left(
\begin{array}{c} -iz^*_2\\ iz^*_1\\ 0  \end{array} \right)
\label{I.47}
\end{equation}
we see that $\sigma_{12}$ has to be imaginary. From (\ref{I.47}) we
get
\begin{eqnarray}
&& zL_{12}z\ =\ 0, \qquad z^*L_{12}z^*\ =\ 0,
\nn
\\
&& z^*L_{12}z\ =\ \frac i2\,(z\times z^*)_3\,,
\qquad
 lL_{12}z\ =\ -\frac i2\,(l\times z)_3\,.
\label{I.48-50}
\end{eqnarray}
Making use of the expression (\ref{I.30}) for $dz$, we can write
\begin{eqnarray}
&& L_{12}z\ =\ \frac{\partial z}{\partial\sigma_{12}}\ =\ -\frac
i2\,z \frac{d\lambda}{d\sigma_{12}} +\frac12\,e^{-i\lambda}(l\times
z^*) \frac{\partial a}{\partial\sigma_{12}}
-\Big(\frac{d\omega}{d\sigma_{12}\times z}\Big),
\nn  \\
&&L_{12}z^*\ =\ \frac{\partial z^*}{\partial\sigma_{12}}\ =\
 \frac i2\,
z^* \frac{d\lambda}{d\sigma_{12}}+ \frac12\,e^{i\lambda}(l\times
z^*) \frac{\partial a}{\partial\sigma_{12}}-\Big(
\frac{d\omega}{d\sigma_{12}} \times z^*\Big),
\label{I.51-52}
\end{eqnarray}
(here $-iz/2$ is $dz/d\lambda$ etc.).   We use Eq.
(\ref{I.48-50}). Substituting in Eq. (\ref{I.51-52})
\begin{eqnarray}
&&  (l\times z^*)\ =\ ie^{i\lambda/2}\Big(\cos\frac a2 l_-
+\sin\frac a2 l_+ \Big),
\nonumber\\
&& (l\times z)\ =\ -ie^{-i\lambda/2}\Big(\cos\frac a2\,l_+ -i\sin
\frac a2\,l_-\Big);
\nn
\\
&& z^2\ =\ ie^{-i\lambda}\sin a\,, \quad z^{*2}\ =\ -ie^{i\lambda}
\sin a\,,
\label{I.53-54}
\end{eqnarray}
we obtain from Eq. (\ref{I.48-50})
\begin{equation}
\frac{\partial a}{\partial\sigma_{12}}\ =\
\frac{d\lambda}{d\sigma_{12}}\ =\ 0.
\label{I.55}
\end{equation}

Similarly, (\ref{I.48-50}) gives
\be
\frac{d\Omega_3}{d\sigma_{12}}\ =\ -\frac i2\,l^{(3)}\,, \quad
  \frac{d\Omega_2}{d\sigma_{12}}\ =\ \frac i2\,l^{(3)}_2\,,
\quad  \frac{d\Omega_1}{d\sigma_{12}}\ =\ -\frac i2\,l^{(3)}_l\,,
\label{I.56-58}
\ee
where $l^{(k)}_i$ stands for the $k$-component of vector $l_i$.
Thus we obtain
\begin{equation}
L_{12}\ =\ -\frac i2\left[l^{(3)}_1
\frac{\partial}{\partial\Omega_1}+l^{(3)}_2
\frac{\partial}{\partial\Omega_2}+l^{(3)}_3
\frac{\partial}{\partial\Omega_3}\right] =\ -\frac
i2\,\frac{\partial}{\partial\omega_3}\,. \label{I.59}
\end{equation}
 and
\begin{eqnarray}
&& L_{23}\ =\ -\frac
i2\left[l^{(1)}_1\frac{\partial}{\partial\Omega_1}+l_2^{(1)}
\frac{\partial}{\partial\Omega_2}+l^{(1)}
\frac{\partial}{\partial\Omega_3}\right] =\ -\frac i2
\frac{\partial}{\partial\omega_1}\,,
\nn \\
&& L_{31}\ =\ -\frac
i2\left[l^{(2)}_1\frac{\partial}{\partial\Omega_1}
+l^{(2)}_2\frac{\partial}{\partial\Omega_2}+
l^{(2)}\frac{\partial}{\partial\Omega_3} \right] =\ -\frac i2
\frac{\partial}{\partial\omega_2}\,. \label{I.60-61}
\end{eqnarray}
Introducing the notations
\begin{equation}
L_1=2L_{23}\,, \quad L_2=2L_{31}\,, \quad L_3=2L_{12}\,,
\label{I.62}
\end{equation}
we can write the general expression for the angular momentum operator
\begin{equation}
L_k\ =\ -i\left[l^{(k)}_1 \frac{\partial}{\partial\Omega_1}
+l^{(k)}_2 \frac{\partial}{\partial\Omega_2} +l^{(k)}
\frac{\partial}{\partial\Omega_3} \right]. \label{I.63}
\end{equation}
It satisfies the commutation relations
\begin{equation}
[L_1,L_2]=-iL_3\,, \quad [L_2,L_3]=-iL_2\,, \quad [L_3L_1]=-iL_2\,.
\label{I.64}
\end{equation}
The square of the angular momentum operator is
\begin{equation}
L^2\ =\ \left(\frac{\partial^2}{\partial\Omega^2_1}
+\frac{\partial^2}{\partial\Omega^2_2}
+\frac{\partial^2}{\partial\Omega^2_3}\right) =\ \Delta_\theta\,.
\label{I.65}
\end{equation}

Let us now turn our attention to the operator $B_{ik}$. We consider
\begin{equation}
B_{12}\ =\ \frac12\Big(iz_1\frac{\partial}{\partial z_2}
+iz_2\frac{\partial}{\partial z_1} -iz^*_1\frac{\partial}{\partial
z^*_2}-iz^*_2\frac{\partial}{\partial z^*_1}\Big) \equiv\
\frac{\partial}{\partial\beta_{12}}\,. \label{I.66}
\end{equation}
From the action of $B_{12}$ on $z$ and $z^*$
\begin{equation}
B_{12}\left(\begin{array}{c}
z_1 \\ z_2\\ z_3 \end{array} \right)=\ \frac12
\left(\begin{array}{c} iz_2\\ iz_1\\ 0 \end{array} \right), \quad
B_{12}\left(\begin{array}{c} z^*_1 \\ z^*_2 \\ z^*_3 \end{array}\right)
=\ \frac12 \left(\begin{array}{c} -z^*_2\\ -z^*_1\\ 0 \end{array}
\right),
\label{I.67}
\end{equation}
it is obvious, that $\beta_{12}$ is real. We make use of the conditions
\begin{eqnarray}
&&  zB_{12}z\ =\ iz_1z_2\,, \qquad z^*B_{12}z^*\ =\ -iz^*_1z^*_2\,,
\label{I.68-70} \nn \\
&& z^*B_{12}z\ =\ \frac i2 (z^*_1z_2+z_1z^*_2)\,,
\qquad
lB_{12}z\ =\ \frac i2(l^{(1)}z_2 +l^{(2)}z_1)
\end{eqnarray}
and of (\ref{I.30}) and (\ref{I.53-54}).

Let us introduce the notation
\begin{equation}
b^{(lm)}_{ik}\ =\ \frac12\Big(l^{(l)}_i l^{(m)}_k +l^{(m)}_i
l^{(l)}_k\Big).
\label{I.71}
\end{equation}
Then, following a procedure similar to that in the case of $L_{ik}$, we
obtain from Eq.~(\ref{I.68-70})
\begin{eqnarray}
&& \frac{da}{d\beta_{12}}\ =\ b^{(12)}_{11} - b^{(12)}_{22}\,,
\label{I.72-72} \nn \\
&& \frac{d\lambda}{d\beta_{12}}\ =\ \Big(b^{(12)}_{11}
+b^{(12)}_{22}\Big)- 2b^{(12)}_{12}\, \frac1{\sin a}\,.
\end{eqnarray}
Equations (\ref{I.68-70}) lead to
\begin{eqnarray}
&& \frac{d\Omega_1}{d\beta_{12}}\ =\ -b^{(12)}_{23} {\tg}\, a
 - b^{(12)}_{13} \frac1{\cos a}\, ,
\nn \\
&& \frac{d\Omega_2}{d\beta_{12}}\ =\ -b^{(12)}_{23}\frac1{\cos a}
-b^{(12)}_{13} {\tg}\, a\,,
\nn \\
&& \frac{d\Omega_3}{d\beta_{12}}\ =\ -b^{(12)}_{12} {\ctg} a\,.
\label{I.74-76}
\end{eqnarray}
Thus the expression for $B_{12}$ can be written as
\begin{eqnarray}
B_{12} &=&
\Big(b^{(12)}_{11}-b^{(12)}_{22}\Big)\frac{\partial}{\partial a}
-\Big(b^{(12)}_{11}+b^{(12)}_{22}\Big)
\frac{\partial}{\partial\lambda} -2b^{(12)}_{12} \left(\frac1{\sin
a}\frac{\partial}{\partial\lambda}+\frac12 {\ctg}
a\frac{\partial}{\partial\Omega_3}\right)
\nonumber\\
&-&\ \left(b^{(12)}_{13} \frac1{\cos a} +b^{(12)}_{23}{\tg}\, a\right)
\frac{\partial}{\partial\Omega_1}-\left(b^{(12)}_{13} {\tg}\, a
+b^{(12)}_{23}\frac1{\cos a}\right)
\frac{\partial}{\partial\Omega_2}\,. \label{I.77}
\end{eqnarray}
The generator $B_{ik}$ of the deformation group of the triangle
obtains the form
\begin{eqnarray}
B_{ik} &=&\Big(b^{(ik)}_{11}-b^{(ik)}_{22}\Big)
\frac\partial{\partial a} -\Big(b^{(ik)}_{11}+b^{(ik)}_{22}\Big)
\frac\partial{\partial \lambda} -2b^{(ik)}_{12} \left(\frac1{\sin a}
\frac\partial{\partial\lambda} +\frac12 {\ctg}
a\frac\partial{\partial\Omega_3}\right)
\nonumber\\
&-&\ b^{(ik)}_{13}\left({\tg}\,
a\frac\partial{\partial\Omega_2}+\frac1{\cos a}
\frac\partial{\partial\Omega_1}\right) -b^{(ik)}_{23} \left({\tg}\, a
\frac\partial{\partial\Omega_1}+\frac1{\cos\, a}
\frac\partial{\partial\Omega_2}\right). \label{I.78}
\end{eqnarray}
Acting in the space of polynomials which include only $z$ (and not
$z^*$), the following identity appears:
\begin{equation}
ie^{i\alpha}\,\frac\partial{\partial\Omega_1}\ =\
\frac\partial{\partial\Omega_2}\,. \label{I.79}
\end{equation}
Thus in the space of polynomials of $z$ $B_{ik}$ might be written as
\begin{eqnarray}
B_{ik} &=& \Big(b^{(ik)}_{11}-b^{(ik)}_{22}\Big)
\frac\partial{\partial a} -\Big(b^{(ik)}_{11}+b^{(ik)}_{22}\Big)
\frac\partial{\partial\lambda}
+2\frac\partial{\partial\lambda}\delta_{ik}\
\nonumber\\
&-&\ 2b^{(ik)}_{12}\left(\frac1{\sin
a}\frac\partial{\partial\lambda} +\frac12 {\ctg}
\,a \frac\partial{\partial\Omega_3}\right) -ib^{(ik)}_{23}
\frac\partial{\partial\Omega_1} +ib^{(ik)}_{13}
\frac\partial{\partial\Omega_2}\,. \label{I.80}
\end{eqnarray}

Let us present also the permutation expressions
\begin{eqnarray}
&& [B_{ik},B_{jl}]\ =\ \frac12(L_{il}\delta_{kj}-L_{jk}\delta_{il})
+\frac i2(L_{ij}\delta_{kl}-L_{lk}\delta_{ij})\,,
\nn \\
&& [B_{ik},L_{jl}]\ =\ \frac i2(B_{il}\delta_{kj}-B_{jk}\delta_{il})
-\frac i2 (B_{ij}\delta_{kl}- B_{ik}\delta_{ij}) \,.
\label{I.81-82}
\end{eqnarray}
In particular,
\begin{eqnarray}
&& [B_{12},B_{11}]=-iL_{12}\,, \quad [B_{12},B_{22}]=iL_{12}\,, \quad
[B_{11},L_{12}]=iB_{12}\,,\nn
\\
&& [B_{22},L_{12}]=-iB_{12}\,, \qquad [B_{12},L_{12}]=-\frac i2
(B_{11}-B_{22}\,.
\end{eqnarray}
From this it follows that $L_{12}$, $B_{12}$ and
1/2$(B_{11}-B_{22})$ form the $SU(2)$ subgroup.

\subsection{The cubic operator \boldmath$\Omega$}

Operators $H_+$ and $H_-$ are the usual raising and lowering operators
in $SU(2)$ taken at the value of the second Euler angle
$-2\Omega_3=2\varphi_1=0$
\begin{eqnarray}
&& H_+\ =\ \frac1{\sqrt2}\left[\frac\partial{\partial
a}+i\frac1{\sin a} \frac\partial{\partial\lambda} +\frac i2{\ctg} a
\frac\partial{\partial\Omega_3}\right],
\label{I.83} \nn \\
&& H_-\ =\ \frac1{\sqrt2}\left[\frac\partial{\partial
a}-i\frac1{\sin a} \frac\partial{\partial\lambda}-\frac i2{\ctg}
a\frac\partial{\partial\Omega_3}\right],
\end{eqnarray}
$\Omega$ can be written in the form
\begin{eqnarray}
\Omega &=& \sum_{i,j,k} L_{ij}L_{jk}B_{ki}=\ -\frac14\Bigg\{ \sqrt2
\Big( -\frac{\partial^2}{\partial\Omega^2_+} H_+
+\frac{\partial^2}{\partial\Omega^2_-}H_-\Big)
+\frac{\partial^2}{\partial\Omega^2_3}
\frac\partial{\partial\lambda}
+\Delta_\theta\frac\partial{\partial\lambda}
\label{I.85}\nn \\
&-& \frac1{\cos
a}\Big(\Delta_\theta-\frac{\partial^2}{\partial\Omega^2_3}
+\frac12\Big) \frac\partial{\partial\Omega_3} +{\tg}\, a \bigg[i\Big(
\frac{\partial^2}{\partial\Omega^2_+}
-\frac{\partial^2}{\partial\Omega^2_-}\Big)
\frac\partial{\partial\Omega^2}- \frac32
\Big(\frac{\partial^2}{\partial\Omega^2_+}
+\frac{\partial^2}{\partial\Omega^2_-} \Big)\bigg]\Bigg\}. \nn \\
\end{eqnarray}
The operator $\Omega$ has a simple meaning in the classical
approximation. Changing the derivative to the velocity and denoting
$\xi=p$ and $\eta=q$, we obtain

\begin{equation}
\frac12\,\Omega\ =\ (\xi L)(qL) - (\eta L)(pL)\,.
\label{I.86}
\end{equation}
The time derivative of this operator is, obviously, zero. If we
direct the axis $z$ along $L$ and introduce two two-dimensional
vectors in the space of permutation
\begin{equation}
x\ =\ (\xi_z\,, \eta_2)\ \mbox{ and  }\ y\ =\ (p_z\,,q_z)\,,
\label{I.87}
\end{equation}
then (I.86) can be written as
\begin{equation}
\label{I.88}
\frac12\,\Omega\ =\ (x\times y)_3\, .
\end{equation}
The operator has the form of the third component of the momentum in
the permutation space. Hence, the symmetry of the problem becomes
obvious: it is spherical in the coordinate space, and axial in the
permutation space.

We do not need the eigenvalues of this operator at small $K$ and
$\nu$ values when the degeneracy is small. Indeed, at a given $K$
and $\nu$ the number of states is determined by the usual $SU(3)$
expression
\begin{equation}
n(K,\nu)\ =\ \frac18\ (K+2)(K+2-2\nu)(K+2+2\nu)\,.
\label{I.89}
\end{equation}
Summing up this formula over $2\nu$ from $-K$ to $K$, we obtain the
well-known expression \cite{9n}:

\begin{equation}
n(K)\ =\ \frac{(K+3)(K+2)^2(K+1)}{12}\,.
\label{I.90}
\end{equation}
The terms for maximal degeneracies go from $\nu=0$ in the case of
even $K$ or from $\nu=1/2$ in the case of odd $K$.

\begin{equation}
n(K,0)\ =\ \left\{ \begin{array}{ll}
\displaystyle\frac 18\,(K+2)^3\,, & K~~-~~ \mbox{ odd },
\\
\displaystyle\frac 18\,(K+1)(K+2)(K+3)\,, & K~~-~~ \mbox{ even }.
\end{array} \right.
\label{I.91}
\end{equation}

\subsection{Solution of the eigenvalue problem}

\begin{equation}
\Phi^L_M\ =\ \sum_\lambda \sum^\lambda_{M'=-\Lambda} a(\Lambda,M')
a(\Lambda,M')\,D^\Lambda_{\nu,M'}(\lambda,a,0)\,D^L_{2M',M}
(\varphi_1,\theta,\varphi_2)\,.
\label{I.92}
\end{equation}

Let us finally consider a few special cases of the solution. As it
is discussed by Dragt \cite{16n}, in the low-dimensional representations
of $SU(3)$ $(L=0,1)$ the $\Omega$ is not needed. Indeed, in the case
of $L=0$ the Laplace operator obtains the form
\begin{equation}
\Delta\ =\ \frac{\partial^2}{\partial a^2}+2{\ctg}2a
\frac\partial{\partial a}
+\frac1{\sin^2a}\frac{\partial^2}{\partial\lambda^2}\,. \label{I.93}
\end{equation}

Obviously, the eigenfunction will be the following
\begin{equation}
\Phi_0\ =\ D^\Lambda_{\nu,0} (\lambda,a,0)\,,
\label{I.94}
\end{equation}
which obeys the equation
\begin{equation}
\Delta\Phi_0\ =\ -\Lambda(\Lambda+1)\Phi_0\,, \qquad
\lambda\ =\ 0,1,\ldots\ .
\label{I.95}
\end{equation}
This solution demonstrates clearly the $SU(2)$ nature of a
non-rotating triangle.

In the case of $L=1$ the solutions are
\begin{eqnarray}
 z_M &=& \sum_{M'=\pm1/2} D^{1/2}_{1/2,M'}(\lambda,a,0)\,
D^1_{2M',M} (\varphi_1,\theta,\varphi_2)\,,
\label{I.96}\nn \\
z^*_M &=& D^{1/2}_{-1/2,-1/2}(\lambda,a,0)\,D^1_{-1,M}
(\varphi_1\theta\varphi_2) -D^{1/2}_{-1/2,1/2}(\lambda,a,0)\,
D^1_{1,M}(\varphi_1\theta\varphi_2)\,,
\quad\quad
\end{eqnarray}
fulfilling the Laplace equation with the value $K=1$. Simultaneously
$z_M$ obeys the equations
\begin{eqnarray}
&& L^2z_M=-2z_M\,, \quad L_3z_M=-Mz_M\,, \quad M=-1,0,1\,,
\label{I.97-99}\nn \\
&& \Delta_a z_M\ =\ -\frac34\,z_M\,, \quad Nz_M\ =\ \frac12\,z_M\,,
\quad \Omega z_M\ =\ -\frac34\, iz_M\,,
\end{eqnarray}
 and, accordingly, $z^*$ obeys

\begin{eqnarray}
&& L^2z^*_M=-2z^*_M\,, \quad L_3z^*_M=-Mz^*_M\,, \quad
\Delta_\alpha z^*_M=-\frac34\,z^*_M\,,
\nonumber\\
&& N\,z^*_M\ =\ -\frac12\,z^*_M\,, \qquad
\Omega z^*_M\ =\ \frac34\,iz^*_M\,.
\label{I.100}
\end{eqnarray}

\section{Eigenfunctions in the Three-Body Problem}

The investigation of the three-particle system leads to the
construction of basis functions in the form of the so-called
K-polynomials, {\it i.e.} harmonic polynomials in the
six-dimensional space. In order to make it possible to work with
such functions which would be a natural generalization of the usual
spherical functions given on the two-dimensional sphere, it is
necessary to find the total system of solutions for the Laplace
equation on the five-dimensional sphere.

In the previous section a method for calculating the generators of
the group of motion on the five-dimensional sphere was found, and
the corresponding system of commuting operators was constructed.

A somewhat unexpected difficulty of the task is due to the fact that
the functions realize a representation of the permutation group of
three particles and at the same time they are eigenfunctions of the
operator of the momentum. If we do not require a permutation
symmetry of the eigenfunction, then, obviously, the problem can be
easily solved. In this case  the simplest way of finding the
solution is via the ``tree''-function method \cite{11n}. The obtained
eigenfunctions (the basis functions) can be characterized by five
quantum numbers:
\be K,\ j_1,\ M_1,\ j_2,\ M_2\,,
\ee
where $K$ -- is the general order of the polynomial,
$j_1,M_1,j_2,M_2$ are the momenta and their projections
corresponding to $\xi$ and $\eta$. Instead of $j_1$ and $j_2$, we
can, of course, introduce the total momentum $J$.

In Ref.~\refcite{1n} a system of functions was built up with certain
permutation symmetries. This system was characterized by the quantum
numbers
 \begin{equation}
 K,\ J,\ M,\ \nu,\ \Omega\,.
 \label{3.2}
 \end{equation}
The last two of them do not coincide with the quantum numbers
corresponding to the ``tree''. The general solution is of the form
 \begin{equation}
 \Phi^J_M,\nu\ =\ \sum a_\nu(\Lambda,M')D^\Lambda_{\nu,M'}(\lambda,a,0)\,D^J_{2M',M}
 (\varphi_1,\theta,\varphi_2)\,.
 \label{3.3}
 \end{equation}
The meaning of the variables will be explained below; the
coefficients $a_\nu(\Lambda,M')$ had to be determined so that the
function (\ref{3.3}) satisfied the Laplace equation on the
five-dimensional sphere and also the equation for the eigenvalues of
the operator $\Omega$. Although the set of equations is simple
enough to solve it in any concrete case, we were not able to find a
general solution.

Let us try to solve the problem in a different way. We will carry
out a transition from the total set of functions constructed by the
``tree''-method to the $K$-harmonics. In this approach the
``tree''-functions will first be transformed into a system with a
given total momentum, {\it i.e.} to that with the quantum numbers
\begin{equation}
 K,\,J,\,M,\,j_1,j_2\,.
 \label{3.4}
 \end{equation}
After that we make a transition to the quantum numbers
 \begin{equation}
  K,\,J,\,M,\,\nu,\,(j_1,j_2)\,.
 \label{3.5}
 \end{equation}
We shall see that this can be done by a simple Fourier
transformation. In fact the pair $(j_1, j_2)$ is not a real quantum
number in the sense that functions with different $(j_1, j_2)$ pairs
do not form an orthonormal system, just remind the genealogy of the
functions. Let us underline here that $j_1$ and $j_2$ ceased to be
eigenvalues after we turned to the Fourier components.

In order to solve our problem, it remains to construct the sum
 \begin{equation}
 C_{(j_1j_2)}\,\Psi^{(j_1j_2)}_{KJM}\,,
 \label{3.6}
 \end{equation}
where $(j_1j_2)$ run through all values of momentum pairs which can
be used for building the total momentum $J\le j_1+j_2\le K$.

We will demonstrate here how the set of functions (\ref{3.5}) can be
produced.

So far we were not able to find the set of eigenfunctions of the
operator $\Omega$ in a closed form. In Ref.~\refcite{12n} an algorithm
was given for calculating such functions in the form of series.
However, the order of the corresponding equation grows with the growth
of the eigenvalue, and, hence, the problem of obtaining the solution in
a general form remains open.

The calculations below lead to such a form of the ``tree''-functions
for which the Fourier transformation becomes simple; formally, we
just show how to carry out the Fourier transformation of the
``tree''-function.

 \subsection{Coordinates and parametrization}

Let us remind the determination of the coordinates used in the
previous section. The three radius vectors $x_i$ $(i=1,2,3)$ which
are connected by the condition $ x_1+x_2+x_3\ =\ 0$ form the the
Jacobi-coordinates $\xi$ and $\eta$ for equal masses:
 \bea
 \label{3.7-9}
 &&
 \xi\ =\ -\sqrt{\frac32}\, (x_1+x_2)\,, \quad \eta\ =\
 \sqrt{\frac12}\, (x_1-x_2)\,,\quad
  \xi^2 +\eta^2\ =   \ \rho^2\,,\nn \\
   &&
 z\ =\ \xi+i\eta\,, \quad z^*\ =\ \xi-i\eta\,.
 \eea
Here $\rho$ is the radius of a five-dimensional sphere; for
simplicity, we suppose it to be unity.

Let us consider a triangle in the vertices of which three particles
are placed. The situation of this triangle in the space is
determined by the complex vectors ${\vec l}_+$ and ${\vec l}_-$
which, together with the third vector ${\vec l}_0 =
[{\vec l}_+ \times {\vec l}_-]$,
form a moving coordinate system. They satisfy the usual
conditions
 \begin{equation}
 l^2_+\ =\ l^2_-\ =\ 0\,, \qquad ({\vec l}_+{\vec l}_-) = 1\,
 \label{3.10}
 \end{equation}
The vectors ${\vec l}_+$ and ${\vec l}_-$ are related to $z$ and
$z^*$ by the expressions
 \begin{eqnarray}
 && z\ =\ e^{-i\lambda/2}\Big(\cos \frac a2\,l_+ +i\sin\frac a2\,l_-\Big),
 \nonumber\\
 &&  z^*\ =\ e^{i\,\lambda/2}\Big(\cos \frac a2\,l_- -i\sin\frac a2\,l_+\Big).
 \label{3.11}
 \end{eqnarray}
Here $\lambda$ and $a$ define the form of the triangle with the
accuracy up to the similarity transition (which we will not
consider, taking the length of the 6-vector $\rho = const$). In
addition, as it was demonstrated in Ref.~\refcite{3n}, the variable $a$
determines the relation between the momenta of inertia of the
triangle.

In the following it will be convenient to return to $\xi$ and $\eta$
and connect them to the coordinates ${\vec l}_+$ and ${\vec{l}}_-$:
 \begin{equation}
 \xi\ =\ \frac12\Big(ul_+ + u^*l_-\Big), \quad \eta\ =\ -
 \frac i2\Big(vl_+-v^*l_-\Big).
 \label{3.12}
 \end{equation}
Here
 \begin{eqnarray}
u=e^{-i\lambda/2} \cos\frac a2 - ie^{i\lambda/2} \sin\frac a2\,, &&
u^*=e^{i\lambda/2} \cos\frac a2 +ie^{-i\lambda/2} \sin\frac a2\,,
 \nonumber\\
 v=e^{-i\lambda/2} \cos\frac a2 +ie^{i\lambda/2} \sin\frac a2\,, &&
 v^*=e^{i\lambda/2} \cos\frac a2 -ie^{-i\lambda/2} \sin\frac a2\,.
 \label{3.13}
 \end{eqnarray}
In these formulae the Euler angles which define the position of the
triangle and the coordinates determining its deformation are
explicitly separated. Note that these expressions are not just
products of the functions of the Euler angles and the functions of
the coordinates connected with the deformation, but sums of the
products of these functions; this in fact reflects the relation of
the deformation and the rotation.

Why the introduction of these coordinates makes sense becomes clear
if we write the expressions for $\xi$ and $\eta$ in the form
\begin{eqnarray}
&& \xi\ =\ \left(\frac12\,uu^*\right)^{1/2} \frac1{\sqrt2}\,
\bigg(e^{i\psi_1}l_+ +e^{-i\psi_1}l_-\bigg)\,,
\label{3.14-15} \nn
\\
&& \eta\ =\ \left(\frac12\,vv^*\right)^{1/2} \frac1{\sqrt2}\,
\bigg(e^{i\psi_2}l_+ + e^{-i\psi_2}l_-\bigg)\,.
\end{eqnarray}
Here we have introduced the phases $\psi_1$ and $\psi$
\begin{equation}
 u=\rho_1 e^{i\psi_1}\,, \qquad v=\rho_2 e^{i\psi}\,,
 \label{3.16}
\end{equation}
from which it follows that
\be
\label{3.17}
\frac u{u^*}\ =\ e^{2i\psi_1}, \qquad \frac v{v^*}\ =\ e^{2i\psi}\,,
\quad
\psi_2\ =\ \psi-\frac\pi2\,,
\ee
while
\begin{equation}
\Theta\ =\ \psi_1-\psi_2
\label{3.18}
\end{equation}
is the angle between the vectors $\xi$ and $\eta$:
\begin{equation}
\xi\eta\ =\ |\xi\|\eta|\cos\Theta\,.
\label{3.19}
\end{equation}
Making use of the relations
\begin{equation}
\xi^2\ =\ \frac12\,uu^*, \qquad \eta^2\ =\ \frac12\,vv^*\,,
\label{3.20}
\end{equation}
we can express the angle $\Theta$ in terms of our variables:
\begin{equation}
\cos\Theta\ =\ \frac{\cos\lambda\sin a}{\sqrt{1-\sin^2\lambda \sin^2
a}}\,.
\label{3.21}
\end{equation}
The irrational connection between the angle $\Theta$ and the angles
$a$ and $\lambda$ forces us to find different ways for the Fourier
transformation.

Let us introduce the unit vectors $n$ and $m$ defined as
\begin{equation}
n\ =\ \frac\xi{|\xi|}\,, \qquad m\ =\ \frac\eta{|\eta|}\,.
\label{3.22}
\end{equation}
From (\ref{3.14-15})  it is clear that
\begin{eqnarray}
n &=& \frac1{\sqrt2}(e^{i\psi_1}l_+ + e^{-i\psi_1}l_- )\,,
\nn
\\
m &=& \frac1{\sqrt2} (e^{i\psi_2}l_+ -e^{-i\psi_2}l_-)\,.
\label{3.23-24}
\end{eqnarray}
It is reasonable to re-write these expressions in the form
\begin{eqnarray}
&& n\ =\ D^1_{01}\Big(0,\frac\pi2,0\Big) (e^{i\psi_1}l_+
+e^{-i\psi_1}l_- )\,,
\nn \\
&& m\ =\ D^1_{01}\Big(0,\frac\pi2,0\Big) (e^{i\psi_2}l_+
+e^{-i\psi_2}l_-)
\label{3.25-26}
\end{eqnarray}
or, for the components of $n$ and $m$,
\begin{eqnarray}
&& n^{(M_1)}\ = \sum_{\mu_1} D^1_{0\mu_1}\Big(0,\frac\pi2,\psi_1\Big)
D^1_{\mu_1M_1}(l_+l_-)\,,
\nn
\\
&& m^{(M_2)}\ =\sum_{\mu_2} D^1_{0\mu_2}\Big(0,-\frac\pi2,\psi_2\Big)
D^1_{\mu_2M_2}(l_+l_-)\,.
\label{3.27-28}
\end{eqnarray}
Let us remind that the components $l_+$ and $l_-$ can be expressed
with the help of the Wigner $D$-functions
\begin{equation}
\label{3.29}
D^1_{mn} (\varphi_1,\theta,\varphi_2)\ =\ l^{(n)}_m\,.
\end{equation}
In fact $D(\varphi_1,\theta,\varphi_2)\equiv D(l_+l_-)$, and we have
introduced this somewhat unusual notation in order to demonstrate
that the Euler angles determine the position of the trihedron given
by the unit vectors ${\vec I}_1$, ${\vec I}_2$, ${\vec I}_3$. The
expression (\ref{3.27-28}) describes the rotation which transforms
the trihedron ${\vec I}_1$, ${\vec I}_2$, ${\vec I}_3$ to one
defined by the unit vector ${\vec n}$ and two perpendicular vectors
${\vec n}_1$, ${\vec n}_2$. As a result of this rotation the vector
${\vec n}$ turns out to be on the ${\vec I}_1$, ${\vec I}_2$ plane.
Similarly, formula (\ref{3.27-28}) gives a description of turning to
a trihedron defined by the unit vectors ${\vec m}_1$, ${\vec m}_2$,
${\vec m}_3$. Thus the expressions (\ref{3.27-28}) can be considered
as formulae for transforming first order Legendre polynomials (of
vectors ${\vec m}$ and ${\vec m}$) which can be generalized to
arbitrary polynomials.

Hence, for the polynomial build up from the unit vector ${\vec n}$,
the following expression can be written:
\begin{equation}
D^{j_1}_{0M_1}(n)\ =\sum^{j_1}_{\mu_1=-j_1} D^{j_1}_{0\mu_1}
\Big(0,\frac\pi2,-\psi_1\Big) D^{j_1}_{\mu_1M_1}
(\varphi_1,\theta,\varphi_2) \label{3.30}
\end{equation}
For the polynomial given by ${\vec m}$ we have, respectively:
\begin{equation}
D^{j_2}_{0M_2}(m)\ =\sum^{j_2}_{\mu_2=-j_2} D^{j_2}_{0\mu_2}
\Big(0,\frac\pi2,-\varphi_2\Big) D^{j_2}_{\mu_2M_2}
(\varphi_1,\theta,\varphi_2)\,. \label{3.31}
\end{equation}
These formulae can be transformed in such a way that
$D^{j_1}_{0M_1}$ and $D^{j_0}_{0M_2}$ turn out to be the functions
of the same arguments:
\begin{eqnarray}
D^{j_1}_{0M_1}(n)&=&
\sum_{\mu_1}\exp\left[i\mu_1\frac{(\psi_1-\psi_2)}2 \right]
D^{j_1}_{0\mu_1}\left(0,\frac\pi2,-\Big(\frac{\psi_1+\psi_2}2\Big)
\right) D^{j_1}_{\mu_1M_1}(\varphi_1,\theta,\varphi_2),
\nn \\
D^{j_2}_{0M_2}(m)&=&\sum_{\mu_2} \exp\left[-i\mu_2
\frac{(\psi_1-\psi_2)}2\right] D^{j_2}_{0\mu_2}\left(0,\frac\pi2,
-\Big(\frac{\psi_1+\psi_2}2\Big)\right)D^{j_2}_{\mu_2M_2}
(\varphi_1,\theta,\varphi_2). \label{3.32-33}  \nn \\
\end{eqnarray}
For the three-body problem one more restriction has to be
introduced. Since there is a definite reflection symmetry with
respect to the moving plane of ${\vec I}_+$ and ${\vec I}_-$,
$\mu_1$ and $\mu_2$ have to be either only even, or only odd. For
example, in the sums (\ref{3.27-28}) there can be only
$\mu_{1,2}=\pm 1$.

We have introduced above quite a diversity of parameters and
coordinates, and carried out a lot of transformations which, so far,
may seem to be superfluous and somewhat artificial. In fact, as we
will see, they simplify the calculations of the Fourier coefficients
of the polynomials.

\subsection{The case of \boldmath $J=0$ }

For states with a total momentum $J=0$ the solution can be easily
obtained. From the equation for this case we got the solution in
the form
\begin{equation}
D^{K/4}_{\nu/2,-\nu/2} (2\lambda,2a,0)\,.
\label{3.34}
\end{equation}
The order $K/4$ of the $D$-function corresponds to the order $K$ of
the polynomial, since the latter is determined by the trigonometric
functions of the argument $a/2$, and so each trigonometric function
of $2a$ increases the degree by four units.

The structure of (\ref{3.34}) can be understood without considering
the equation. Indeed, the polynomial at $J=0$ can not contain
vectors ${\vec l}_+$ and ${\vec l}_-$, and, hence, has to be the
function of $z^2$ and $z^{*2}$. We know that
\begin{eqnarray}
&&  z^2\ =\ i\sin ae^{-i\lambda}\,,
\nonumber\\
&&  z^{*2}\ =\ -i\sin ae^{i\lambda}\,.
\label{3.35}
\end{eqnarray}
Having a look at the tables for $D$-functions it becomes clear that
\begin{eqnarray}
&&  z^2\ =\ D^{1/2}_{-1/2,1/2} (2\lambda,2a,0)\,,
\nonumber\\
&&  z^{*2}\ =\ -D^{1/2}_{1/2,-1/2} (2\lambda,2a,0)\,,
\label{3.36}
\end{eqnarray}
and that the sum of the lower indices is zero for each $D$-function.
When constructing harmonic polynomials from functions (\ref{3.36}),
this feature, obviously, remains valid also for higher order
$D$-functions (which is a reason for (\ref{3.34}). Consequently, a
specific property of our problem is the absence of the diagonal
elements
\begin{equation}
D^{1/2}_{1/2,1/2}\,, \quad D^{1/2}_{-1/2,-1/2}
\label{3.37}
\end{equation}
in the basis. The interesting task of the expansion of function
(\ref{3.34}) over the ``tree''-functions, {\it i.e.} over functions
with pair angular momenta arises; the state of the system at $J=0$
is described as a rotation of vectors $\xi$ and $\eta$ in opposite
directions, with different momenta. This leads to a connection with
the theory of Clebsch-Gordan coefficients,
\begin{equation}
\left( \begin{array}{ccc} K/4 & K/4 & j\\  \nu/2 & -\nu/2 & 0
\end{array} \right).
\label{3.38}
\end{equation}
Let us note that the amplitudes of states with momenta $j_1=j_2=j$
turn out to be proportional to the Wigner coefficient, the
respective calculations are given below.

\subsubsection{States with momenta $j_1=j_2=j$
and the Wigner coefficients}

To obtain the contribution of the partial momenta in the $J=0$
state, we have to calculate the Fourier coefficient of the
function
\begin{equation}
\Phi_0(\xi,\eta)\ =\ (\cos\Phi)^J(\sin\Phi)^j\,
P^{(j+1/2,j+1/2)}_{K/2-j}(\cos2\Phi)P_j(n,m)\,. \label{3a.1}
\end{equation}
As it was said already, at  $\xi^2+\eta^2=1$ we suppose
$\cos^2\Phi=\xi^2$ and $\sin^2\Phi=\eta^2$, and re-write
(\ref{3a.1}) in the form

\begin{equation}
\Phi_0(\xi,\eta)\ =\ (\xi^2)^{j/2}P^{(j+1/2,j+1/2)}_{K/2-j}
(\xi^2-\eta^2)P_j(n,m)\,.
\label{3a.2}
\end{equation}
A state with zero angular momentum can be constructed from two
partial momenta $j_1$ and $j_2$ which are equal to each other
($j_1=j_2=j$). Since for such a state the quantum number $\Omega$
plays no role, it differs from the states presented above only by
the substitution of $\nu$ by $j$. Hence, the function (\ref{3a.2})
is a superposition
\begin{equation}
\sum_\nu C(j,\nu)\,D^{K/4}_{\nu/2,-\nu/2}(2\lambda,2a,0)\,.
\label{3a.3}
\end{equation}
The coefficient $C(j,\nu)$ has to be calculated. The Fourier
coefficient of (\ref{3a.2}) will be obtained having an additional
condition, ${\vec m}{\vec n}=1$; this corresponds to $a=\pi/2$ in
(\ref{3.21}). On the other hand,
\[
\cos2\Phi\ =\ \sin a\sin\lambda\,,
\]
which, if $a=\pi/2$, gives

\begin{equation}
\cos2\Phi\ =\ \sin\lambda, \qquad \sin2\Phi\ =\ \cos\lambda\,.
\label{3a.4}
\end{equation}
In order to be able to use the standard formulae, we take $\sin
\lambda = \cos \Lambda$, and change from the Jacobi polynomial to
the Gegenbauer polynomial:
\begin{equation}
P^{(j+1/2,j+1/2)}_{K/2-j}(\cos2\Phi)\ =\
\frac{\Gamma(2j+2)\Gamma(K+3/2)}{\Gamma((K/2)+j+2)\Gamma(j+3/2)}\,
C^{j+1}_{K/2-j}(\cos2\Phi)\,.
\label{3a.5}
\end{equation}
If so, (\ref{3a.1}) can be re-written in the form
\begin{equation}
(\cos\Phi)^j(\sin\Phi)^J\ =\ \frac{\Gamma(2j+2)\Gamma(K+3/2)}{\Gamma(
(K/2)+j+2)\Gamma(j+3/2)}\,C^{j+1}_{K/2-j}(\cos2\Phi)\,.
\label{3a.6}
\end{equation}
Let us make use now of the integral representation of the Gegenbauer
polynomial \cite{16n}:
\begin{eqnarray}
\frac1{2^j}(\sin\lambda)^j C^{j+1}_{K/2-j}(\cos\lambda)&=&
\frac{j^j}{2^{2j+1}}\ \frac{\Gamma(2+j+(K/2))}{(K/2)!\Gamma(j+1)}\
\nn \\
& \times& \int\limits^\pi_0 (\cos\Lambda-i\sin\Lambda\cos\vartheta
)^{K/2}C^{1/2}_j(\cos\vartheta)\sin\vartheta d\vartheta. \qquad
\label{3a.7}
\end{eqnarray}
Taking into account $C^{1/2}_j(\cos\vartheta=P_j(\cos\vartheta)$, we
have
\bea
\Phi_0(\xi,\eta)
&=&\frac{i^j}{2^{2j+1}} \frac{\Gamma(2j+2)\Gamma(K+3/2)}{
(K/2)!\Gamma(j+1)\Gamma(j+3/2)}
\nn \\
&\times& \int\limits^\pi_0
(\cos\lambda-i\sin\Lambda\cos\vartheta)^{\frac K2}
P_j(\cos\vartheta)\sin\vartheta\,d\vartheta.
\label{3a.8}
\eea
The expansion into a series can be easily obtained directly, if we
first expand $\cos\Lambda$ and $\sin\Lambda$ in exponents and open
the brackets. Extracting the term with the exponent
$e^{-i\nu\Lambda}$, we get
\begin{eqnarray}
&& \hspace*{-1cm}
\left( {\frac K2 \atop \frac K4-\frac\nu2}\right) \int\limits^\pi_0
\Big(\frac{1-\cos\vartheta}2\Big)^{\frac{K/2-\nu}2}
\Big(\frac{1+\cos\vartheta}2\Big)^{\frac{K/2+\nu}2} P_j(\cos\vartheta)
e^{-i\nu\lambda} \sin\vartheta\,d\vartheta\ =
\nonumber\\
&& = \left({\frac K2 \atop \frac K4-\frac\nu2} \right)\int\limits^\pi_0
\Big(\sin\frac\vartheta2\Big)^{K/2-\nu}\Big(\cos\frac\vartheta2
\Big)^{K/2+\nu} P_j(\cos\vartheta)e^{-i\nu\Lambda}
\sin\vartheta\,d\vartheta.
\label{3a.9}
\end{eqnarray}
This can be re-written as

\begin{equation}
i^{-K/2}\int\limits^\pi_0 P^{K/4}_{K/4,\nu/2}(\cos\vartheta)
P^{K/4}_{-K/4,-\nu/2} (\cos\vartheta) P^j_{00}(\cos\vartheta)
e^{-i\nu\Lambda} \sin\vartheta\,d\vartheta\,,
 \label{3a.10}
\end{equation}
leading to the expression
\begin{eqnarray}
\Phi_0(\xi,\eta) &=& \sum_\nu\, \frac{i^{j-K/2}}{2^{2j}\sqrt{2j+1}} \,
\frac{\Gamma(2j+2)\,\Gamma(K+3/2)}{\Gamma(j+1)\,\Gamma(j+3/2)}\ \times
\nonumber\\
&&\times \left(\frac K4,\frac\nu2,-\frac\nu2\Big|j0\right)
e^{-i\nu\lambda}\left[\Big(\frac K2-j\Big)!\Big(\frac K2+j+1\Big)!
\right]^{-1/2}.
\label{3a.11}
\end{eqnarray}
If, instead of (\ref{3a.1}), we begin with the orthonormal function
\begin{equation}
\Phi^{j+1/2,j+1/2}_0 (\xi,\eta)\ =\ \left(N^{j+1/2,j+1/2}_{K/2-j}
\right)^{-1/2} \Phi_0(\xi,\eta)\,
\label{3a.12}
\end{equation}
where
\begin{equation}
N^{j+1/2,j+1/2}_{K/2-j}\ =\ \frac{[i\Gamma(K/2+3/2)]^2}{2(K+2)\,
\Gamma(K/2-j+1)\Gamma(K/2+j+2)}\,,
\label{3a.13}
\end{equation}
the calculation leads to the following expansion of the
``tree''-functions into $K$-harmonics:

\begin{eqnarray}
\Phi^{j+1/2,j+1/2}_0(\xi,\eta) &=& -\sum_\nu \frac{i^j}{2^{2j-1/2}}
\Big(\frac{K+2}{2j+1}\Big)^{1/2} \frac{\Gamma(2j+2)\,\Gamma(K+3/2)}{
\Gamma(j+1)\Gamma(j+3/2)\Gamma((K+3)/2)}\
\nonumber\\
&& \times\ \left(\frac K4,\frac\nu2,\frac K4,-\frac\nu2\Big|j0\right)
D^{K/4}_{\nu/2,-\nu/2}(2\lambda,2a,0)\,.
\label{3a.14}
\end{eqnarray}
Here we skipped the condition $a=\pi/2$. Obviously, the expansion of
the $K$-harmonics into the ``tree''-functions is determined by the
same coefficients as the expansion (\ref{3a.14}).

\subsection{The ``tree''-functions }

\begin{figure}[h]
\centerline{\epsfig{file=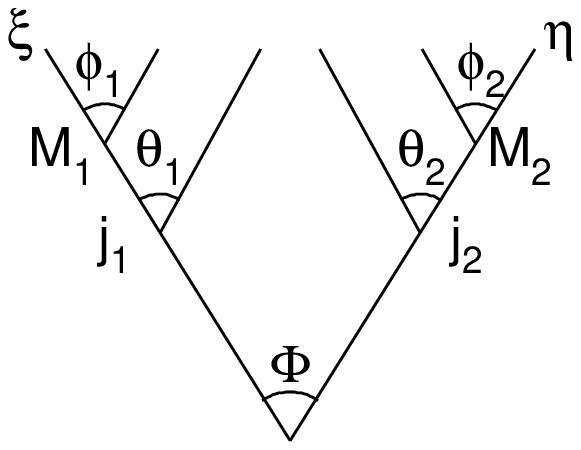,width=0.25\textwidth}}
\caption{"Tree"
\label{fig-tree}}
\end{figure}

The solution of the Laplace equation on the five-dimensional sphere
can be written in a coordinate system corresponding to the ``tree''
(see Fig.~\ref{fig-tree}).
In this system we presume
\begin{equation}
\xi\ =\ \cos\Phi n\,, \qquad \eta\ =\ \sin\Phi m\,,
\label{3.39}
\end{equation}
taking into account that $\xi^2+\eta^2=1$. The eigenfunction related
to this ``tree'' can be build up in the following way \cite{11n}. The
junction $n$ (where $2n=K-j_1-j_2$) corresponds to the function
\begin{equation}
(\cos\Phi)^{j_1}\,(\sin\Phi)^{j_2} P_n^{(j_1+1/2,j_2+1/2}(\cos2\Phi).
\label{3.40}
\end{equation}
Due to (\ref{3.39})
\be
\cos^2\Phi\ =\ \xi^2, \qquad \sin^2\Phi\ =\ \eta^2\,,
\ee
and, hence, the expression (\ref{3.40}) can be re-written in the
form
\begin{equation}
\xi^{j_1}\eta^{j_2} P^{(j_1+1/2,j_2+1/2)}_{(K-j_1-j_2)/2}(\xi\eta)\,,
\label{3.41}
\end{equation}
where
\begin{equation}
P^{(j_1+1/2,j_2+1/2)}_{K-j_1-j_2)/2}(\xi,\eta)=\!\sum^N_{m=0}\!
\Big({n+j_1+1/2 \atop m}\Big)\Big({n+j_2+1/2 \atop n-m}\Big)
(\xi^2)^m (\eta^2)^{n-m} (-1)^{n-m}.
 \label{3.42}
\end{equation}
The ``branches'' $\xi$ and $\eta$ characterized by $j_1M_1$ and
$j_2M_2$ correspond, accordingly, to the functions
\begin{eqnarray}
&& P_{j_2M_1}(n)\ =\ P^{M_1}_{j_1}(\vartheta_1)\,e^{-iM_1\varphi_1}\,,
\nonumber \\
&&
P_{j_2M_2} (m)\ =\ P^{M_2}_{j_2}(\vartheta_2)\,e^{-M_2\varphi_2}\,.
\label{3.43}
\end{eqnarray}
Multiplying the functions (\ref{3.42}) and (\ref{3.43}), we arrive
at an eigenfunction which corresponds to the ``tree'' as a whole:
\begin{equation}
\Phi(\xi,\eta)\ =\
P_J,M(\xi,\eta)\,P^{(j_1+1/2,\,j_2+1/2)}_{K-j_1-j_2)/2}(\xi\eta)\,,
\label{3.44}
\end{equation}
where
\begin{eqnarray}
&&  P_{J_1M}(\xi,\eta)\ =\ P_{j_1M_1}(\xi)P_{j_2M_2}(\eta)\,,
\label{3.45} \nn \\
&&  P_{j_1M_1}(\xi)\ =\ \xi^{j_1} P_{j_1M_1}(n)\,,
\quad
  P_{j_2M_2}(\eta)\ =\ \eta^{j_2}P_{j_2M_2}(m)\,.
\end{eqnarray}
Making use of the relations (\ref{3.17}) and (\ref{3.20}), we turn
now to the variables $u,v,u^*$ and $v^*$ and re-write the
expressions (\ref{3.45})  in the form
\begin{eqnarray}
P_{j_1M_1}(\xi)&=&i^{M_1}\left[\frac{(j_1+M_1)!}{(j_1-M_1)!}
\right]^{1/2}\nn \\
&\times& \sum_{\mu_1}\frac1{2^{j_1}}u^{(j_1+\mu_1)/2}
(u^*)^{(j_1-\mu_1)/2}\Delta^{(j_1)}_{0\mu_1}D^{j_1}_{\mu_1M_1}
(\varphi_1,\theta,\varphi_2),
\label{3.48}
\nn \\
P_{j_2M_2}(\eta)&=&i^{M_2}\left[\frac{(j_2+M_2)!}{(j_2-M_2)!}
\right]^{1/2} \nn \\
&\times&
\sum_{\mu_2} \frac{(-i)^{1/2}}{2^{-j_2}}v^{(j_2+\mu_2)/2}
(v^*)^{(j_2-\mu_2)/2} \Delta^{(j_2)}_{0\mu_2}D^{j_2}_{\mu_2M_2}
(\varphi_1,\theta\varphi_2).
\end{eqnarray}
Here the notation
\begin{equation}
\label{3.50}
\Delta^{(m)}_{kl}\ =\ D^m_{kl}\Big(0,\frac\pi2,0\Big)\
\end{equation}
 is used.

Expanding the product
$D^{j_1}_{\mu_1M_1}(\varphi_1,\theta,\varphi_2)D^{j_2}_{\mu_1M_2}
(\varphi_1,\theta,\varphi_2)$ over the functions $D^j_{\mu
M}(\varphi_1,\theta,\varphi_2)$ (where $M=M_1+M_2$,
$\mu=\mu_1+\mu_2$) and extracting from the sum one term with a
definite $J$ (where $|j_1-j_2|\le J\le j_1+j_2$), we obtain
\begin{eqnarray}
P_{J,M}(\xi,\eta) &=& i^M\left[
\frac{(j_1+M_1)!(j_2+M_2)!}{(j_1-M_1)!
(j_2-M_2)!}\right]^{1/2} (j_1j_2,00|J0)(j_1j_1,M_1M_2|JM)
\nonumber \\
&\times&\ \sum_{\mu_1,\mu_2}\frac{(-i)^{\mu^2}}{2^{j_1+j_2}}
(u^*)^{j_1-\mu_1)/2} v^{(j_2+\mu_2)/2} u^{(j_1+\mu_1)/2}
(v^*)^{(j_2-\mu_2)/2}
\nonumber \\
& \times&\ (j_1j_2\mu_1\mu_2|J\mu)^2 \Delta^{(J)}_{0\mu}
D^J_{\mu M} (\varphi_1,\theta,\varphi_2)\,.
\label{3.51}
\end{eqnarray}
With the help of (\ref{3.17}) and (\ref{3.20}), the expression
(\ref{3.42}) can be re-written in the form
\begin{eqnarray}
P^{(j_1+1/2,j_2+1/2)}_{(K-j_1-j_2)/2} (\xi,\eta) &=& \sum_m \frac1{2^n}
\Big( {n+j_1+1/2 \atop m}\Big) \Big({n+j_2+1/2 \atop n-m} \Big)
\nonumber \\
& \times&\ (-1)^{n-m} (uu^*)^m(vv^*)^{n-m}\,.
\label{3.52}
\end{eqnarray}
Inserting the formulae (\ref{3.51}) and (\ref{3.52}) into
(\ref{3.44}), we obtain for the eigenfunction $\Phi(\xi,\eta)$ the
following expression:
\begin{eqnarray}
\Phi(\xi,\eta)&=& A'_{JM}\sum_m\sum_{\mu_1\mu_2}
\frac{(-i)^{\mu_2}}{2^{j_1+j_2+n}}(-1)^{n-m}(j_1j_2,\mu_1\mu_2|J\mu)^2
\Delta^{(J)}_{0\mu} \Big({n+j_1+\frac12 \atop m} \Big)
\nonumber \\
&\times& \Big({n+j_2+\frac12 \atop n-m}\Big)
(u^*)^{\frac{j_1-\mu_1}2+m}
v^{\frac{j_2+\mu_2}2+n-m} u^{\frac{j_1+\mu_1}2+n-m} D^J_{\mu M}
(\varphi_1,\theta,\varphi_2),\quad\quad\;
\label{3.53}
\end{eqnarray}
where
\be
A'_{JM}\ =\ (-1)^{M/2}\left[ \frac{(j_1+M_1)!(j_2+M_2)!}{(j_1-M)!
(j_2-M_2)!} \right]^{1/2} (j_1j_2,00|J0)(j_1j_2,M_1M_2|JM)\,.
\ee
As it was already mentioned, all the further calculations are
necessary in order to obtain the function $\Phi(\xi,\eta)$ in a form
for which the Fourier transformation becomes relatively simple.
Let us use the formulae given in Ref.~\refcite{14n}:
\begin{eqnarray}
&& \frac{1}{\sqrt{(j-k)!(j+k)!}}
\Big(\cos\frac\alpha2 \,e^{i\gamma/2}
+i\sin\frac\alpha2\,e^{-i\gamma/2}\Big)^{j-k}\ \times
\nonumber\\
&&\times\ \Big(i\sin\frac\alpha2\,e^{i\gamma/2}\cos\frac\alpha2\,
e^{-i\gamma/2}\Big)^{j+k}=\sum^j_{l=-j}
\frac{P^j_{lk}(\cos\alpha)}{\sqrt{(j-l)!(j+l)!}}\,e^{-il\gamma},
\nn
\\
&&
\frac{1}{\sqrt{(j-k)!(j+k)!}}\Big(\cos\frac\alpha2\,e^{-i\gamma/2}
-i\sin\frac\alpha2\,e^{i\gamma/2}\Big)^{j-k}\ \times
\nonumber\\
&& \times\ \Big(-i\sin\frac\alpha2\,e^{-i\gamma/2}+\cos\frac\alpha2\,
e^{i\gamma/2}\Big)^{j+k}=\sum^j_{l=-j} \frac{\bar P^j_{lk}(\cos\alpha)
}{\sqrt{(j-l)!(j+l)!}}\,e^{il\gamma}. \qquad
\label{3.55}
\end{eqnarray}
Comparing these expressions with the formulae (\ref{3.13}) for
$u,u^*,v$ and $v^*$ and remembering that
\begin{equation}
\bar P^j_{lk}(\cos\alpha)\ =\ (-1)^{l-k}P^j_{-l-k}(\cos\alpha)\,,
\label{3.56}
\end{equation}
we can write
\begin{eqnarray}
&&(u^*)^{\frac{j_1-\mu_1}2+m} v^{\frac{j_2+\mu_2}2+n-m}=\left[
\Big(\frac{j_1-\mu_1}2+m\Big)!\Big(\frac{j_2+\mu_2}2 +n-m\Big)!
\right]^{1/2}
\nn \\
&& \times \sum^{(K-\delta)/4}_{\nu_1=-(K-\delta)/4}
P^{(K-\delta)/4}_{\nu_1,W+\mu/4}(\cos a)e^{-i\nu_1\lambda} \left[
\Big(\frac{K-\delta}4-\nu_1\Big)!\Big(\frac{K-\delta}4+\nu_1\Big)!
\right]^{-1/2},
\nn \\
&&
u^{\frac{j_1+\mu_1}2+m}(v^*)^{\frac{j_2-\mu_2}2+n-m}=\left[
\Big(\frac{j_1+\mu_1}2+m\Big)!\Big(\frac{j_2-\mu_2}2+n-m\Big)!
\right]^{1/2}
\nn \\
&&\times \sum^{(K+\delta)/4}_{\nu_2=-(K+\delta)/4}\!\!\!
P^{(K+\delta)/4}_{\nu_2,-W+\mu/4}(\cos a)(-1)^{\nu^2-W+\mu/4}
e^{i\nu_2\lambda}\nn \\
&&\times \left[\Big(\frac{K+\delta}2-\nu_2\Big)!\Big(
\frac{K+\delta}4+\nu_2\Big)!\right]^{-1/2},
\end{eqnarray}
where $\delta=\mu_1-\mu_2$, $W=\frac14 (j_2-j_1)+\frac12 n-m$.

 Then,
expanding $P^{\frac14 (K-\delta)}_{\nu_1,W+\mu/4}
(\cos a)$ and
$P^{(K+\frac14 \delta)}_{-\nu_2,\, -W+\frac14 \mu}(\cos a)$
over the functions
$P^{K/2-\kappa}_{\nu,\mu/2}(\cos a)$  (where $\nu=\nu_1-\nu_2$), the
eigenfunction $\Phi(\xi,\eta)$ can be re-written in the form
\begin{eqnarray}
\Phi(\xi,\eta)&=&A'_{JM}\sum_{m,\mu,\delta,\nu,\varepsilon,\kappa}
\frac1{2^{j_1+j_2+n}}
(-1)^{\frac{K-\delta}4+\frac{\varepsilon-\nu}2+\frac\mu2-\frac{j_2}2}
\Big(j_1,j_2,\frac{\mu+\delta}2, \frac{\mu-\delta}2\Big|J,\mu\Big)^2
\nn \\
&\times&\left(\!\frac{K-\delta}4,\frac{K+\delta}4,
\frac{\nu+\varepsilon}2,\frac{\nu-\varepsilon}2\Big|\frac K2-\kappa,\nu
\right)\nn \\
&\times&\left(\frac{K-\delta}4,\frac{K+\delta}4,W+\frac\mu4,-W+\frac\mu4
\Big|\frac K2-\kappa,\frac\mu2\right)
\nonumber\\
&\times& \Delta^{(J)}_{0,\mu}\Big( {n+j_1+1/2 \atop m}\Big)
\Big( {n+j_2+1/2 \atop n-m}\Big)
\nn \\
&\times&\bigg[\Big(\frac{j_1}2
-\frac{\mu+\delta}4 +m\Big)!\Big(\frac{j_1}2+\frac{\mu+\delta}4+m\Big)!
\nn \\&&\times
\Big(\frac{j_2}2+\frac{\mu\!-\!\delta}4+n-m\Big)!\Big(\frac{j_2}2
-\frac{\mu\!-\!\delta}4+n-m\Big)!\bigg]^{\frac12}
\nn \\
&\times& \bigg[ \Big(
\frac{K\!-\!\delta}4-\frac{\nu\!+\!\varepsilon}2\Big)!\Big(
\frac{K\!-\!\delta}4 +\frac{\nu\!+\!\varepsilon}2\Big)!
\nonumber\\
&&\times\ \Big(\frac{K+\delta}4-\frac{\nu-\varepsilon}2\Big)!
\Big(\frac{K+\delta}4+\frac{\nu-\varepsilon}2 \Big)! \bigg]^{-\frac12}
\nn \\
&\times&
P^{K/2-\kappa}_{\nu,\mu/2} (\cos a)D^J_{\mu,M}(\cos a) D^J_{\mu,M}
(\varphi_1,\theta,\varphi_2) e^{-i\nu\lambda}.
\label{3.59}\nn\\
\end{eqnarray}
Making use of several relations for the Clebsch-Gordan coefficients,
we can change (\ref{3.59}) so that, introducing
\begin{equation}
\label{3.60} P^{(\alpha,\beta)}_k(0)\ =\ \frac1{2^k} \sum^k_{m=0}
\Big( {k+\alpha \atop m}\Big)\Big( {k+\beta \atop k-m}\Big)
(-1)^{k-m}\,,
\end{equation}
the sum over $\epsilon=\nu_1+\nu_2$ can be easily taken.

In the following we shall consider only one definite term of the sum
over $\nu$. After some rather cumbersome algebraic transformations
we obtain the final result
\begin{eqnarray}
\Phi(\xi,\eta)&=& A_{JM}\sum_m \sum_{\mu\delta} \sum_\kappa \left(
j_1,j_2, \frac{\mu+\delta}2,\frac{\mu-\delta}2\Big| j\mu\right)^2
\nonumber\\
&\times& \left(\frac{K-\delta}4,\frac{K+\delta}4, W+\frac\mu4,
-W+\frac\mu4 \Big|\frac K2-\kappa,\frac\mu2\right)
\nonumber\\
&\times&\Big(\frac{j_1}2+m,\frac{j_2}2+n-m,\frac{\mu+\delta}4,
\frac{\mu-\delta}4\Big|\frac K2,\frac\mu2\Big)^{-1} \nn \\
&\times&\frac{(-1)^{\frac{K+\mu-\delta}4-\frac\nu4+\kappa}}{2^{K/4}}
\frac{\Delta^{(J)}_{0\mu}
\Delta^{(K/2-\kappa)}_{\delta/2,\nu}}{\Delta^{K/2}_{K/2,\mu/2}}
\nonumber\\
&\times&  \sqrt{\frac{(K-2\kappa+1)}{(K+\kappa+1)!\kappa!}}
\Big( {n+j_1+1/2 \atop m}\Big) \Big({n+j_2+1/2 \atop n-m} \Big)
\nonumber\\
& \times& \sqrt{(j_1+2m)!(j_2+2n-2m)!}\ D^{K/2-\kappa}_{\nu,\mu/2}
(\lambda,a,0)\,D^J_{\mu,M}\ (\varphi_1,\theta,\varphi_2)\,,\quad
\label{3.61}
\end{eqnarray}
where $A_{JM}\ =\ [(-1)^{-j_2/2}/2^{(j_1+j_2)/2}]A'_{JM}$\,.

Thus we obtained an expression for the general solution of the
problem in the form (\ref{3.3}). (In our case the notations
$M'=\mu/2,\Lambda/2-\kappa$ are used.) From the structure of the
coefficients at
$D^{K/2-\kappa}_{\nu,\mu/2}(\lambda,a,0)D^j_{\mu,M}(\varphi_1,\theta,\varphi_2)$
it becomes clear why in Ref.~\refcite{3n} the explicit form of $a_\nu(\Lambda,M')$
could not be determined.

\subsection{A different way of obtaining the eigenfunction
\boldmath$\phi(\xi,\eta)$}

Calculating the explicit form of the eigenfunction we have noticed
that, instead of the product of two $D$-functions, there exists a
probably more convenient form of $\Phi(\xi,\eta)$.

Let us start with the formula (\ref{3.53}). Taking the explicit
expressions of $u$, $v$, $u^*$ and $v^*$ (\ref{3.13}), we expand
them into a series over the degrees of $\sin(a/2)$ and $\cos(a/2)$:
\begin{eqnarray}
u^A &=& \sum^{A/2}_{s=-A/2}\left({A \atop
\displaystyle\frac{A+s}2}\right)\Big( \cos\frac
a2\Big)^{(A+s)/2}(-i)^{(A+s)/2} e^{i\lambda s/2}\,
\nonumber\\
v^B &=&
\sum^{B/2}_{t=-B/2}\left({B \atop \displaystyle\frac{B+t}2 }\right)
\Big(\cos\frac a2\Big)^{(B-t)/2}\Big(\sin\frac a2\Big)^{(B+t)/2}
(i)^{(B+t)/2} e^{s\lambda t/2}\,,
\nonumber\\
u^{*C}&=& \sum^{C/2}_{u=-C/2}\left({C \atop \displaystyle\frac{C+u}2}
\right)\Big(\cos\frac a2\Big)^{(C+u)/2}\Big(\sin\frac a2\Big)^{(C-u)/2}
 e^{i\lambda u/2}\,,
\nonumber\\
v^{*D}&=& \sum^{D/2}_{v=-D/2} \Big({D \atop
\displaystyle\frac{D+v}2}\Big) \Big(\cos\frac a2\Big)^{(D+v)/2}
\Big(\sin\frac a2\Big)^{(D-v)/2} (-i)^{(D-v)/2} e^{i\lambda v/2}.
\qquad
\label{3.62}
\end{eqnarray}
With the help of these relations we can write
\begin{eqnarray}
&&
(u^*)^{\frac{j_1-\mu_1}2+m}v^{\frac{j_2+\mu_2}2+n-m}
u^{\frac{j_1+\mu_1}2+m} (v^*)^{\frac{j_2-\mu_2}2+n-m}\quad =
\nonumber \\
&&= \sum_{{s,t,u,v \atop s+t+u+v=-2\nu}} \left( \begin{array}{c}
\displaystyle\frac{j_1+\mu_1}2+m \\
\displaystyle\frac{(j_1+\mu_1)/2+m+s}2 \end{array}\right)
\left(\begin{array}{c} \displaystyle\frac{j_2+\mu_2}2 +n-m \\
\displaystyle\frac{(j_2+\mu_2)/2+n-m+t}2 \end{array}\right)
\nonumber\\
&& \times\ \Big(\cos\frac a2\Big)^{\frac{E-s-t+u+v}2}
\Big(\sin\frac a2\Big)^{\frac{K+s+t-u-v}2}
(i)^{\frac{-\delta-s+t-u+v}2} e^{i[(s+t+u+v)/2]\lambda}.
\label{3.63}
\end{eqnarray}
Making use of (\ref{3.60}), let us introduce
$P^{(\alpha,\beta)}_k(0)$. If so, the two sums can be easily
calculated, and we have
\begin{eqnarray}
\Phi(\xi,\eta)&=& A'_{JM}\sum_m\!\sum_{\mu\delta}\,
\frac{(-i)^{(\mu-\delta)/2}}{2^{j_1+j_2+n}}(-1)^{n-m} \Big(j_1,j_2,
\frac{\mu+\delta}2,\frac{\mu-\delta}2 \Big| J\mu\Big)^2
\nonumber\\
&\times& \Delta^{(J)}_{0\mu}\Big({n+j_1+1/2 \atop n-m}\Big)
\Big({n+j_2+1/2 \atop n-m} \Big)
\sum_{{t,u \atop s+t+u+v=-2\nu}}\!\! (i)^{-\frac K2-\frac\delta2}
2^{\frac K2} e^{-i\nu\lambda}
\nonumber\\
&\times&  \Bigg[\left(\frac{j_1}2+\frac{\mu+\delta}4+m\right)!
\left(\frac{j_1}2-\frac{\mu+\delta}4+m\right)!
\left(\frac{j_2}2+\frac{\mu-\delta}4+n-m\right)!
\nonumber\\
&\times& \left(\frac{j_2}2-\frac{\mu-\delta}4+n-m\right)!\Bigg]^{1/2}
\Bigg[\left(\frac K4+\frac\mu4+\frac{s+t}2\right)!
\left(\frac K4+\frac\mu4-\frac{s+t}2\right)!
\nonumber\\
&\times&  \left(\frac K4-\frac\mu4+\frac{u+v}2\right)!
\left(\frac K4-\frac\mu4-\frac{u+v}2\right)!\Bigg]^{-1/2}
\nn \\
&\times& \Delta^{(\frac K4+\frac\mu4)}_{-\frac{s+t}2,W-\frac\delta4}
\Delta^{(\frac K4-\frac\mu4)}_{-\frac{u+v}2,-W-\frac\delta4}
\nn \\
&\times&\Big(\cos\frac a2\Big)^{\frac K2-\frac{(s+t)-(u+v)}2}
\Big(\sin\frac a2\Big)^{\frac K2+\frac{(s+t)-(u+v)}2} D^J_{\mu M}
(\varphi_1,\theta,\varphi_2)\,.
\label{3.64}
\end{eqnarray}
This expression can be re-written in the form
\begin{eqnarray}
\Phi(\xi,\eta) &=&\sum_{\mu,\delta,\sigma,W}
N(K,\nu,j_1,j_2|\mu,\delta,\sigma,W)\quad \times
\nonumber\\
&\times& (\cos a+1)^{\frac K4-\frac\sigma4}(\cos a-1 )^{\frac
K4+\frac\sigma4} D^J_{\mu M}(\varphi_1,\theta\varphi_2)
e^{-i\nu\lambda},
\label{3.65}
\end{eqnarray}
where
\begin{eqnarray}
&& \hspace*{-1.2cm}
N(K,\nu,j_1,j_2|\mu,\delta,\sigma,W)\ =\ A_{JM}(-1)^{n-m}
\Big(j_1,j_2,\frac{\mu+\delta}2,\frac{\mu-\delta}2\Big|J_\mu\Big)^2\
\times \nonumber\\
&& \times\ \Delta^{(J)}_{0\mu}\Big({n+j_1+1/2 \atop m}\Big)
\Big({n+j_2+1/2 \atop n-m}\Big) \widetilde\Delta^{(\frac
K4+\frac\mu4)}_{\frac\nu2-\frac\sigma4, W-\frac\sigma4}
\widetilde\Delta^{(\frac K4-\frac\mu4)}_{\frac\nu2
+\frac\sigma4,-W-\frac\delta4}\,
 \label{3.66}
\end{eqnarray}
and we have introduced the notations
 $\sigma=s+t-(u+v)$
and
\begin{equation}
\sqrt{\frac{(l-n)!(l+n)!}{(l-m)!(l+m)!}}\, \Delta^{(l)}_{mn}\ =\
\widetilde\Delta^{(l)}_{mn}\,.
\label{3.67}
\end{equation}
$A_{JM}$ is a normalization factor which we will not calculate. The
summation limits in (\ref{3.65}) are given by

\begin{equation}
-K\le\sigma\le K\,, \qquad -\frac{K-2j_2}4\le W\le\frac{K-2j_1}4\,.
\label{3.68}
\end{equation}
The limits in $\mu$ and $\delta$ at a definite $\sigma$ can be
obtained when the factorials in the denominator of
$\widetilde\Delta$ turn into zero.

\subsection{Mini-conclusion}

The construction of a basis for three free particles which realizes
the representation of the group of rotation in the three-dimensional
space and of the permutation group, turned out to be unexpectedly
difficult. The set of equations determining the eigenfunctions of
the problem appeared to be very complicated, and its solution could
be found only by an unconventional method. We succeeded in building
functions which satisfied only four of the five equations. Hence,
the final solution (including the quantum number $\Omega$) has to
be calculated by inserting the linear combination of the solutions
with different $(j_1j_2)$ into the equation for $\Omega$, obtained in
Ref.~\refcite{1n}. Let us underline, however, that the last, fifth equation
can be easily solved in every concrete case. In order to find the
coefficients it is sufficient to compare the higher orders of
$\cos(a/2)$ in the polynomials, this was done in Ref.~\refcite{12n}.

Still, we think that the orthogonalization of the polynomials can be
carried out in a more efficient way, may be even without the
operator $\Omega$.

The wave functions which were obtained here allow to consider
another problem, namely: how the rotational spectrum of the system
appears. To do this, the characteristics of the superposition over
the quantum numbers $K$ and $\Omega$ have to be investigated. It
would be also interesting to find out whether our method can be
applied to the motion of a heavy top and especially to the case of
the Kowalewski top \cite{kowa} (for recent discussions see for
example \cite{koma,horo} and references therein). A possible
application of the presented technique is the classification of the
Dalitz plots \cite{3n} and the calculation of the matrix elements of
pair interactions.

\section{A Complete Set of Functions in the Quantum Mechanical
Three-Body Problem }

A complete set of basis functions for the quantum mechanical three
body problem is here chosen in the form of hyperspherical functions.
These functions are characterized by quantum numbers corresponding
to the chain $O(6)\supset SU(3)\supset O(3)$. Equations are derived
to obtain the basis functions $m$ in an explicit form.

Elementary processes involving three interacting particles exhibit
an extremely complicated structure. It is therefore important to
have at least a complete understanding of systems consisting of
non-interacting particles.

In any classification of multiparticle states it is important to
diagonalize those variables which are known to be constants of
motion from general invariance principles, one usually takes the
total energy and the angular momentum. We can deal with the case of
equal masses because of the evident changeover:
\be
 \frac{1}{\sqrt{ m_i}}\frac{\partial}{\partial x_i}=
 \frac{\partial}{\partial X_i}.
\ee

Our aim here is to present a complete set of orthonormal functions,
corresponding to three free particles. Doing so, we introduce
hyperspherical functions, {\it i.e.} functions, which are defined on
the five-dimensional sphere, and are eigenfunctions of the angular
part of the six-dimensional Laplacian. They have to describe states
with given angular momenta and definite permutation symmetry
properties. This choice of functions is due to the invariance of the
free Hamiltonian under the $O(6)$ or the $SU(3)$ group.

The classification of the three-particle states based on these
groups according to the chain $O(6)\supset SU(3)\supset O(3)$ gives
four quantum numbers, namely: $K$ -- the six-dimensional momentum,
corresponding to the eigenvalue of the six-dimensional Laplacian;
$J$ -- the angular momentum and its projection $M$, and a number
$\nu$, which characterizes the permutation symmetry. On the other
hand the motion of a system of $n$ particles in a given energy and
momentum state can be defined by $3n-4$ parameters, and requires for
its quantum-mechanical description $3n-4$ quantum numbers, {\it
i.e.} five in the case of three particles.  That means that the
states labeled according to the chain above might be degenerate;
this degeneracy can be eliminated either by the straightforward
orthogonalization of the functions, or with the help of a hermitian
operator $\hat\Omega$, which we take from the group $O(6)$, and
which commutes with the $O(3)$ generators. The eigenvalue of this
operator is the fifth -- missing -- quantum number $\Omega$.
Unfortunately, since $\hat\Omega$ is a cubic operator, it leads to
rather complicated eigenvalue equations.

\subsection{Casimir operators and eigenfunctions}

Let us re-calculate the operators, the eigenvalues of which we are
trying to find. They are
\bea && \Delta\ =\ |A_{ik}|^2\,, \qquad
A_{ik}\ =\ iz_i\frac\partial{\partial
z_k}-iz^*_k\frac\partial{\partial z^*_i}
\eea
the $SU(3)$ generators;
\bea
J_{ik}=\ \frac12(A_{ik}-A_{ki})\ =\
\frac12\Big(iz_i\frac\partial{\partial z_k}
-iz_k\frac\partial{\partial z_i} +iz^*_i\frac\partial{\partial
z^*_k} -iz^*_k \frac\partial{\partial z^*_i}\Big),
\eea
 Here $\Delta$ is the Laplace operator on the five-dimensional
sphere;
$A_{ik}$ are the generators of the three-dimensional rotation group.
The scalar operator reads
\bea
N=\frac12\sum_k\left(z_k\frac\partial{\partial z_k}-z^*_k
\frac\partial{\partial z^*_k}\right) =\ \frac1{2i}\mbox{ Sp }A\,,
\eea
the eigenvalue of which is $\nu$. Finally, the operator
\bea
\hat\Omega&=&\sum_{i,k,l} J_{ik}B_{kl}J_{li}\,,
\nn \\
B_{ik}&=&\frac12(A_{ik}+A_{ki})\ =\ \frac12\left(
iz_i\frac\partial{\partial z_k}+iz_k\frac\partial{\partial z_i}-iz^*_i
\frac\partial{\partial z^*_k}-iz^*_k\frac\partial{\partial z^*_i}\right)
\quad
\eea
is the generator of the group of deformations of the triangle.

The explicit expressions for these five commuting operators are the
following. Using
\begin{eqnarray}
ds^2 = |dz|^2=\ g_{ik}q^iq^k&=&\varrho^2\Big[ \frac14 da^2
+\frac14d\lambda^2+\frac12d\Omega^2_1+\frac12d\Omega^2_2\ +
\nn \\
& +&d\Omega^2_3-\sin a\,d\Omega_1d\Omega_2 -\cos
a\,d\Omega_3d\lambda\Big] +d\varrho^2\,, \qquad
\end{eqnarray}
where $d\Omega_i$ are infinitesimal rotations about the moving axes,
 we obtain the Laplacian:
\begin{eqnarray}
\Delta &=& g^{-1/2} \frac\partial{\partial
q^i}\,g^{ik}g^{1/2}\frac\partial{\partial q^k}
\nn \\
&=&\Bigg\{\frac\partial{\partial a^2}+2{\ctg}2a\frac\partial{\partial
a} +\frac1{\sin^2a} \left(\frac{\partial^2}{\partial\lambda^2}+{\cos}
a \frac{\partial^2}{\partial\lambda\partial\Omega_3}
+\frac14\frac{\partial^2}{\partial\Omega^2_3}\right)\ +
\nn \\
&&+\
\frac1{2\cos^2a}\left[\frac{\partial^2}{\partial\Omega^2_1}+{\sin}\, a
\Big(\frac{\partial^2}{\partial\Omega_1\partial\Omega_2}
+\frac{\partial^2}{\partial\Omega_2\partial\Omega_1}\Big)
+\frac{\partial^2}{\partial\Omega^2_2}\right]\Bigg\}.
\end{eqnarray}
The explicit form of $N$ is $N=i(\partial/\partial\lambda)$. If a
harmonic function of $\Phi$ is an eigenfunction of $\Delta$, it has
to fulfil
\be
\Delta\Phi\ =\ -K(K+4)\Phi, \qquad
N\Phi\ =\ \nu\Phi\,.
\ee
Let us note here that if a harmonic function belongs to the $SU(3)$
representation $(p,q)$, then $K=p+q,$ $\nu=1/2(p-q)$.

The operator $J_{ik}$ has the form

\be
J_{ik}\ =\ -\frac i2\,\varepsilon_{lkl} \left[l^{(1)}_1
\frac\partial{\partial\Omega_1} +l^{(1)}_2
\frac\partial{\partial\Omega_2} +l^{(1)}
\frac\partial{\partial\Omega_3} \right].
\ee

We obtain
\begin{eqnarray}
\hat\Omega &=& -\frac14 \Bigg\{2^{1/2}\left(
\frac{\partial^2}{\partial\Omega^2_+}H_+
+\frac{\partial^2}{\partial\Omega^2_-} H_- \right)
+\frac{\partial^2}{\partial\Omega^2_3}
\frac\partial{\partial\lambda} +\Delta\ominus
\frac\partial{\partial\lambda}\ -
\nn\\
&& -\ \frac1{{\cos} a} \left(\Delta\ominus
-\frac{\partial^2}{\partial\Omega^2_3} + \frac12\right)
\frac\partial{\partial\Omega_3}\ +
\nn\\
&&+\ {\tg}\, a\left[1\left(\frac{\partial^2}{\partial\Omega^2_+}
-\frac{\partial^2}{\partial\Omega^2_-}\right)
\frac\partial{\partial\Omega_3} -\frac32
\left(\frac{\partial^2}{\partial\Omega^2_+}
+\frac{\partial^2}{\partial\Omega^2_-} \right)\right]\Bigg\},
\end{eqnarray}
where
\begin{eqnarray}
H_{\pm} &=& 2^{-1/2}\left[\frac\partial{\partial a}\pm1 \frac{1}{{\sin}
a}\, \frac\partial{\partial\lambda} \pm\frac i2{\ctg}\, a
\frac\partial{\partial\Omega_3}
\right],\nn \\
\frac\partial{\partial\Omega_{\pm}} &=&
2^{-1/2}\left(\frac\partial{\partial\Omega_1} \pm
i\,\frac\partial{\partial\Omega_2}\right).
\end{eqnarray}
Before writing the eigenfunctions of these five operators, we have
to make a few remarks. One can show that for $K<4$ all states are
simple; in the interval $4\le K<8$ doubly degenerated states show up;
as $K$ is growing, the number of degenerated states grows too,
and at the value $K=4n$ an $n$-fold degeneration appears.
Besides, states with $J=0$ and $J=0$ values are not degenerated.
Consequently, for practical purposes it is enough to deal with four
quantum numbers.

Let us look for the harmonic functions $\Phi$ which satisfy
the eigenvalue equations of the Laplace operator and the operator
$N$ with eigenvalues $K(K+4)$ and $\nu$, respectively. The general
form is the following:
\be
\Phi'_{M\nu}\ =\ \sum_\nu\sum_\mu a_\nu(\kappa,\mu)\,
D^{(K/2)-\nu}_{\nu(\mu/2)}\,(\lambda,a,0)\,D^J_{\mu,M}
(\varphi_1\ominus\varphi_2)\,.
\ee
It is easy to understand the meaning of this solution. One can
consider the second $D$-function -- which is the eigenfunction of $J^2$
and $J_3$ -- as an eigenfunction of a rotating rigid top with the
projection of the angular momentum on the moving axis equal to $\mu$. This
projection is not conserved in our case, that is why we have to sum
over different values of $\mu$. That is just the point where we need an
additional operator to orthogonalize the obtained functions. The
coefficients $a_\nu(\kappa,\mu)$ have to be defined from the equations
\be
\Delta\,\Phi'_{M\nu}\ =\ -K(K+4)\,\Phi^J_{M\nu}\,,
\qquad
\hat\Omega\,\Phi'_{M\nu}\ =\ \Omega\,\Phi'_{M\nu}\,.
\ee
These equations are unfortunately somewhat complicated:
\begin{eqnarray}
&&
\sum_{\kappa,\mu} \Bigg\{ \bigg[ a_\nu(\kappa,\mu-2)\frac12
\sqrt{\Big(\frac K2-\frac\mu2-\kappa+1 \Big)\Big(
\frac K2+\frac\mu2 -\kappa\Big)}\ \times
\nn\\
&&\times\ \sqrt{(J-\mu+2)(J-\mu+1)(J+\mu-1)(J+\mu)}
+a_\nu(\kappa,\mu+2)\frac12\ \times
\nn\\
&& \times\ \sqrt{\Big(\frac K2\!+\!\frac\mu2\!-\!\kappa\!+\!1\Big)
\Big(\frac K2\!-\!\frac\mu2\!-\!\kappa\Big)}\sqrt{(J+\mu+2)(J+\mu+1)
(J\!-\!\mu\!-\!1)(J\!-\!\mu)}\ +
\nn\\
&&+\ a_\nu(\kappa,\mu)(\nu\mu^2+J(J+1)\nu\!-4i\Omega)\bigg]
D^{K/2\,\nu}_{\nu\,(\mu/2)}(\lambda\,a,0)D^J_{\mu M}
(\varphi_1\ominus\varphi_2)+a_\nu(\kappa,\mu)\ \times
\nn\\
&& \times \bigg[\frac\mu{\cos a}\Big(J(J+1)-\mu^2+\frac12\Big)
D^{K/2-\kappa}_{\nu(\mu/2)} (\lambda,a,0) D^J_{\mu,M}
(\varphi_1\ominus\varphi_2) +i{\tg} a\ \times
\nn\\
&& \times\ \bigg(\Big(\frac\mu2+\frac34\Big)
\sqrt{(J\!-\mu)(J+\mu+1) (J\!-\mu\!-1)(J+\mu+2)}
D^{K/2-\kappa}_{\nu,\mu/2} (\lambda,a,0)\ \times
\nn\\
&& \times\ D^J_{\mu+2M}(\varphi_1\ominus\varphi_2)\!-\Big(\frac\mu2
-\frac34\Big) \sqrt{(J\!+\mu)(J\!-\mu+1)(J+\mu-1)(J\!-\mu+2)}\ \times
\nn\\
&& \times\ D^{K/2-\nu}_{\nu(\mu/2)} (\lambda,a,0)\,D^J_{\mu-2,M}
(\varphi_1\ominus\varphi_2)\bigg)\bigg]\Bigg\}=\ 0,
\end{eqnarray}
and
\begin{eqnarray}
&&
\sum_{\kappa,\mu} \Bigg\{a_\nu(\kappa,\mu)\bigg[-\Big(\frac K2
-\kappa\Big)\Big(\frac K2-\kappa+1\Big)+\frac14K(K+4)-\frac\mu2
+\frac\nu{\cos a}\ -
\nn \\
&&-\ \frac1{2\cos^2a}\,J(J+1)+\frac{\mu^2}{2\cos^2a}\bigg]
D^{K/2-\kappa}_{\nu(\mu/2)}(\lambda,a,0)\,D^J_{\mu,M}
(\varphi_1\ominus\varphi_2)\ -
\nn \\
&&-\ a_\nu(\kappa,\mu\!-\!2)i{\tg} a\sqrt{\Big(\frac K2-\frac\mu2
-\kappa+1\Big)\Big(\frac K2+\frac\mu2-\kappa\Big)}
D^{K/2-\nu}_{\nu(\mu/2)} (\lambda,a,0)\ \times
\nn \\
&& \times\ D^J_{\mu-2,M}(\varphi_1\ominus\varphi_2) +a_\nu(\kappa,\mu)
\bigg[- \frac i4\,\frac{\sin a}{\cos^2a}\quad \times
\nn \\
&& \times\ \sqrt{(J-\mu)(J+\mu+1)(J-\mu-1)(J+\mu+2)} \quad \times
\nn \\
&&\times\ D^{K/2-\kappa}_{\nu,(\mu/2)}(\lambda,a,0)\,D^J_{\mu+2\,M}
(\varphi_1\ominus\varphi_2)+\frac i4\, \frac{\sin a}{\cos^2a}\quad
\times
\nn \\
&&\times \sqrt{(J\!+\!\mu)(J\!-\!\mu\!+\!1)(J\!+\!\mu\!-\!1)
(J\!-\!\mu\!+\!2)}
\nn \\
&&\times
D^{K/2\!-\!\kappa}_{\nu,(M/2)} (\lambda,a,0)
D^J_{\mu-2,M} (\varphi_1\ominus\varphi_2)\bigg]\!\Bigg\}=0,
\end{eqnarray}
and, although it is quite easy to solve this set of equations for every
particular case, we have not been able so far to obtain a general
solution.

\subsection{Another way of constructing a set of eigenfunctions}

There is another way to find this complete set of functions. In
fact, the problem becomes complicated because of the requirement of
definite permutation symmetry properties. Without them it would be
simple to construct the wanted functions with the help of the
graphical method of the so-called ``tree-functions" \cite{11n}, which was
proposed by Vilenkin and Smorodinsky. We have to modify these
functions, {\it i.e} we have to find a transformation from the
complete set of ``tree-functions'' to the $K$ harmonics. ($K$
harmonics are hyperspherical functions possessing definite
permutation symmetry properties). Thus we first construct the
``tree-functions'', which are characterized by quantum numbers
\[
K, j_1, M_1, j_2, M_2
\]
($j_1,M_1$ and $j_2,M_2$ are angular momenta and their projections
conjugated to $\vec\xi$ and $\vec\eta$). We have to transform these
functions to a set of  $K$ harmonics which is described by the
quantum numbers $K,J,M,\nu,(j_1j_2)$. In order to do this it is
necessary to carry out a simple Fourier transform. To be correct,
$(j_1j_2)$ is not a real quantum number in the sense that functions
corresponding to different pairs $(j_1j_2)$ do not form an
orthogonal set, but this notation demonstrates where we get these
functions from. Their explicit expression is the following:
\begin{eqnarray}
&&
\Phi^{j_1j_2}_{J\,M\nu}(\vec\xi,\vec\eta)\ =\ A_{JM}\sum_m
\sum_{n,\delta}\sum_\kappa \left(j_1,\frac{\mu+\delta}2; j_2,
\frac{\mu-\delta}2\Big|J;\mu\right)^2\ \times
\nn \\
&& \times\ \frac{\Big(\frac{K\!-\delta}4,W\!+\!\frac\mu4;
\frac{K\!+\!\delta}4, -W\!+\!\frac\mu4\Big|\frac
K2\!-\!\kappa;\frac\mu2\Big)}{\Big(
\frac{i_1}2\!+\!m,\frac{\mu\!+\!\delta}4;\frac{j_2}2\!+\!n\!-\!m,
\frac{\mu\!-\!\delta}4\Big|\frac K2;\frac\mu2\Big)}
\frac{(-1)^{\frac{K+n-\delta}4-\frac\nu2+\kappa}}{2^{K/4}}\ \times
\nn \\
&& \times\ \frac{\Delta^{(J)}_{0\mu}
\Delta^{((K/2)-\kappa)}_{\delta/2\,\nu}}{\Delta^{(K/2)}_{K/2,\mu/2}}
\left[ \frac{(j_1\!+\!2m)!(j_2+2n\!-\!2m)!}{(K+\kappa+1)!\kappa!}
\right]^{1/2}
\Big({n\!+\!j_1\!+\!\frac12 \atop m}\Big)\Big({n\!+\!j_2\!+\!\frac12
\atop n-m}\Big)\ \times
\nn \\
  && \times\ D^{(K/2)-\kappa}_{\nu,\mu/2}\,(\lambda,a,0)\
D^J_{\mu,M}(\varphi_1\ominus\varphi_2)\,,
\end{eqnarray}
where $A_{JM}$ consists of normalization constants and
Clebsch--Gordan coefficients.

The solutions of the eigenvalue equations for $K$ and $\Omega$
have to be linear combinations of these functions:
\be
\Phi^J_{M\nu}\ =\ \sum_{j_1j_2}c_{(j_1j_2)} \Phi^{j_2j_2}_{JM\nu}
(\vec\xi,\vec\eta)\,,
\ee
where $(j_1j_2)$ will run over each pair of values which can give such a
total angular momentum $J$ that
\be
J\ \le\ j_1+j_2\ \le\ K\,.
\ee
Looking at the structure of the coefficient, it is easy to
understand that our attempt to determine $a_\nu(\kappa,\mu)$
directly could not be successful.

\section{A Symmetrical Basis in the Three-Body Problem}

In the theory of representation of the $O(n)$ group one usually
considers the canonical Gelfand-Zeitlin chain
$O(n)\,\supset\,O(n-1)\,\supset ...$. Such a chain is, however,
inconvenient, when the subject is the many-body problem where the
chain $O(n)\,\supset\,O(n-3)\,\supset ...$ is more natural. This
chain corresponds to the decrease of the number of particles by one
( {\it i.e} that of the degree of freedom by three). In addition, in
the many-body problem it is reasonable to follow the schemes in the
framework of which the angular momenta are summed up, and make sure
that the total angular momentum $J$ and its projection $M$ remain
conserved at all rotations. Of course, the number of group
parameters ({\it i.e.} that of the Euler angles) will decrease.

Indeed, there is no need to consider all the rotations. It is
sufficient to take into account only those which do not mix up the
components of different vectors.

In the three particle problem the two vectors $\eta$ and $\xi$ form
the $O(6)$ group. Here, in order to include the angular momentum $J$
into the number of observables, one has to separate the $O(3)$
group. If so, there remain only 5 angles of the 15 of the $O(6)$
group; they characterize the position of vectors $\eta$ and $\xi$ in
the six-dimensional space. Three angles define the situation of the
plane of the vectors $\eta$, $\xi$ in the three-dimensional space,
two angles determine the angle between these two vectors and their
lengths (with the condition $\xi^2+\eta^2=\rho^2$).  In several
works, \cite{1n,2n,4n,12n,16n,20n} attempts were made to construct
total sets of eigenfunctions for the three-body problem. This was
based on the invariance of the Laplacian under $O(6)$. The explicit
calculations were carried out differently. One of the possibilities
is that the eigenfunctions correspond to the classification of the
three-particle states according to the chain $ O(3)\, \supset\,
SU(3)\, \supset\, O(6)$ which is characterized by five quantum
numbers. Four of them -- $K$, $J$, $M$, $\nu$ -- are the
six-dimensional angular momentum, corresponding to the eigenvalue of
the 6-Laplacian, the usual three-dimensional momentum and its
projection, and a number, characterizing the permutation symmetry.
Generally speaking, a number of states of the system belongs to the
given set $K$, $J$, $M$, $\nu$. The fifth quantum number $\Omega$
which solves this problem is the eigenvalue of the hermitian
operator $\hat{\Omega}$, commuting with the $O(3)$ generators. The
operator was introduced by Racah \cite{13n}; its explicit form was
obtained in Descartes coordinates by Badalyan, in polar coordinates
it is given in Ref.~\refcite{2n}. In order to find the eigenfunctions
$\Phi^{KJM}_{\nu\Omega}$ corresponding to the above choice of
quantum numbers, we have to carry out the common solution of the
complicated equations $\Delta\Phi = -K(K+4)\Phi$ and
$\hat{\Omega}\Phi=\Omega\Phi$.

A different way of constructing the set of functions is based on the
graphical tree method. With the help of a simple algorithm a
function with quantum numbers $K, J, M, j_1, j_2$ is built up. If
so, $\Phi^{j_1j_2}_{KJM}$ are, by definition, eigenfunctions of the
six-dimensional Laplacian $O(6)$; they do not have, however,
definite permutation symmetry properties. Hence, we have to turn
from the obtained set of eigenfunctions to functions which change
simply when the coordinates of the particles are permutated,
 {\it i.e.} to the system with the quantum numbers $K, J, M, \nu, \Omega$.

We could not calculate explicitly the transformation coefficients
yet. However, making use of the result obtained in Ref.~\refcite{4n}, one
can carry out a transformation which leads to a system with quantum
numbers $K, J, M, \nu, (j_1j_2)$. Generally speaking, this system is
not orthonormalized yet. It can be orthonormalized either by a
standard method, or by the construction of eigenfunctions of the
operator $\hat{\Omega}$ from the function
$\Phi^{\nu(j_1j_2)}_{KJM}$. The solution of the corresponding
equations is not a difficult task. However, these equations turn out
to be high order equations. Thus the solutions of these equations
can hardly be presented explicitly.

The obtained solution provides a natural description for the basis
of three particles.

\subsection{Basis functions}

Systems of basis functions characterized by different quantum
numbers correspond to different parametrizations. What concerns the
eigenfunctions symmetrized over permutations, it is convenient to
consider them in the $z$, $z^*$ space, while the functions
$\Phi^{j_1j_2}_{KJM}$ are determined in the space of $\eta$ and
$\xi$. The latter are well known as the tree functions, see the
previous sections. Let us recall notations:
\bea \label{5.15} &&
 \Phi^{j_1j_2}_{KJM}(\eta,\xi)\ = \sum_{m_1+m_2=M}
C^{JM}_{j_1m_1j_2m_2} \Phi_K^{j_1j_2m_1m_2} (\eta,\xi)\ =\
Y^{j_1j_2}_{JM}(m,n) \Psi_{Kj_1j_2}(\Phi).
\nn \\
&& Y^{j_1j_2}_{JM}(m,n)\ =\ \sum_{m_1m_2}
C^{JM}_{j_1m_1j_2m_2} Y_{j_1m_1}(m) Y_{j_2m_2}(n)\,, \quad
%
\eea
 Introducing $\cos\Phi\ =\ \xi, \quad \sin\Phi =\ \eta$ we can
write:
\bea &&
 \Psi_{Kj_1j_2}(\Phi)\ =\ N_{Kj_1j_2}(\sin\Phi)^{j_1}(\cos\Phi)^{j_2}
P^{(j_1+1/2,j_2+1/2)}_{(K-j_1-j_2)/2} (\cos2\Phi),
\nn \\
&&
 N_{Kj_1j_2}=\left[ \frac{2(K+2)\Gamma((K-j_1-j_2)/2+1)
\Gamma((K+j_1+j_2)/2+2)}{\Gamma((K-j_1+j_2)/2+3/2)\,
\Gamma((K+j_1-j_2)/2+3/2)} \right]^{1/2}.\qquad
\eea
Instead of the Jacobi polynomial we can introduce the Wigner
$d$-function.
Carrying out the transition from the Jacobi
polynomial to the Wigner $d$-function, it is worth mentioning that
there exist different notations. Indeed, in Ref.~\refcite{14n} the definition
\be
P^{\alpha,\beta}_k (\cos2\theta)=i^{b-1}\left[
\frac{(l-b)!(l+b)!}{(l-a)!(l+a)!}\right]^{1/2} (\sin\theta)^{b-a}
(\cos\theta)^{-b-a} d^l_{ab}(\cos2\theta).
\ee
is given. Here $i$ appears because  the unitary matrices are
defined as
\be
u(\varphi,\theta,\psi)=\left( \begin{array}{cc}
e^{i\varphi/2} & 0\\  0 & e^{-i\varphi/2} \end{array} \right)
\left(\begin{array}{cc} \cos\frac\theta2 & i\sin\frac\theta2\\
i\sin\frac\theta2 & \cos\frac\theta2 \end{array}\right)
\left( \begin{array}{cc} e^{i\varphi/2} & 0\\
0 & e^{-i\varphi/2} \end{array} \right),
\ee
where the second matrix is a unitary one while in Ref.~\refcite{17n}
it is
orthogonal:
\be
u(\varphi,\theta\psi)=\left( \begin{array}{cc}
e^{i\varphi/2} & 0\\ 0 & e^{-i\varphi/2} \end{array} \right)
\left( \begin{array}{cc} \cos\frac\theta2 & \sin\frac\theta2 \\
-\sin\frac\theta2 & \cos\frac\theta2 \end{array} \right)
\left( \begin{array}{cc}
e^{i\varphi/2} & 0\\ 0 & e^{-i\varphi/2} \end{array} \right).
\ee
According to the definition in Ref.~\refcite{17n},
\begin{eqnarray}
\label{5.19}
 &&
P^{\alpha,\beta}_k(\cos2\theta)\ =\ (-1)^{b-a} \left[
\frac{(l-b)!(l+b)!}{(l-a)!(l+a)!}\right]^{1/2} (\sin\theta)^{b-a}
(\cos\theta)^{-b-a} d^l_{ab}(\cos2\theta)\,,
\nonumber\\
&& l\ =\ k+\frac{\alpha+\beta}2\,, \quad a\ =\ \frac{\alpha+\beta}2\,,
\quad b\ =\ \frac{\beta-\alpha}2\,.
\end{eqnarray}
In terms of the $d$-function the eigenfunction (\ref{5.15}) obtains
the form
\begin{eqnarray}
\label{5.20}
 &&
\Phi^{j_1j_2}_{KJM}(\eta,\xi)\ =\ 2(K+2)^{1/2}\,(-1)^{-j_1-1/2}\
\nn \\
&&\times \sum_{m_1m_1} C^{JM}_{j_1m_1j_2m_2}Y_{j_1m_1}(m) Y_{j_2m_2}(n)
\frac1{(\sin2\Phi)^{1/2}} d^{(K+1)/2}_{(j_1+j_2+1)/2,(j_2-j_1)/2}
(\cos2\Phi).\nn \\
\end{eqnarray}
The factor $(\sin 2\Phi)^{-1/2}$ appears due to the different
normalizations over the angle $\Phi$ in the Jacobi polynomial and in
the $d$-function. The Jacobi polynomial is normalized in the
six-dimensional space, the element of the volume is
$
\cos^2\Phi \sin^2\Phi\,d\Phi\ =\ \frac14 \sin^22\Phi\,d\Phi\,
$
while for the $d$-function, normalized in the usual space, we have
$\sin 2\Phi d \sin 2\Phi$.

The transition to the basis function constructed on the vector pair
$\eta'$, $\xi'$ which is related to $\eta$, $\xi$ by the
transformation
\begin{equation}
\label{5.21}
 \Big({\eta'\atop\xi'}\Big)\ =\ \left( \begin{array}{cc}
\cos\varphi & \sin\varphi \\ -\sin\varphi & \cos\varphi \end{array}
\right) \Big({\eta\atop\xi}\Big),
\end{equation}
can be carried out with the help of the coefficient
$\langle\,j'_1j'_2|j_1j_2\rangle^{\varphi}_{KJM}$ obtained in
Ref.~\refcite{18n} and Ref.~\refcite{19n}:
\begin{equation}
\label{5.22}
 \Phi^{j_1j_2}_{KJM}(\eta',\xi')\ =\ \sum_{j'_1j'_2} \,
\langle\,j'_1j'_2|j_1j_2\,\rangle^\varphi_{KJM}\, \Phi^{j'_1j'_2}_{KJM}
(\eta,\xi)\,.
\end{equation}
We see that for the
transition of the basis function from $\eta$, $\xi$ to the function
of the vector $z$, $z^*$ the coefficient
$\langle\,j'_1j'_2|j_1j_2\rangle^{\varphi}_{KJM}$ has to be used at
the value $\varphi=\pi/4$, after substituting $\xi$ by $\zeta$. In
the following it will be shown how the obtained function
$\Phi^{j_1j_2}_{KJM}(z,z^*)$ can be transformed into
$\Phi^{\nu(j_1j_2)}_{KJM}(z,z^*)$.

But let us first investigate the transformation coefficient
(\ref{5.22}) in detail.

\subsection{Transformation coefficients
\boldmath$\langle j'_1j'_2|j_1j_2\rangle^\varphi_{KJM}$ }

In Ref.~\refcite{18n} the transformation coefficient is defined as the overlap
integral of $\Phi^{j'_1j'_2}_{KJM}(\eta,\xi)$  and
$\Phi^{j_1j_2}_{KJM}(\eta',\xi')$ :
%
\bea
\label{5.23}
&&\langle j'_1j'_2|j_1j_2\rangle^\varphi_{KLM}= \nn \\
&&=\sum_{{m_1m_2\atop m'_1m'_2}} C^{JM}_{j_1m_1j_2m_2}
C^{JM}_{j'_1m'_1j'_2m'_2} \int\!\!
\Big(\Phi_K^{j'_1j'_2m'_1m'_2}(\eta,\xi)\Big)^*
\Phi_K^{j_1j_2m_1m_2}(\eta',\xi')\,d\eta\,d\xi.\qquad
\eea
As it was already mentioned, the transformation takes place at given
$J$ and $M$ values. The explicit form of the coefficient is
calculated with the help of the production function for
$\Phi^{j_1i_2m_1m_2}_K (\eta\xi)$.
The calculation is carried out in two steps \cite{19n}. First
 we consider
the coefficient corresponding to the representation in which the
vector $\eta$ is expanded over the ``new'' vectors $\xi'$ and
$\eta'$. After that, we do the same for $\xi$. As it is presented in
Ref.~\refcite{19n}, the transformation coefficient, corresponding to the
transition $\Phi^{j_10}_{K_1j_1m1}(\eta',0)$ to
$\Phi^{pq}_{K_1j_1m_1}(\eta',\xi')$, has the form:
\begin{eqnarray}
\label{5.24}
&&
\langle\,pq|j_10\,\rangle^\varphi_{Kj_1m_1}\ =\ (-1)^{(K_1+j_1)/2}
2^{(K_1-j_1)/2}\quad \times
\nn \\
&&\times \left[\frac{\Big(\frac{K_1-p-q}2\Big)!\Big(
\frac{K_1+p+q}2\Big)!\Big(\frac{K_1-j_1}2\Big)! (K_1+1)
!!}{(K_1\!-\!p\!+\!q\!+\!1)!!(K_1\!+\!p\!-\!q\!+\!1)!!
\Big(\frac{_1\!+\!j_1}2 +1\Big)!(K_1\!-\!j_1\!+\!1)!!}\!\!\right]^{1/2}
\nonumber \\
&& \times\ \Big({p~q~j_1\atop0~0~0}\Big)[(2p+1)(2q+1)]^{1/2}
(\cos\varphi)^p (\sin\varphi)^q\,P^{(p+1/2,q+1/2)}_{(K_1-p-q)/2}
(-\cos2\varphi)\qquad
\end{eqnarray}
and is in fact the tree function in $O(6)$. The expression
(\ref{5.24}) is
similar to the formula describing the $O(2)$-rotation, which
transforms the Legendre-polynomial with the help of the function
$D^J_{0M}=P_{JM}$, being the tree function in $O(2)$. The same
procedure has to be carried out for the coefficient of the
transformation of $\Phi^{0j_2}_{K_2j_2m_2}(0,\xi)$ into
$\Phi^{0r}_{K_2j_2m_2}(\eta'\xi')$; after that, collecting the
momenta $p+r=j'_1$ and $q+s=j'_2$ (with the total momentum $J$ and a
given $K=K_1+K_2$), we obtain the general expression for the
coefficient
\begin{eqnarray}
\label{5.25}
 &&
\langle\,j'_1j'_2|j_1j_2\,\rangle^\varphi_{KJM}\ =\
\frac\pi4\,(-1)^{J+(K+j_1+j_2)/2}\quad \times
\nonumber
\\
&& \times\ \Big(\frac{K_1-j_1}2\Big)!\Big(\frac{K_1+j_1+1}2\Big)!
\Big(\frac{K_2-j_2}2\Big)!\Big(\frac{K_2+j_2+1}2\Big)!
\nonumber
\\
&&\times \left[\frac{\Big(\frac{K-j'_1-j'_2}2\Big)!
\Big(\frac{K+j'_1+j'_2}2+1\Big)!(K+j'_1+j'_2+1)!!(K+j'_1-j'_2+1)!!}{
\Big(\frac{K-j_1-j_2}2\Big)!\Big(\frac{K-j_1+j_2}2+1\Big)!
(K-j_1+j_2+1)!!(K+j_1-j_2+1)!!}\right]^{1/2}
\nonumber  \\
&&\times\ \sum_{pr\,qs}\left\{\!\!\left\{\! \begin{array}{ccc}
p & r & j'_1\\  q & s & j'_2\\ j_1 & j_2 & J\end{array}\!
\right\}\!\!\right\}
\bigg[ \Gamma\left(\!\frac{K_1-p+q}2+\frac32\!\right)\Gamma
\left( \!\frac{K_1+p-1}2+\frac32\!\right)
\nonumber\\
&& \times\ \Gamma\left(\frac{K_2-r+s}2 +\frac32\right)\Gamma
\left(\frac{K_2+r-s}2 +\frac32\right)\bigg]^{-1} (\cos\varphi)^{p+s}
(\sin\varphi)^{q+r}
\nonumber\\
&& \times\ P^{(p+1/2,q+1/2)}_{(K_1-p-q)/2} (-\cos2\varphi)
P^{(s+1/2,r+1/2)}_{(K_2-s-r)/2} (-\cos2\varphi).
\end{eqnarray}
Here we introduce the notation
\begin{eqnarray}
\label{5.26}
\left\{\!\!\left\{\!\! \begin{array}{ccc}
p & r & j'_1\\ q & s & j_2\\  j_1 & j_2 & J \end{array}\!\!\right\}\!\!
\right\}\!&=&[ (2j_1+1)(2j_2+1)(2j'_1+1)(2j'_2+1)]^{1/2}
(2p+1)(2q+1)(2s+1)
\nonumber\\
&\times&
(2r+1)\Big(\begin{array}{ccc}p & q & j_1\\0 & 0 & 0\end{array}\Big)
\Big(\begin{array}{ccc}s & r & j_2\\0 & 0 & 0\end{array}\Big)
\Big(\begin{array}{ccc}p & r & j'_1\\0 & 0 & 0\end{array}\Big)
\Big(\begin{array}{ccc}q & s & j'_2\\0 & 0 & 0\end{array}\Big)
\left\{\begin{array}{ccc} p & r & j'_1\\ q&s&j'_2\\ j_1&j_2&J
\end{array}\right\},
\end{eqnarray}
which underlines the way how the momenta are transformed.

In Ref.~\refcite{18n} and Ref.~\refcite{19n} not only the technique of the calculations differs
but also the form of the final expressions. Thus it is reasonable to
find the explicit connection between the two forms of the
transformation coefficients
$\langle\,j'_1j'_2|j_1j_2\,\rangle^{\phi}_{KJM}$ and prove their
analogousness. Just this was done in Ref.~\refcite{4n}, where a simpler way
of obtaining this coefficient was presented, namely, the substitution
of the overlap integral by the matrix element of an exponential
function which in fact equals unity.

Instead of calculating the matrix element from the exponent, we
expand it in a series over the orders of $\cos\phi$ and $\sin\phi$
and transform this series into one over the Jacobi polynomials,
{\it i.e.} the eigenfunctions of the Laplacian. It is convenient to
carry
out this expansion in two steps: first expand the exponent over the
spherical Bessel functions and then collect these functions in
Jacobi polynomials. Since this way of calculating the transformation
coefficient may turn out to be useful also for other coefficients in
the theory of angular momenta, we present it below in detail.

Let us consider the six-dimensional vector
\be
\left( {P_i \atop i\,Q_i}\right).
\ee
We calculate the matrix element of the exponential function
\begin{equation}
\label{5.27}
 \exp \Big[-2(P^2_i+Q^2_i)\Big]
\end{equation}
over $J$ and $M$, with the condition $P_i^2-Q_i^2=0$. Let us
transform one of the multiplication factors in $P_i^2$ and $Q_i^2$,
expressing $P_i$ and $Q_i$ in terms of $P_k$ and $Q_k$ with the help
of the transformation (\ref{5.21}). Then
\begin{eqnarray}
\label{5.28}
 &&\langle\,j'_1j'_2\,|\exp\Big[-2(P^2_i+Q^2_i)\Big]
j_1j_2\, \rangle_{KJM}\ =
\nn \\
&&=\,\langle j'_1j'_2|\exp\!\Big[\!-2P_iP_k\cos\varphi\!
-\! 2Q_iP_k\cos\varphi\!
+2iQ_iP_k\sin\varphi\!+2iP_iQ_k\sin\varphi\Big]|j_1j_2\rangle_{KJM}.
\nn \\
\end{eqnarray}
The matrix element can be written as
\begin{eqnarray}
\label{5.29}
 &&
\sum_{m_1m_2} \! C^{JM}_{j_1mj_2m} C^{JM}_{j'_1m'_1j'_2m'_2}
\nn \\&\times&\int\!\!\exp\Big[\!-2P_iP_k\cos\varphi-2Q_iQ_k\cos\varphi
+2iQ_iP_k\sin\varphi +2iP_iQ_k\sin\varphi\Big]
\nonumber\\
&\times& Y^*_{j'_1m'_1}(\hat P_k) Y^*_{j'_2m'_2}
(\hat Q_i)d\hat P_id\hat Q_i d\hat P_k d\hat Q_k\,,
\end{eqnarray}
where $\hat{P}_i$, $\hat{Q}_i$, $\hat{P}_k$ and $\hat{Q}_k$ are the
usual three-dimensional polar angles. (Let us mention that the
expression (\ref{5.29}) coincides up to the normalization with the
formula for the transformation coefficient in Ref.~\refcite{18n}.) If we now
extract in this sum the term with eigenfunctions characterized by
the quantum numbers $j_1, j_2, j'_1, j'_2$, it turns out to be
equal, up to the normalization, to the coefficient we are looking
for. Indeed, the overlap integral is, by definition, a matrix
element of unity in a mixed representation.

Further, we apply the well-known expansion of the plane wave
\be
e^{ipx}\ =\ \sum_{\lambda_\mu} i^\lambda
j_\lambda(px)\,Y^*_{\lambda_\mu}(p)\,Y_{\lambda_\mu}(x)
\ee
and, expanding into a series the exponent in the integrand
(\ref{5.29}), we arrive at a rather long, but simple expression
\begin{eqnarray}
\label{5.30}
 &&
\sum_{{m_1m_2 \atop m'_1m'_2}} C^{JM}_{j_1m_1j_2m_2}
C^{JM}_{j'_1m'_1j'_2m'_2}\int\!\sum_{{pr\,qs\atop \pi\rho\kappa\sigma}}
(-1)^{(p+r+q+s)/2} j_p(2iP_iP_k \cos\varphi)j_r
(2Q_iP_k\sin\varphi)
\nonumber\\
&& \times\ j_q(2P_iQ_k\sin\varphi) j_s(2iQ_iQ_k\cos\varphi)
Y^*_{p\pi}(\hat P_i) Y^*_{q\kappa}(\hat P_i) Y_{j_1m_1}(\hat P_1)
Y^*_{r\rho} (Q_i)Y^*_{s\sigma}(Q_i)\ \times
\nonumber\\
&& \times\ Y_{j_2m_2}(Q_i)Y_{p\pi}(\hat P_k)Y_{r\rho}(\hat P_k)
Y^*_{j'_1m'_1}(\hat P_k)Y_{q\kappa}(\hat Q_k) Y_{s\sigma}(\hat Q_k)
Y^*_{j'_2m'_2}(\hat Q_k) d\hat P_i d\hat Q_i d\hat P_k d\hat Q_k\,.
\nn \\
\end{eqnarray}
Making use of the well-known features of the spherical functions
\be
Y^*_{lm}(\vartheta,\varphi)\ =\ (-1)^m Y_{l-m}(\vartheta,\varphi)
\ee
and
\be
\int\limits^{2\pi}_0\! d\varphi\!\int\limits^\pi_0\! d\vartheta
\sin\vartheta Y_{l_1m_1}(\vartheta,\varphi)
Y_{l_2m_2}(\vartheta,\varphi)=\left[
\frac{(2l_1+1)(2l_2+1)}{4\pi(2l_3+1)}\right]^{1/2}C^{l_30}_{l_10l_20}
C^{l_3m_3}_{l_1m_1l_2m_2}
\ee
(see Ref.~\refcite{17n}), we re-write (\ref{5.30}) in the form

 \begin{eqnarray}
 \label{5.31}
&&\frac{\pi^2}{16} \sum_{{pr\,qs\atop
\pi\rho\kappa\sigma}} \sum_{{m_1m_2 \atop m'_1m'_2}}
C^{JM}_{j_1m_1j_2m_2} C^{JM}_{j'_1m'_1j'_2m'_2}
C^{j_1m_1}_{p\pi\,q\kappa} C^{j_2m_2}_{r\rho s\sigma}
C^{j'_1m'_1}_{p\pi r\rho} C^{j'_2m'_2}_{q\kappa s\sigma}
C^{j_10}_{p0q0} C^{j_20}_{r0s0} C^{j'_10}_{p0r0} C^{j'_20}_{q0s0}
\nonumber \\
&& \times\ \frac{ (2p+1)(2r+1)(2q+1)(2s+1)}{[(2j_1+1)(2j_2+1)
(2j'_1+1)(2j'_2+1)]^{1/2}} j_p(2iP_iP_k\cos\varphi)\ \times
\nn \\
&&\times\ j_r(2Q_iP_k\sin\varphi) j_p(2P_iQ_k\sin\varphi)
j_s(2iQ_iQ_k\cos\varphi) (-1)^{m_1+m_2-j_1-j_2+1/2(p+r+q+s)}.\nn \\
\end{eqnarray}
The summation of the Clebsch-Gordan coefficients leads to the
$9j$-coefficient

\begin{eqnarray}
&&
\sum_{m_sm_{sk}} C^{j_{12}m_{12}}_{j_1m_1j_2m_2}
C^{j_{34}m_{34}}_{j_3m_3j_4m_4} C^{jm}_{j_{12}m_{12}j_{34}m_{34}}
C^{j_{13}m_{13}}_{j_1m_1j_3m_3} C^{j_{24}m_{24}}_{j_2m_2j_4m_4}
C^{j'm'}_{j_{13}m_{13}j_{24}m_{24}}\ =
\nn \\
&&=\ \delta_{jj'}\delta_{mm'}\Big[(2j_{12}+1)(2j_{13}+1) (2j_{24}+1)
(2j_{34}+1)\Big]^{1/2} \left\{ \begin{array}{ccc}
j_1 & j_2 & j_{12}\\ j_3 & j_4 & j_{34} \\ j_{13} & j_{24} & j
\end{array} \right\}.\nn \\
\end{eqnarray}
Hence, in (\ref{5.31}) we can get rid of the series of sums,
including the $9j$-coefficient, separating the quantum numbers
$j_1$, $j_2$, $j'_1$, $j'_2$, $J$:

\begin{eqnarray}
\label{5.32}
 &&
 \frac{\pi^2}{16} \sum_{pr\,qs} \left\{ \begin{array}{ccc}
p & r & j'_1\\  q & s & j_2\\ j_1 & j_2 & J \end{array} \right\}
C^{j'_10}_{p0r0} C^{j'_20}_{q0s0} C^{j_10}_{p0q0} C^{j_20}_{p0s0}
(2p+1)(2r+1)(2q+1)(2s+1)
\nn \\
& \times&(-1)^{1/2(p+r+q+s)}j_p(2iP_iP_k \cos\varphi)
j_r(2Q_iP_k\sin\varphi) j_q(2P_iQ_k\sin\varphi)
j_s(2iQ_iQ_k\cos\varphi). \nonumber \\
\end{eqnarray}
In the following, we apply the expression for the expansion of the
product of two Bessel functions into the sum of hypergeometric
functions (see Ref.~\refcite{21n}). Here we substitute in the standard
formulae the hypergeometric functions by Jacobi polynomials (or
Wigner's $d$-functions). Substituting also the lengths of the
vectors $P_i$, $P_k$, $Q_i$, $Q_k$ by unity, we can write
\begin{eqnarray}
\label{5.33}
 && \hspace*{-0.5cm} j_p(2i\cos\varphi)\,j_q(2\sin\varphi)\ =\
\frac\pi4 \sum_{K_1} P^{(p+1/2,q+1/2)}_{(K_1-p-q)/2}
(-\cos2\varphi)\ \times \nonumber
\\
&& \times\ \frac{(i)^{K_1-q}(\cos\varphi)^p(\sin\varphi)^q}{\Gamma
\Big(\frac{K_1+p-q}2+\frac32\Big(\Gamma\Big(\frac{K_1-p+q}2
+\frac32\Big)}\,j_s(2i\cos\varphi)j_r(2\sin\varphi)\ =
\nonumber \\
&&=\ \frac\pi4 \sum_{K_2} P^{(s+1/2,r+1/2)}_{(K_2-s-r)/2}
(-\cos2\varphi) \frac{(i)^{K_2-r}(\cos\varphi)^s(\sin\varphi)^r}{
\Gamma\Big(\frac{K_2+s-r}2+\frac32\Big)\Gamma\Big(\frac{K_2-p+q}2
+\frac32\Big)}.\quad
\end{eqnarray}
Selecting from these sums only terms with definite $K=K_1+K_2$, we
obtain the expression
\begin{eqnarray}
\label{5.34}
 && \hspace*{-0.5cm} \frac1{(16)^2} \sum_{pr\,qs} \left\{\!\!\left\{
\begin{array}{ccc} p & r & j'_1\\ q & s & j_2\\ j_1 & j_2 & J
\end{array}
\right\}\!\!\right\}\frac{(-1)^{K/2}}{\Gamma\Big(\frac{K_1-p+q}2
 +\frac32\Big)\Gamma\Big(\frac{K_1+p-q}2+\frac32\Big)}\
\times \nonumber
\\ && \times\
\left[\Gamma\left(\frac{K_2-r+s}2+\frac32\right)
\Gamma\left(\frac{K_2+r-s}2+\frac32\right)\right]\ \times
\nonumber
\\
&&\times\ (\cos\varphi)^{p+s}(\sin\varphi)^{q+r}
P^{(p+1/2,q+1/2)}_{(K_1-p-q)/2} (-\cos 2\varphi)
P^{(s+1/2,r+1/2)}_{(K_2-s-r)/2} (-\cos2\varphi). \qquad
\end{eqnarray}
The formula (\ref{5.34}) coincides with the general form of the coefficient
$\langle\,j'_1j'_2|j_1j_2\,\rangle^{\phi}_{KJM}$, given in Ref.~\refcite{19n}.

Let us, finally, present the orthonormalized transformation
coefficient in terms of the $d$-function which makes the
interpretation of different expressions easier with the help of the
six-dimensional rotations:
\begin{eqnarray}
\label{5.35}
 &&\langle
j'_1j'_2|j_1j_2\rangle^\varphi_{KJM}\!=\!\frac\pi2(-1)^{J+1}\!
\left(\frac{K_1\!-\!j_1}2\right)!\left(\frac{K_1\!+\!j_1\!+\!1}2\right)!
\left(\frac{K_2\!-\!j_2}2\right)!\left(\frac{K_2\!+\!j_2\!+\!1}2\right)!
\nonumber\\
&&\times\! \frac{a_{Kj'_1j'_2}}{a_{Kj_1j_2}} \sum_{pr\,qs}\!\left[\!
\left(\!\frac{K_1\!-\!p\!+\!q\!+\!1}2\!\right)!\!
\left(\!\frac{K_1\!+\!p\!-\!q\!+\!1}2\right)!\!
\left(\!\frac{K_1\!-\!p\!-\!q}2\right)!\!
\left(\!\frac{K_1\!+\!p\!+\!q+\!2}2\right)!\!
\right]^{-\frac12}
\nonumber\\
&&\times \left[\left(\frac{K_2\!-\!s\!+\!r\!+\!1}2\right)!
\left(\frac{K_2\!+\!s\!-\!r\!+\!1}2\right)!
\left(\frac{K_2\!-\!r\!-\!s}2\right)!
\left(\frac{K_2\!+\!r\!+\!s}2\right)!\right]^{1/2} \times
\nonumber\\
&&\times \left\{\left\{ \begin{array}{ccc}
p & r & j'_1\\ q & s & j'_2\\ j_1 & j_2 & J \end{array}\right\}\right\}
\frac1{\sin2\varphi} d^{(K_1+1)/2}_{q+p+1)/2,(p-q)/2} (2\varphi)
d^{(K_2+1)/2}_{(r+s+1)/2,(s-r)/2} (2\varphi)\,,
\end{eqnarray}
where
\bea
&&a_{Kj_1j_2}=
\nn \\
&&=\left[\left(\frac{K\!-\!j_1\!-\!j_2}2\right)!\left(
\frac{K\!+\!j_1\!+\!j_2}2+\!1\right)!\left(\frac{K\!-\!j_1\!+\!j_2\!+\!1}2
\right)!\left( \frac{K\!+\!j_1\!-\!j_2\!+\!1}2+\!1
\right)!\right]^{\frac12}. \nn \\
\eea
Since the normalizations of the Jacobi polynomial and the
$d$-function are known, we do not have to think about it while
carrying out the calculations. (Let us remind once more that the
factor $(\sin 2\phi)^{-1}$ is related to the six-dimensional
normalization.)

The expression (\ref{5.35}) is similar to the formula of the ``three
$d$-functions'' in $SU(2)$; it shows the expansion of one of the
$O(6)$ $d$-functions over the products of two $d$-functions of a
special form. As usual in such expressions, there is a freedom in
the choice of $K_1$ or $K_2$ ($K_1+K_2=K$).

\subsection{Applying
\boldmath$\langle\,j'_1j'_2|j_1j_2\,\rangle^{\Phi}_{KLM}$ to the
three-body problem}

Let us use the obtained formulae for the calculation of the
coefficients of the transition from $\Phi^{j'_1j'_2}_{KJM}(\eta,\xi)$
to $\Phi^{j_1j_2}_{KJM}(z,z^*)$. As it was shown already, this can
be done by the application of the transformation coefficient at the
$2\phi = \pi/2$ value, first substituting $\eta$ by $i\eta=\zeta$.
{The argument of the $d$-functions in (\ref{5.35}) is $2\phi$. The
turns by $\pi/2$ -- the so-called Weyl-coefficients -- are included
in many formulae of the theory of $O(n)$ representation.) The
function of the new arguments
\begin{equation}
\label{5.36}
 \Phi^{j_1j_2}_{KJM} (z,z^*)\ =\ \sum_{j'_1j'_2}\
\langle\,j'_1j'_2|j_1j_2\,\rangle^{\pi/4}_{KJM} \Phi^{j'_1j'_2}_{KJM}
(\zeta,\xi)
\end{equation}
can be written in the form
\begin{eqnarray}
\label{5.37}
 &&
\Phi^{j_1j_2}_{KJM} (z,z^*)\,=\, \frac\pi2(-1)^{J+1} \left(\!
\frac{K_1\!-\!j_1}2\!\right)!\left(\!\frac{K_1\!+\!j_1\!+\!1}2\!
\right)!\left(\!\frac{K_2\!-\!j_2}2\!\right)!\left(\!
\frac{K_2\!+j_2\!+\!1}2\!\right)!
\nonumber\\
&&\times\ \sum_{{j'_1j'_2 \atop m_1m_2}} C^{JM}_{j'_1m'_1j'_2m'_2}
N_{Kj'_1j'_2} \frac{a_{Kj'_1j'_2}}{a_{Kj_1j_2}} \sum_{pr\,qs}\bigg[\!
\left(\frac{K_1\!-\!p\!+\!q\!+\!1}2\!\right)!\left(\! \frac{K_1\!+\!p
\!-\!q\!+\!1}2 \right)!\ \times
\nonumber\\
&&\times\,\left(\!\frac{K_1\!-\!p\!-\!q}2\!\right)!\left(\!
\frac{K_1\!+\!p\!+\!q}2 +1\!\right)!\!\bigg]^{-\frac12}\!\bigg[\left(
\frac{K_2\!-\!s\!+\!r\!+\!1}2\!\right)!\left(\!
\frac{K_2\!+\!s\!-\!r\!+\!1}2 \right)!\ \times
\nonumber\\
&& \times\ \left(\frac{K_2\!-\!r\!-\!s}2\!\right)!\left(\!
\frac{K_2\!+\!r\!+\!s}2 +1\! \right)!\!\bigg]^{-\frac12}
\left\{\!\!\left\{\!\! \begin{array}{ccc}
p & r & j'_1\\ q & s & j_2\\ j_1 & j_2 & J\end{array}\!\!\right\}\!\!
\right\} d^{(K_1+1)/2}_{(q+p+1)/2,(p-q)/2}\Big(\frac\pi2\Big)\ \times
\nonumber\\
&& \times\ d^{(K_2+1)/2}_{(r+s+1)/2,(s-r)/2} \Big(\frac\pi2\Big)
Y_{j'_1m'_1}(z) Y_{j'_2m'_2}(z^*) \frac1{z^2\! -\!z^{*2}}
d^{(K+1)/2}_{(j'_1+j'_2+1)/2,(j'_1-j'_2)/2}(z^2\!+\!z^{*2}),\nn \\
\end{eqnarray}
where the relations
 \begin{eqnarray}
\label{5.38}
 && \frac12\,(z^2+z^{*2})\ =\ \xi^2-\eta^2\,, \quad
 zz^*\ =\ \xi^2+\eta^2\,.
\nonumber\\
&& z^2-z^{*2}\ =\ 2i\sin2\Phi\ =\ 4i\xi\eta\,,
\end{eqnarray}
are taken into account.

As it was mentioned in the previous section, it is convenient to use
$K_2=r+s$ in (\ref{5.37}).

The expression (\ref{5.37}) presents a series over the degrees of
$z$ and $z^*$. In each term of this series the degree of $z$ is
$p+q$, that of $z^*$ -- $r+s$. Considering the parametrization $z$,
 we see that each term of $z$ and $z^*$ introduces a
factor $exp(-i\lambda/2)$ and $exp(i\lambda/2)$, respectively.
Because of this, any term in the series (\ref{5.37}) will contain a
factor $exp(-i\lambda(p+q-r-s)/2)$. The series can be changed into a
Fourier series over $exp(-i\nu\lambda)$, if collecting all terms of
the series with a given
\begin{equation}
\label{5.39}
 p+q-r-s\ =\ 2\nu\,.
\end{equation}
In this case each term of the new series will have a definite value
of $\nu$ and, hence, the Fourier series will be at the same time a
series over the eigenfunctions of the operator $N$. We arrive at
functions characterized by the set $K,J,M,\nu, (j_1j_2)$. Their
normalization can be easily obtained from that of the $d$-functions.
The only remaining problem is the transition from $(j_1j_2)$ to
$\Omega$, which we have already mentioned above. Although a
multiplicity of the equations appears practically only at large $K$,
the construction of convenient expressions deserves further efforts.

\subsection{The $d$-function of the $O(6)$ group}

The coefficient $\langle j'_1j'_2|j_1j_2\rangle^{\phi}_{KJM}$
describes the rotations in the six-dimensional space. From these
rotations one can, obviously, construct an arbitrary rotation. The
$O(n)$ rotations are usually composed of $n(n-1)/2$ rotations on all
coordinate planes \cite{22n}. With the help of the calculated coefficients
rotations can be constructed in any plane, characterized by two
arbitrary vectors $\eta$ and $\xi$. In other words, these
coefficients lead to simultaneous rotations in three two-dimensional
planes $(\eta_x,\xi_x)$, $(\eta_y,\xi_x)$ and $(\eta_z,\xi_z)$ and
simplify the calculations seriously.

It would be rather interesting to generalize all this to the group
$O(n)$ $(n>6)$, and consider tensors of higher dimensions instead of
vectors.

\section{Symmetries in the Classical Three-Body Problem}

Quantum mechanical three-body systems possess the symmetry of motion
of a five-dimensional sphere with respect to both the free motion
and elastic forces (see for example Refs.~\refcite{2n,7n,10n}). As the
quantum mechanical problem obviously has the same symmetry as the
classical one, it seems to be worthwhile to consider the classical
equations of motion from this point of view.

Arbitrary motions of a three-body system can be described as
rotations and deformations of a triangle formed by the three
particles: the equations of motion of the triangle turn out to be
very similar to the equations of two coupled tops, one of them
reflecting the hidden (non-geometrical) symmetry of the deformative
motion of the triangle.

We collect here different types of equations and formulae connected
with the classical three-body problem.

\subsection{Examples of non-rotating triangles }

In dealing with a three-particle system, let us first recall the
used here system of coordinates. The radius vectors $\vec
x_l$ $(i=1,2,3)$ of the three particles are fixed by the condition
\begin{equation}
\vec x_1+\vec x_2+\vec x_3\ =\ 0.
\label{1.1}
\end{equation}
The Jacobi coordinates $\vec\xi$ and $\vec\eta$ are given in the case
of equal masses in the form
\begin{eqnarray}
&& \vec\xi\ =\ -\sqrt{\frac32}\, (\vec x_1+\vec x_2),
\nonumber\\
&& \vec\eta\ =\ \frac1{\sqrt2}\,(\vec x_1-\vec x_2),
\nonumber\\
&& \xi^2+\eta^2\ =\ x^2_1+x^2_2+x^2_3\ =\ \varrho^2\,,
\label{1.2}
\end{eqnarray}
where $\varrho$ is the radius of the five-dimensional sphere. Further,
we introduce the complex vector
\begin{eqnarray}
&& \vec z\ =\ \vec\xi+i\vec\eta\,,
\nonumber\\
&& \vec z^*\ =\ \vec\xi-i\vec\eta\,.
\label{1.3}
\end{eqnarray}
Consider now a triangle with vertices $x_1,x_2,x_3$. The position of
this triangle in space is characterized by the vectors $\vec l_1$ and
$\vec l_2$, which together with the vector
$\vec l=\vec l_1\times\vec l_2$ form the moving coordinate system. They
are connected with vectors $z$ and $z^*$ in the following way:
\begin{eqnarray}
&& \vec z\ =\ \frac\varrho{\sqrt2}\,e^{-i(\lambda/2)}
\Big(e^{i(a/2)}\vec l_1+ie^{-i(a/2)}\vec l_2\Big),
\nonumber\\
&& \vec z^*\ =\ \frac\varrho{\sqrt2}\,e^{i(\lambda/2)}
\Big(e^{-i(a/2)}\vec l_1-ie^{i(a/2)}\vec l_2\Big),
\label{1.4}
\end{eqnarray}
where $0\le a\le\pi$, $0\le\lambda\le2\pi$. The variables $\lambda$ and
$a$ determine the form of the triangle. Using expressions (\ref{1.4}),
we can write $\vec\xi$ and $\vec\eta$ in the form
\begin{eqnarray}
\label{1.5}
\vec\xi &=& \frac\varrho{\sqrt2}\left(\cos\frac{a-\lambda}2\vec l_1
+\sin\frac{a+\lambda}2\vec l_2\right),
\nonumber\\
\vec\eta &=& \frac\varrho{\sqrt2}\left(\sin\frac{a-\lambda}2\vec l_1
+\cos\frac{a+\lambda}2\vec l_2 \right).
\end{eqnarray}
This means that we can consider $\vec\xi$ and $\vec\eta$ as a result of
two transformations
\begin{equation}
\left[
\begin{array}{c}  \vec\xi\\  \vec\eta  \end{array}
\right] =\ \frac\varrho{\sqrt2}  \left[
\begin{array}{c}
\cos\frac\lambda2\,\sin\frac\lambda2 \\
-\sin\frac\lambda2\, \cos\frac\lambda2  \end{array} \right]
\left[ \begin{array}{c}
\cos\frac a2\,\sin\frac a2\\
\sin\frac a2\,\cos\frac a2 \end{array}  \right]
\left[ \begin{array}{c}  \vec l_1 \\ \vec l_2  \end{array} \right].
\label{1.6}
\end{equation}
To make the picture clearer, let us consider the case of a
non-rotating triangle. We need for this purpose the expressions
\begin{eqnarray}
&& \xi^2\ =\ \frac{\varrho^2}2\,(1+\sin a\sin\lambda)\,,
\nonumber\\
&& \eta^2\ =\ \frac{\rho^2}2\,(1-\sin a\sin\lambda)\,,
\label{1.7} \nn  \\
&&\vec\xi\vec\eta\ =\ \frac{\rho^2}2 \sin a\cos\lambda\,.
\label{1.8}
\end{eqnarray}
The angle $\theta$ between vectors $\vec\xi$ and $\vec\eta$
\begin{equation}
\vec\xi\vec\eta\ =\ |\xi|\,|\eta| \cos\theta
\label{1.9}
\end{equation}
can be written in terms of our variables as
\begin{equation}
\cos\theta\ =\ \frac{\cos\lambda
\sin a}{\sqrt{1-\sin^2a\sin^2\lambda}}\,.
\label{1.10}
\end{equation}
Note that the components of the moment of inertia are
\begin{equation}
\varrho^2\sin^2\Big(\frac a2-\frac\pi2\Big), \qquad
\varrho^2\cos^2\Big(\frac a2 - \frac\pi4\Big), \qquad\varrho^2.
\label{1.11}
\end{equation}
It is obvious that if $a=$const, the variations of $\lambda$ lead to
deformations of the triangle which do not affect the values of the
momenta of inertia. We can write
\begin{eqnarray}
&& |\xi|\ =\ \frac\varrho{\sqrt2}\,\sqrt{1+C\sin\lambda}\,,
\nonumber\\
&& |\eta|\ =\ \frac\varrho{\sqrt2}\,\sqrt{1-C\sin\lambda}\,,
\nonumber\\
&& \cos\theta\ =\ \frac{C \cos\lambda}{\sqrt{1-C^2\sin^2\lambda}}\,,
\label{1.12}
\end{eqnarray}
where $C =\sin a$.

For example, if $a=0$ ({\it i.e.} $C=0$), we have
\begin{equation}
|   \xi|=\frac\varrho{\sqrt2}\,, \quad |\eta|=\frac\varrho{\sqrt2}\,,
\quad \cos\theta=0\,.
\label{1.13}
\end{equation}
In this case vectors $\vec\xi$ and $\vec\eta$ are orthogonal
independently of the value of $\lambda$, and only similarity
transformations of the triangle are possible. On the other hand, if
$a=\pi/2$, then
\begin{eqnarray}
&& |\xi|\ =\ \frac\varrho{\sqrt2}\,\sqrt{1+\sin\lambda}\,,
\nonumber\\
&& |\eta|\ =\ \frac\varrho{\sqrt2}\,\sqrt{1-\sin\lambda}\,,
\nonumber\\
&& \cos\theta\ =\ 1\,,
\label{1.14}
\end{eqnarray}
{\it i.e.} the system is linear and the ends of the vectors $\vec\xi$
and $\vec\eta$ are oscillating about the point $\varrho/\sqrt2$.
Expressing the positions of all three particles in the c.m. system
in terms of $\vec\xi$ and $\vec\eta$:
\begin{eqnarray}
&& \vec x_1\ =\ \frac1{\sqrt6}\vec\xi +\frac1{\sqrt2}\vec\eta\ =\
\sqrt{\frac23}\Big[\cos\frac{2\pi}2\vec\xi+\sin\frac{2\pi}3\vec\eta
\Big],
\nonumber\\
&& \vec x_2\ =\ -\frac1{\sqrt6}\vec\xi -\frac1{\sqrt2}\vec\eta\ =\
\sqrt{\frac23}\Big[\cos\frac{4\pi}3\vec\xi+\sin\frac{4\pi}3\vec\eta
\Big],
\nonumber\\
&& \vec x_3\ =\ \sqrt\frac23\,\vec\xi
\label{1.15}
\end{eqnarray}
it will be easy to represent the position of the particles by their
radius vectors. As an illustration, we consider the case $a=\pi/2$ for
different values of $\lambda$:

\begin{figure}[h]
\centerline{\epsfig{file=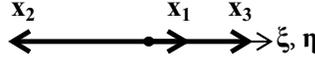,width=0.35\textwidth}}
\caption{$\alpha=\pi/2,\quad\lambda=0$}
\end{figure}
\begin{figure}[h]
\centerline{\epsfig{file=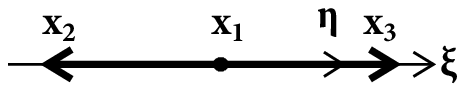,width=0.35\textwidth}}
\caption{$\alpha=\pi/2,\quad\lambda=\pi/6$}
\end{figure}

Consider now those deformations which are connected with the change
of $a$ {\it i.e.} those which do not leave the moment of inertia
unaltered.

Let $\lambda=0$, then
\begin{equation}
|   \xi|=\frac\varrho{\sqrt2}\,, \quad |\eta|=\frac\varrho{\sqrt2}\,,
\quad  \sin a=\cos\theta\,.
\label{1.16}
\end{equation}
The angle between $\vec\xi$ and $\vec\eta$ is
\[
\theta\ =\ \frac\pi2 -a\,.
\]
We list here a few special cases:
\begin{figure}[h]
\centerline{\epsfig{file=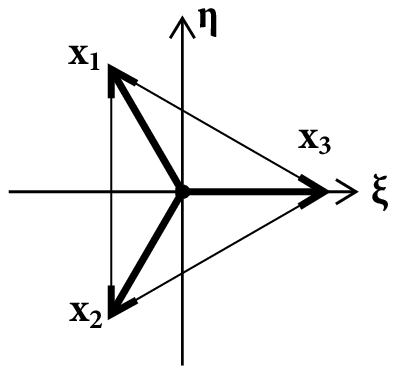,width=0.35\textwidth}\hspace{10mm}
            \epsfig{file=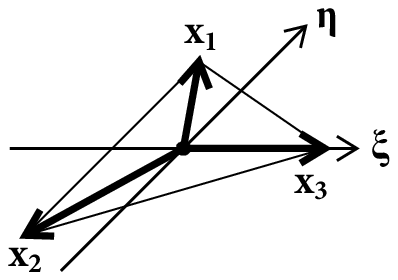,width=0.35\textwidth}}
\caption{$\alpha=0,\quad\lambda=0
\qquad\qquad\qquad\alpha=\pi/4,\quad\lambda=0$}
\end{figure}
\begin{figure}[h]
\centerline{\epsfig{file=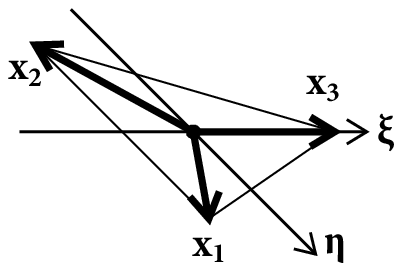,width=0.35\textwidth}\hspace{10mm}
            \epsfig{file=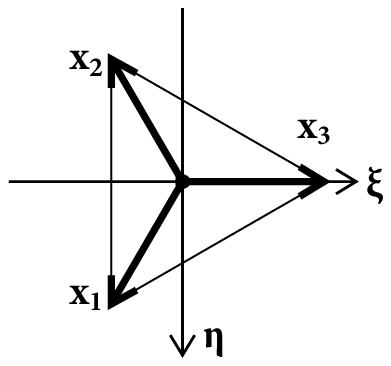,width=0.35\textwidth}}
\caption{$\alpha=3\pi/4,\quad\lambda=0
\qquad\qquad\qquad\alpha=\pi,\quad\lambda=0$}
\end{figure}

Considering the case $\lambda=\pi/2$, from the formulae
\begin{eqnarray}
&& |\xi|\ =\ \frac\varrho{\sqrt2}\,\sqrt{1+\sin a}\,,
\nonumber\\
&& |\eta|\ =\ \frac\varrho{\sqrt2}\,\sqrt{1-\sin a}\,,
\label{1.17} \\
&& \cos\theta\ =\ 0\,,
\nonumber
\end{eqnarray}
it can be easily seen that $\vec\xi$ and $\vec\eta$ are orthogonal and
their lengths can oscillate between zero and $\varrho$.

\begin{figure}[h]
\centerline{\epsfig{file=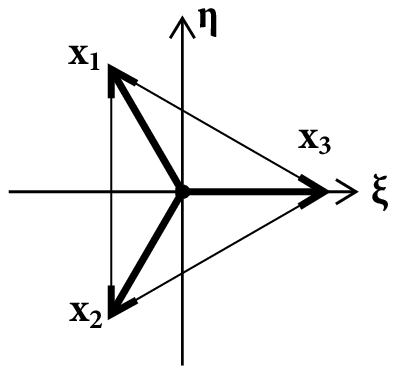,width=0.35\textwidth}\hspace{10mm}
            \epsfig{file=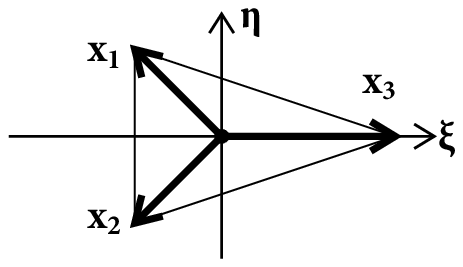,width=0.35\textwidth}}
\caption{$\alpha=0,\quad\lambda=\pi/2
\qquad\qquad\qquad\alpha=\pi/6,\quad\lambda=\pi/2$}
\end{figure}
\begin{figure}[h]
\centerline{\epsfig{file=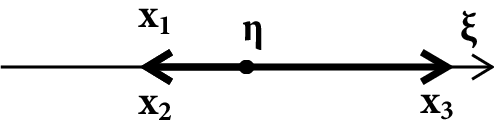,width=0.35\textwidth}}
\caption{$\alpha=\pi/2,\quad\lambda=\pi/2$}
\end{figure}


\subsection{The free Lagrangian}

In this subsection we present the Euler equations. First of all, we
have to construct the Lagrangian $L=T-U$. For free particles we have
\begin{equation}
L\ =\ T\ =\ \frac12\left[\begin{array}{c} ds\\ dt \end{array}
\right]^2.
\label{1.18}
\end{equation}

Let us begin with the expression
\begin{equation}
\label{1.19}
d\vec z\ =\ \frac1\varrho\,\vec zd\varrho-\frac i2\,\vec zd\lambda
+\frac12\,e^{?\lambda} (\vec l\times\vec z)^*da-(d\vec\omega\times
\vec z)\,,
\end{equation}
where $d\vec\omega$ is the infinitesimal rotation with projections
$d\omega_l$ onto the fixed axes. The rotations about the moving axes
are defined as
\begin{equation}
\label{1.20}
d\Omega_l\ =\ \vec l_l\,d\vec\omega\,.
\end{equation}
They can he expressed in terms of the Euler angles in the form
\begin{eqnarray}
&& d\Omega_1\ =\ -\cos\varphi_1\sin\theta d\varphi_2
+\sin\varphi_1 d\theta\,,
\nonumber\\
&& d\Omega_2\ =\ \sin\varphi_1\sin\theta d\varphi_2+\cos\varphi_1
d\theta\,,
\nonumber\\
&& d\Omega_3\ =\ -d\varphi_1-\cos\theta d\varphi_2\,.
\label{1.21}
\end{eqnarray}
 From (19) we get
\begin{eqnarray}
ds^2\ =\ |dz|^2 &=& \varrho^2\bigg[\frac14 da^2+\frac14 d\lambda^2
+\frac12 d\Omega^2_1+\frac12 d\Omega^2_2\ +
\nonumber\\
&& +\  d\Omega^2_3-\sin ad\,\Omega_1d\Omega_2-\cos ad\Omega_3 d\lambda
\bigg] +d\varrho^2.
\label{1.22}
\end{eqnarray}
Obviously, the wanted expression will be
\begin{eqnarray}
T &=& \frac12\varrho^2\bigg[\frac14\dot a^2+\frac14\dot\lambda^2
+\frac12\dot\Omega^2_1+\frac12\dot\Omega^2_2+\dot\Omega^2_3\ -
\nonumber\\
&&- \sin a\dot\Omega_1\dot\Omega_2-\cos a\dot\Omega_3\dot\lambda
\bigg]+\frac12\,\dot\varrho^2\,.
\label{1.23}
\end{eqnarray}
Due to the formula
\begin{equation}
p_t\ =\ \frac{\partial T}{\partial\dot q_t}\,,
\label{1.24}
\end{equation}
we can write the momenta
\begin{eqnarray}
&& p_a\ =\ \frac14\,\varrho^2\dot a\,,
\nonumber\\
&& p_\lambda\ =\ \frac12\,\varrho^2\left(\frac12\,\dot\lambda-\cos
a\dot\Omega_3\right),
\nonumber\\
&& p_{\Omega_1}\ =\ \frac12\,\varrho^2(\dot\Omega_1-\sin
a\dot\Omega_2),
\nonumber\\
&&  p_{\Omega_2}\ =\ \frac12\,\varrho^2(\dot\Omega_2-\sin
a\dot\Omega_1),
\nonumber\\
&& p_{\Omega_3}\ =\ \frac14\,\varrho^2(2\dot\Omega_3-\cos
a\dot\lambda),
\nonumber\\
&& p_\varrho\ =\ \dot\varrho,
\label{1.25}
\end{eqnarray}
and the corresponding $\dot p_i$:
\begin{eqnarray}
&& \dot p_a\ =\ \frac14\,\varrho\dot\varrho\dot
a+\frac12\,\varrho^2\ddot a,
\nonumber\\
&& \dot p_\lambda\ =\ \frac12\,\varrho^2\left(\frac14\ddot\lambda
+\sin a\dot a\dot\Omega_3-\cos a\ddot\Omega_3\right) +\varrho \left(
\frac12\dot\lambda\dot\varrho-\cos a\dot\varrho\dot\Omega_3\right),
\nonumber\\
&& \dot p_{\Omega_1}\ =\ \frac12\varrho^2(\ddot\Omega_1-\cos a\dot a
\dot\Omega_2-\sin a\ddot\Omega_2)+\varrho(\dot\varrho\dot\Omega_1
-\sin a\dot\varrho\dot\Omega_2),
\nonumber\\
&& \dot p_{\Omega_2}\ =\ \frac12\varrho(\ddot\Omega_2-\cos a\dot a
\dot\Omega_1-\sin a\ddot\Omega_1)+\varrho(\dot\varrho\dot\Omega_2
-\sin a\dot\varrho\dot\Omega_1),
\nonumber\\
&& \dot p_{\Omega_3}\ =\ \frac12\varrho^2(2\ddot\Omega_3+\sin a\dot a
\dot\lambda-\cos a\ddot\lambda)+\varrho(2\dot\varrho\dot\Omega_3
-\cos a\dot\varrho\dot\lambda),
\nonumber\\
&& \dot p_\varrho\ =\ \dot\varrho\,.
\label{1.26}
\end{eqnarray}
We can now construct the equations of motion
\begin{equation}
\frac d{dt}\,\frac{\partial T}{\partial\dot q_i} -\frac{\partial
T}{\partial q_i}\ =\ 0\,.
\label{1.27}
\end{equation}

To obtain the equations explicitly, we have to return to the Euler
angles. Indeed, as $\dot\Omega_i$ are not derivatives of any angles
$\Omega_i$ (that is why they are called quasi-coordinates), the Euler
equation in terms of these quasi-coordinates must be written in another
form.

Thus, instead of (\ref{1.23}) we have to take the Lagrangian expressed
in terms of the Euler angles:
\begin{eqnarray}
T &=& \frac12\,\varrho^2\Bigg[\frac14\dot a^2+\frac14\dot\lambda^2
+\dot\varphi^2_1+\frac12\dot\theta^2+\frac12\dot\varphi^2_2+\frac12
\cos^2\theta\dot\varphi^2_2+2\cos\theta\dot\varphi_1\dot\varphi_2
\nonumber\\
&& +\ \sin a\bigg(\frac12
\sin2\varphi_1\sin^2\theta\dot\varphi^2_2+\cos2\varphi_1\sin\theta
\dot\varphi_2\dot\theta-\frac12\sin2\varphi_1\dot\theta^2\bigg)
\nonumber\\
&&+\  \cos a(\dot\varphi_1 \dot\lambda+\cos\theta
\dot\varphi_2\dot\lambda)\Bigg]+\frac12\dot\varrho^2\,.
\label{1.28}
\end{eqnarray}
The equations of free motion are as follows:
\begin{eqnarray}
&& \hspace*{-0.5cm}
\frac12\ddot a-\cos a\left(\frac12\sin2\varphi_1\sin^2\theta
\dot\varphi^2_2+\cos2\varphi_1\sin\theta\dot\varphi_2\dot\theta
-\frac12\sin2\varphi_1\dot\theta^2\right)
\nonumber\\
&& +\ \sin a\Big(\dot\varphi_1\dot\lambda+\cos\theta\dot\varphi_2
\dot\lambda\Big) + \frac12\,\dot\varphi\dot a\ =\ 0\,,
\label{1.29}    \\
&& \hspace*{-0.5cm}
\frac12\ddot\lambda-\sin a(\dot a\dot\varphi_1+\cos\theta \dot a
\dot\varphi_2)+\cos a(\dot\varphi_1\sin\theta\dot\theta\dot\varphi_2
+\cos\theta\ddot\varphi_2)
\nonumber\\
&&+\ \frac1\varrho\,\dot\varrho\dot\lambda+\frac2\varrho\cos a
(\dot\varphi_1\dot\varrho+\cos\theta\dot\varphi_2\dot\varrho)\ =\ 0\,,
\label{1.30} \\
&&\hspace*{-0.5cm}
\dot\varphi_1-\sin\theta\dot\theta\dot\varphi_2+\cos\theta\dot\varphi_2
+\frac12 \cos a\dot\lambda\ -
\nonumber\\
&& -\ \frac12\sin a\Big[\dot a\dot\lambda+\cos2\varphi_1\sin^2\theta
\dot\varphi^2_2 -2\sin2\varphi_1\sin\theta\dot\varphi_2\dot\theta
-\cos2\varphi_1\dot\theta^2\Big]
\nonumber\\
&& +\ \frac1\varrho\Big(2\dot\varphi_1\dot\varrho+2\cos\theta
\dot\varphi_2\dot\varrho+ \cos a\dot\lambda\dot\varrho\Big)\ =\ 0,
\label{1.31} \\
&& \hspace*{-0.5cm}
\frac12\sin a\cos2\varphi_1\ddot\theta+\frac12\sin\theta(1+\sin a
2\varphi_1)\ddot\varphi_2 -(1+\sin a\sin2\varphi_1)\dot\varphi_1
\dot\theta
\nonumber\\
&& +\ \sin a\Big(\cos2\varphi_1\sin\theta\dot\varphi_1\varphi_2
-\frac12\ctg\theta\cos2\varphi_1\dot\theta^2+\sin2\varphi_1\cos\theta
\dot\varphi_2\dot\theta\Big)
\nonumber\\
&& +\ \cos a\left(\frac12\sin2\varphi_1\theta\dot a\dot\varphi_2
+\frac12 \cos 2\varphi_1\dot\theta\dot a-\frac12\dot\lambda\dot\theta
\right)
\nonumber\\
&&+\ \frac1\varrho\Big[\sin\theta\dot\varphi_2\dot\varrho(1+\sin a
\sin2\varphi_1)+\sin a\cos2\varphi_1\dot\theta\dot\varrho\Big]=\ 0,
\label{1.32}\\
&& \hspace*{-0.5cm}
(1-\sin a\sin\varphi_1)\ddot\theta+\frac12\sin2\theta\dot\varphi^2_2
+2\sin\theta\dot\varphi_1\dot\varphi_2
+\sin a\Big(\!\! -2\cos2\varphi_1\dot\varphi_1\theta
\nonumber\\
&& +\ \cos2\varphi_1
\sin\theta\ddot\varphi_2\,2\sin2\varphi_1\sin\theta\dot\varphi_1
\dot\varphi_2-\frac12\sin2\varphi_1\sin2\theta\dot\varphi^2_2\Big)
\nonumber\\
&& +\ \cos a\Big(\cos2\varphi_1\sin\theta\dot a\dot\varphi^2_2
-\sin2\varphi_1\dot a\dot\theta +\sin\theta\dot\varphi_2\dot\lambda
\Big)
\nonumber\\
&& +\ \frac2\varrho\Big(\dot\theta\dot\varrho+\sin a\cos2\varphi_1
\sin\theta\dot\varrho\dot\varphi_2-\sin a\sin2\varphi_1\dot\varrho
\dot\theta\Big)\ =\ 0;
\label{1.33}
\end{eqnarray}
and finally,
\begin{eqnarray}
&& \hspace*{-0.5cm}
\dot\varrho-\varrho\bigg[\frac14\dot a^2+\frac14\dot\lambda^2+
\dot\varphi^2_1+\frac12\dot\theta^2+\frac12\dot\varphi^2_2+\frac12
\cos^2\theta\dot\varphi^2_2 +2\cos\theta\dot\varphi_1\dot\varphi_2
\nonumber\\
&& +\ \sin a\left(\frac12\sin2\varphi_1\sin^2\theta\dot\varphi^2_2
+\cos2\varphi_1\sin\theta\dot\varphi_2\dot\theta
-\frac12\sin2\varphi_1\dot\theta^2\right)
\nonumber\\
&& +\ \cos a (\dot\varphi_1\dot\lambda
+\cos\theta\dot\varphi_2\dot\lambda)\bigg]\ =\ 0.
\label{1.34}
\end{eqnarray}

A few particular cases are considered below.

1) {\em Motion of the triangle in the plane:}
\[
\dot\theta\ =\ \dot\varphi_2\ =\ 0\,.
\]
In this case the free Lagrangian can be written
\begin{equation}
T\ =\ \frac12\varrho^2\left[\frac14\,\dot a^2+\frac14\,\dot\lambda^2
+\cos a\dot\varphi_1\dot\lambda+\dot\varphi^2_1\right]
+\frac12\,\dot\varrho^2\,,
\label{1.35}
\end{equation}
or, remembering that
\begin{eqnarray}
p_{\varphi_1} &=& \frac12\,\varrho^2\Big(2\dot\varphi_1+2\cos\theta
\dot\varphi_2+\cos a\dot\lambda\Big),
\nonumber\\
p_{\varphi_2} &=& \frac12\,\varrho^2\Big(\dot\varphi_2+\cos^2\theta
\dot\varphi_2-\sin a\sin2\varphi_1\sin^2\theta\dot\varphi_2
\nonumber\\
&& +\ 2\cos\theta\dot\varphi_1+\sin a\cos2\varphi_1\sin\theta\dot\theta
+\cos a\cos\theta\dot\lambda\Big),
\nonumber\\
p_\theta &=& \frac12\,\varrho^2\Big(\dot\theta+\sin a\cos2\varphi_1
\sin\theta\dot\varphi_2-\sin a\sin2\varphi_1\dot\theta\Big),
\label{1.36}
\end{eqnarray}
in the form
\begin{equation}
T\ =\ \frac1{\varrho^2}\left[2p^2_a+\frac1{\sin^2a}\Big(
\frac12p^2_{\varphi_1} -2p_{\varphi_1}p_\lambda\cos
a+2p^2_\lambda\Big)\right]+\frac12p^2_\varrho\,.
\label{1.37}
\end{equation}
The equations of motion are in this case
\begin{eqnarray}
&& \frac12\,\ddot a+\frac1\varrho\,\dot\varrho\dot a+\sin a\dot\varphi_1
\dot\lambda\ =\ 0\,,
\nonumber\\
&& \frac12\dot\lambda+\frac1\varrho\dot\varrho\dot\lambda-\sin a\dot a
\dot\varphi_1+\cos a\ddot\varphi+\frac2\varrho\cos a\dot\varphi_1
\dot\varrho\ =\ 0\,,
\nonumber\\
&& \ddot\varphi_1+\frac12\cos a\lambda-\frac12\sin a\dot a\dot\lambda
+\frac1\varrho(2\dot\varphi_1\dot\varrho+\cos a\dot\lambda\dot\varrho)\
=\ 0\,,
\nonumber\\
&& \ddot\varrho-\varrho\left(\frac14\,\dot a^2+\frac14\,\dot\lambda^2
+\dot\varphi^2_1+\cos a\dot\varphi_1\dot\lambda\right) =\ 0\,.
\label{1.38}
\end{eqnarray}

2) {\em Deforming triangle: }
\[
\dot\theta\ =\ \dot\varphi_2\ =\ 0, \qquad p_{\varphi_1}\ =\ 0.
\]

The free Lagrangian obtains the form
\begin{equation}
T\ =\ \frac12\varrho^2\left(\frac14\dot a^2+\frac14\dot\lambda^2
\sin^2 a\right)+\frac12\dot\varrho^2=\ \frac2{\varrho^2}\left(p^2_a
+\frac1{\sin^2a}p^2_\lambda\right)+\frac12p^2_\varrho\,.
\label{1.39}
\end{equation}

It should be noted that for $p_\varrho=0$ this expression has the same
form as the Lagrangian of the rotator. We see here an example of the
hidden symmetry, which can be generalized to the case of a deforming
rotator.

The equations of motion corresponding to the Lagrange function (39) are
\begin{eqnarray}
&& \frac12\ddot a+\frac1\varrho \dot\varrho\dot a -\frac12\sin a \cos
a\dot\lambda^2\ =\ 0,
\nonumber\\
&& \sin a\left(\frac12\ddot\lambda+\frac1\varrho\dot\varrho\dot\lambda
\right)+\cos a\dot a\dot\lambda\ =\ 0,
\nonumber\\
&& \varrho-\varrho\left(\frac14\dot
a^2+\frac14\sin^2a\dot\lambda^2\right) =\ 0\,.
\label{1.40}
\end{eqnarray}
If we add to the right-hand side forces depending only on $\varrho$,
 we get the equations of a non-rigid rotator.

\subsection{Potentials for three-body systems}

Two examples of interacting particles will be investigated.

\subsubsection{The harmonic oscillator potential}

The equilibrium state of the three-particle system is an equilateral
triangle of side $\varrho_0$ which can be described by the vectors
$\vec\xi_0$ and $\vec\eta_0$:
\begin{eqnarray}
&& \vec\xi_0\ = \left(\begin{array}{c}
0\\ \frac{\varrho_0}{\sqrt2} \end{array} \right),  \qquad
\vec\eta\ =\ \left( \begin{array}{c}
\frac{\varrho_0}{\sqrt2} \\  0  \end{array} \right),
\nonumber\\
&& \vec\xi_0 \vec\eta_0\ =\ 0, \qquad \xi^2_0-\eta^2_0\ =\ 0.
\label{1.41}
\end{eqnarray}
The parameters $a_0$ and $\lambda_0$ corresponding to the equilibrium
state will have the values
\[
a_0\ =\ \pi, \qquad \lambda_0\ =\ 0,
\]
and consequently
\begin{equation}
\vec z_0\ =\ \frac{\varrho_0}{\sqrt2}\Big(e^{i(\pi/2)}\vec l_1
+ie^{-i(\pi/2)} \vec l_2\Big).
\label{1.42}
\end{equation}

Consider the motion of the three particles with the potential energy
\begin{equation}
U\ =\ \frac12\Big[(\xi-\xi_0)^2+(\eta-\eta_0)^2\Big]=\ \frac12 \left(
\varrho^2+\varrho^2_0\,2\varrho\varrho_0\sin\frac a2 \cos\lambda2
\right).
\label{1.43}
\end{equation}
From the expression
\begin{equation}
F_t\ =\ - \frac{\partial U}{\partial q_i}\,,
\label{1.44}
\end{equation}
we obtain
\begin{eqnarray}
&& F_a\ =\ \frac12\,\varrho\varrho_0\cos\frac a2 \cos\frac\lambda2\,,
\nonumber\\
&& F_\lambda\ =\ -\frac12\,\varrho\varrho_0\sin\frac a2
\sin\frac\lambda2\,,
\nonumber\\
&& F_\varrho\ =\ -\varrho+\varrho_0\sin\frac a2\cos\frac\lambda2\,,
\nonumber\\
&& F_{\varphi_1}\ =\ F_{\varphi_2}\ =\ F_\theta\ =\ 0\,.
\label{1.45}
\end{eqnarray}
Constructing $L=T-U$, it is now easy to get the equations of motion
\begin{equation}
\label{1.46}
\frac d{dt}\,\frac{\partial L}{\partial\dot q}
-\frac{\partial L}{\partial q_l}\ =\ 0\,.
\end{equation}
The equations
\be
\frac d{dt}\,\frac{\partial L}{\partial\dot\varphi_l}
-\frac{\partial L}{\partial\varphi_l}\ =\ 0
\ee
(where $\varphi_3=\theta$) stay unchanged; instead of Eqs.
(\ref{1.29}), (\ref{1.30}) and (\ref{1.34}) we obtain
\begin{eqnarray}
&& \hspace*{-0.5cm}
\frac12\ddot a-\cos a\left(\frac12\sin2\varphi_1\sin^2\theta
\dot\varphi^2_2+\cos2\varphi_1 \sin\theta\dot\varphi_2\dot\theta
-\frac12\sin2\varphi_1\dot\theta^2\right)
\nonumber\\
&&+\ \sin a(\dot\varphi_1\dot\lambda+\cos\theta\dot\varphi_2
\dot\lambda)+\frac1\varrho\dot\varrho\dot a-\frac{\varrho_0}\varrho
\cos\frac a2\cos\frac\lambda2\ =\ 0,
\label{1.47} \\
&&
\frac12\ddot\lambda-\sin a(\dot a\dot\varphi_1+\cos\theta\dot a
\dot\varphi_2)+\cos a(\varphi_1-\sin\theta\dot\theta\dot\varphi_2
+\cos\theta\dot\varphi_2)\ +
\nonumber\\
&&+\ \frac1\varrho\dot\varrho\dot\lambda+\frac2\varrho \cos a
(\dot\varphi_1\dot\varrho+\cos\theta\dot\varphi_2\dot\varrho)
+\frac{\varrho_0}\varrho \sin\frac a2 \sin\frac\lambda2\ =\ 0,
\label{1.48}
\end{eqnarray}
and
\begin{eqnarray}
&&
\varrho-\varrho\bigg[\frac14\dot a^2+\frac14\dot\lambda^2
+\dot\varphi^2_1+\frac12\dot\theta^2+\frac12\dot\varphi^2_2+\frac12
\cos^2\theta\dot\varphi^2_2+2\cos\theta\dot\varphi_1\dot\varphi_2
\nonumber\\
&&+\ \sin a\left(\frac12\sin2\varphi_1\sin^2\theta\dot\varphi^2_2
+\cos2\varphi_1\sin\theta\dot\varphi_2\dot\theta -\frac12 \sin
2\varphi_1\dot\theta\right)
\nonumber\\
&&+\ \cos a\Big(\dot\varphi_1\dot\lambda+\cos\theta\dot\varphi_2
\dot\lambda\Big)\bigg]+\varrho-\varrho_0\sin\frac a2
\cos\frac\lambda2\ =\ 0.
\label{1.49}
\end{eqnarray}

Let us consider again the case of a non-rotating triangle:
\be
\dot\theta=\dot\varphi_2=0\ \mbox{ and }\ p_{\varphi_1}=0,\
\mbox{ i.e. }\  \dot\varphi_1=-\frac12\cos a\dot\lambda.
\ee
The equations of motion assume the form
\begin{eqnarray}
&& \frac12\dot a+\frac1\varrho\dot\varrho\dot a-\frac12\sin a\cos a
\dot\lambda^2\ \frac{\varrho_0}\varrho\cos\frac a2\frac\lambda2\ =\ 0,
\nonumber\\
&&
\sin^2a\left(\frac12\lambda+\frac1\varrho\dot\varrho\dot\lambda\right)
+\sin a\cos a\dot a\dot\lambda+\frac{\varrho_0}\varrho\sin\frac a2
\sin\frac\lambda2\ =\ 0,
\nonumber\\
&& \dot\varrho-\varrho\left(\frac14\dot a^2+\frac14\sin^2a\dot\lambda^2
\right)+\varrho-\varrho_0\sin\frac a2\cos\frac\lambda2\ =\ 0.
\label{1.50}
\end{eqnarray}
If in this case we take $\varrho_0=0$, we will have
\begin{eqnarray}
&& \frac12\ddot a+\frac1\varrho\dot\varrho\dot a-\frac12\sin a
\cos a \dot\lambda^2\ =\ 0,
\nonumber\\
&& \sin a\left(\frac12\lambda+\frac1\varrho\dot\varrho\dot\lambda
\right) +\cos a\dot a\dot\lambda\ =\ 0,
\nonumber\\
&& \ddot\varrho-\varrho\left(\frac14\dot a^2+\frac14\sin^2
a\dot\lambda^2\right)+\varrho\ =\ 0.
\label{1.51}
\end{eqnarray}
As an example, let us consider solutions for constant $\lambda$ and
$a$, illustrating a few cases of the deformed triangle in detail. The
projections of the radius vectors $\vec x_l$; onto the axes $\vec l_1$
and $\vec l_2$ are as follows:
\begin{eqnarray}
&& x^{(1)}_1\ =\ \frac\varrho2\left(\sin\frac{a-\lambda}2
-\frac1{\sqrt3} \cos \frac{a-\lambda}2\right),
\nn \\
&& x^{(1)}_2\ =\ -\frac\varrho2\left(\sin\frac{a-\lambda}2
+\frac1{\sqrt3} \cos\frac{a-\lambda}2\right),
\nn \\
&& x^{(1)}_3\ =\ \frac\varrho{\sqrt3} \cos \frac{a-\lambda}2\,,
\end{eqnarray}
and
\begin{eqnarray}
&& x^{(2)}_1\ =\ \frac\varrho2\left(\cos\frac{a+\lambda}2
-\frac1{\sqrt3} \sin \frac{a+\lambda}2 \right),
\nonumber\\
&& x^{(2)}_2\ =\ -\frac\varrho2\left( \cos\frac{a+\lambda}2
+\frac1{\sqrt3} \sin \frac{a+\lambda}2\right),
\nonumber\\
&& x^{(2)}_3\ =\ \frac\varrho{\sqrt3} \sin\frac{a+\lambda}2\,.
\label{1.52}
\end{eqnarray}
If $a=$const, $\lambda=$const and $\varrho_0\neq0$, we obtain from
(\ref{1.50})
\begin{eqnarray}
&& \frac{\varrho_0}\varrho \cos\frac a2\cos\frac\lambda2\ =\ 0,
\nonumber\\
&& \frac{\varrho_0}\varrho \sin\frac a2 \sin\frac\lambda2\ =\ 0,
\nonumber\\
&& \varrho+\varrho-\varrho_0\sin\frac a2 \cos\frac\lambda2\ =\ 0.
\label{1.53}
\end{eqnarray}
It can be easily seen that in this case $a$ and $\lambda$ are multiples
of $\pi$, and only similarity transformations are possible; for
example:
\begin{figure}[h]
\centerline{\epsfig{file=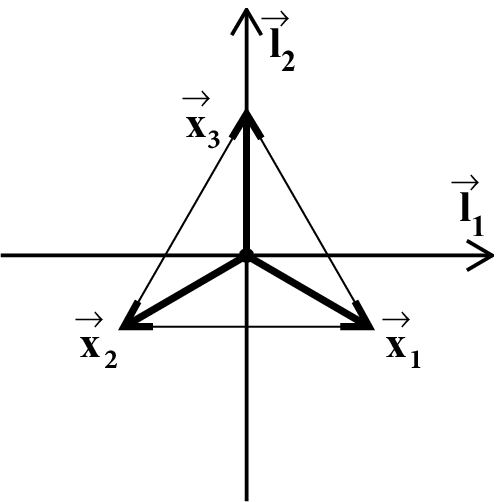,width=0.30\textwidth}\hspace{10mm}
            \epsfig{file=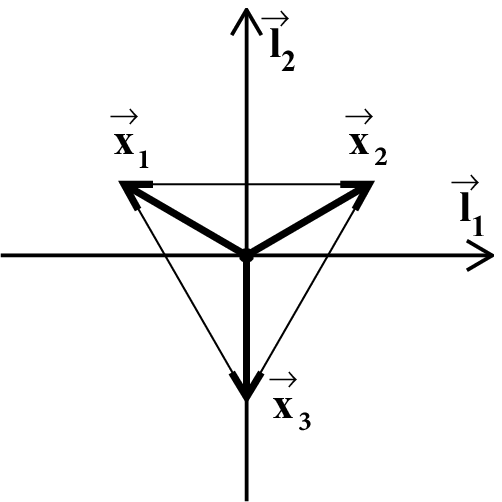,width=0.30\textwidth}}
\end{figure}
\be
\begin{array}{lcl}
\lambda=0,\ a=\pi,  &\qquad\qquad\qquad&  \lambda=0,\ a=3\pi
\\
x^{(1}_1\ =\ \displaystyle\frac\varrho2, &&
x^{(1)}_1\ =\ \displaystyle\frac\varrho2,
\\
x^{(1)}_2\ =\ \displaystyle\frac\varrho2, &&
x^{(1)}_2\ =\ \displaystyle\frac\varrho2,
\\
x^{(1)}_3\ =\ 0 && x^{(1)}_3=0,
\\
x^{(2)}_1\ =\ \displaystyle\frac1{2\sqrt3}\,\varrho, &&
x^{(2)}_1\ =\ \displaystyle\frac \varrho{2\sqrt3},
\\
x^{(2)}_2\ =\ -\displaystyle\frac 1{2\sqrt3}\varrho,  &&
x^{(2)}_2\ =\ \displaystyle\frac \varrho{2\sqrt3},
\\
x^{(2)}_3\ =\ \displaystyle\frac \varrho{\sqrt3}, &&
   x^{(2)}_3\ =\ \displaystyle\frac \varrho{\sqrt3}\,.
\end{array}
\ee

If, on the contrary, $\varrho=0$, then the fixed value of $a$ still
does not determine the value of $\lambda$, so that arbitrary
deformations are possible. We give in the following a few examples:
\[
\begin{array}{lcl}
\mbox{\bf a)  }\lambda = a\qquad
x^{(1)}_1=-\displaystyle\frac\varrho{2\sqrt3}, &\qquad&
x^{(2)}_1=\displaystyle\frac\varrho2\Big(\cos a
-\displaystyle\frac1{\sqrt3} \sin a\Big),
\\
\qquad \qquad \qquad x^{(1)}_2=-\displaystyle\frac\varrho{2\sqrt3}, &&
x^{(2)}_2=- \displaystyle\frac\rho2
\Big(\cos a+\displaystyle\frac1{\sqrt3}\sin a\Big),
\\
\qquad \qquad \qquad x^{(1)}_3=\displaystyle\frac\varrho{\sqrt3}, &&
x^{(2)}_3 =\displaystyle\frac\varrho{\sqrt3} \sin a.
\end{array}
\]
\begin{figure}[h]
\centerline{\epsfig{file=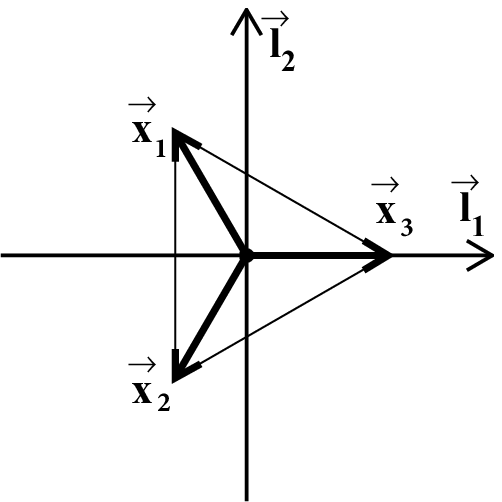,width=0.30\textwidth}\hspace{10mm}
            \epsfig{file=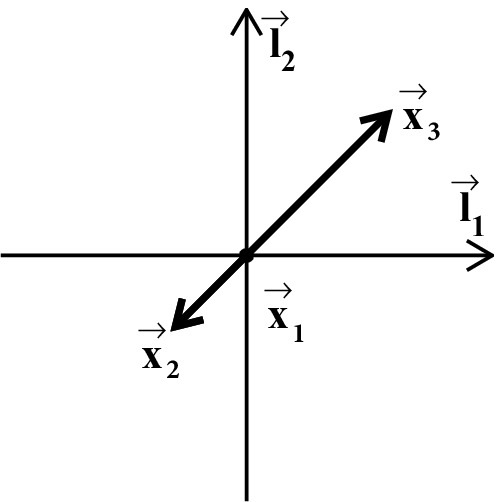,width=0.30\textwidth}}
$ \qquad\qquad\qquad  \qquad\qquad
\lambda=a=0, \qquad\qquad\qquad\qquad \lambda=a=\frac\pi2,$
\end{figure}
\bea
x^{(2)}_1=\frac\rho2\,,\ x^{(2)}_2=\frac\rho2\,,\ x^{(2)}_3=0\,,\
\qquad
\qquad
x^{(2)}_1=-\frac\varrho{2\sqrt3}\,,\ x^{(2)}_2=-\frac\varrho{2\sqrt3},\
x^{(2)}_3=\frac\varrho2. \nn
\eea
\begin{figure}[h]
\centerline{\epsfig{file=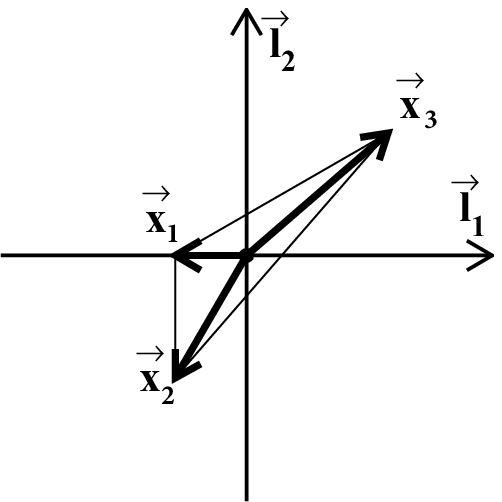,width=0.30\textwidth}\hspace{10mm}
            \epsfig{file=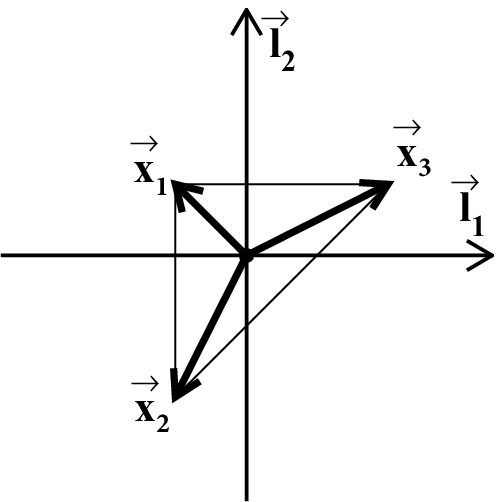,width=0.30\textwidth}}
$ \qquad\qquad\qquad  \qquad\qquad
\lambda=a=\frac\pi3, \qquad\qquad\qquad\qquad \lambda=a=\frac\pi6,$
\vspace{-5mm}
\end{figure}
\[
x^{(2)}_1=0, \quad x^{(2)}_2=\frac\varrho2, \;
x^{(2)}_3=\frac\varrho2,
\qquad
\qquad
x^{(2)}_1=\frac\varrho{2\sqrt3}, \;
x^{(2)}_2=\frac\varrho{\sqrt3}, \;
x^{(2)}_3=\frac\varrho{2\sqrt3}\,.
\]

\begin{figure}[h]
\centerline{\epsfig{file=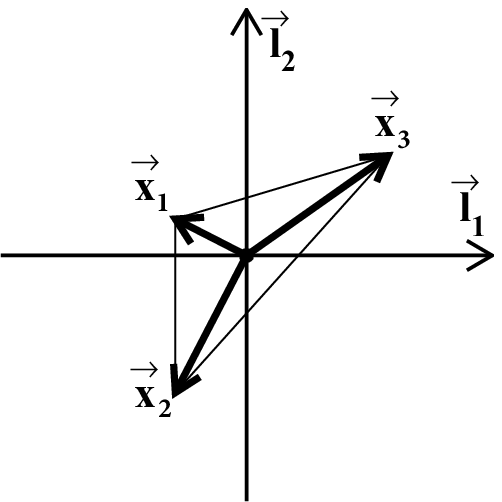,width=0.30\textwidth}\hspace{10mm}
            \epsfig{file=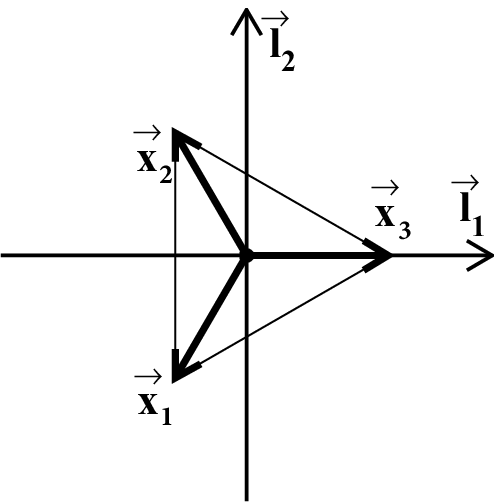,width=0.30\textwidth}}
$ \qquad\qquad\qquad  \qquad\qquad
\lambda=a=\frac\pi4, \qquad\qquad\qquad\qquad \lambda=a=\pi,$
\vspace{-10mm}
\end{figure}
\[
\begin{array}{lcl}
x^{(2)}_1=\frac\varrho{2\sqrt3}\left(1-\frac1{2\sqrt3}\right),
&\qquad&
x^{(2)}_1=-\frac\varrho2\,, \\
x^{(2)}_2=- \frac\varrho{2\sqrt2}\left(1+\frac1{\sqrt3}\right),
&\qquad&
x^{(2)}_2=\frac\varrho2\,,  \\
x^{(2)}_3=\frac\varrho{\sqrt6},
&\qquad&
x^{(2)}_3=0.
\end{array}
\]
\[
\mbox{\bf b)  }
a-\lambda=\frac\pi3, \quad x^{(3)}_1=0, \quad
x^{(2)}_1=\frac\varrho2\bigg(\cos\Big(\lambda+\frac\pi6\Big)
\frac1{\sqrt3} \sin\Big(\lambda+\frac\pi6\Big)\bigg) ,
\]
\vspace{-8mm}
\begin{eqnarray*}
&&\qquad\qquad\qquad\qquad\quad
x^{(1)}_2=-\frac\varrho2, \quad
x^{(2)}_2=\frac\varrho2\bigg(\cos\Big(\lambda+\frac\pi6\Big)
+\frac1{\sqrt3}\sin\Big(\lambda+\frac\pi6\Big)\bigg),
\\
&&\qquad\qquad\qquad\qquad\quad
x^{(1)}_3=\frac\varrho2, \qquad x^{(2)}_3=\frac\varrho{\sqrt3}
\sin\Big(\lambda+\frac\pi6\Big),
\end{eqnarray*}
\newpage
\begin{figure}[h]
\centerline{\epsfig{file=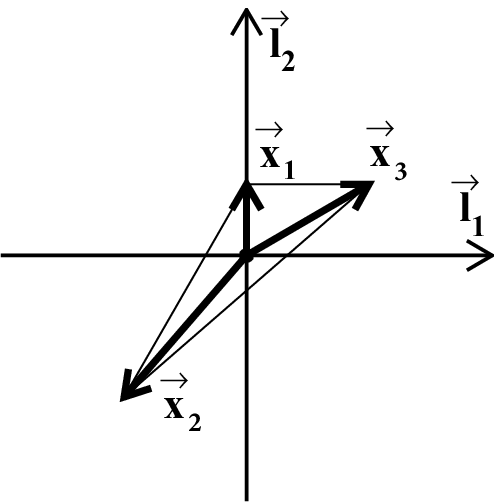,width=0.30\textwidth}\hspace{10mm}
            \epsfig{file=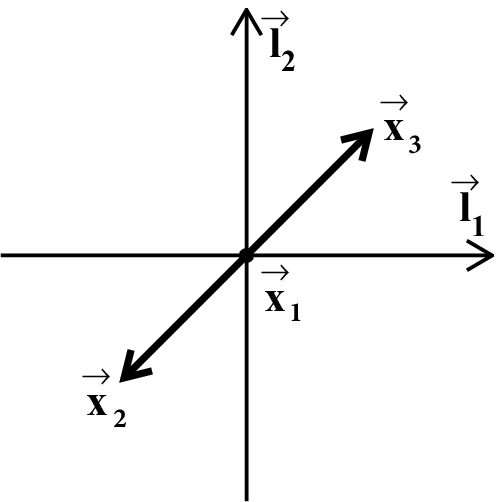,width=0.30\textwidth}}
$ \qquad\qquad\qquad
\lambda\ =\ 0, \quad a\ =\ \frac\pi3, \qquad\qquad\qquad\qquad
\lambda\ =\ \frac\pi6, \quad a\ =\ \frac\pi2,$
\vspace{-9mm}
\end{figure}
\[
\begin{array}{lcl}
x^{(2)}_1=\frac\varrho{2\sqrt3}, \;
x^{(2)}_2=-\frac\varrho{\sqrt3}, \;
x^{(2)}_3=\frac\varrho{2\sqrt3},
&\qquad&
x^{(2)}_1=0, \;
x^{(2)}_2=\frac\varrho2, \;
x^{(2)}_3=\frac\varrho2.
\end{array}
\]
\begin{figure}[h]
\centerline{\epsfig{file=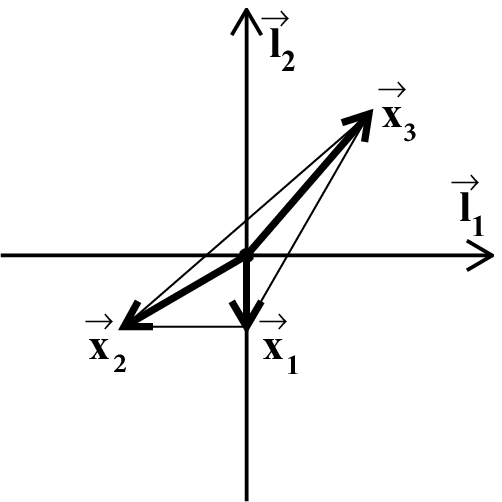,width=0.30\textwidth}\hspace{10mm}
            \epsfig{file=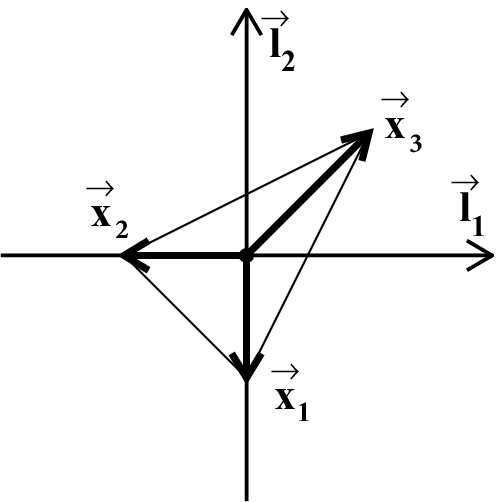,width=0.30\textwidth}}
$ \qquad\qquad\qquad
\lambda\ =\ \frac\pi3, \quad a\ =\ \frac{2\pi}3,
\qquad\qquad\qquad\qquad \lambda\ =\ \frac\pi3, \quad a\ =\
\frac{5\pi}6,$
\vspace{-8mm}
\end{figure}
\[
\begin{array}{lcl}
x^{(2)}_1= -\frac\varrho{2\sqrt3}, \;
x^{(2)}_2=\frac\varrho{2\sqrt3}, \;
x^{(2)}_3=\frac\varrho{\sqrt3},
&\qquad&
x^{(2)}_1=- \frac\varrho2, \;
x^{(2)}_2=0, \;
x^{(2)}_3=\frac\varrho2.
\end{array}
\]
\begin{figure}[h]
\centerline{\epsfig{file=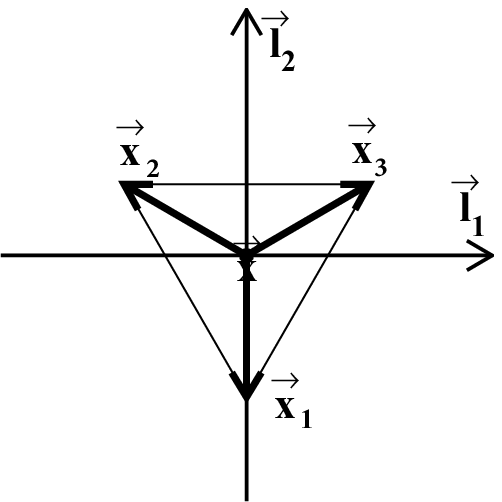,width=0.30\textwidth}\hspace{10mm}
            \epsfig{file=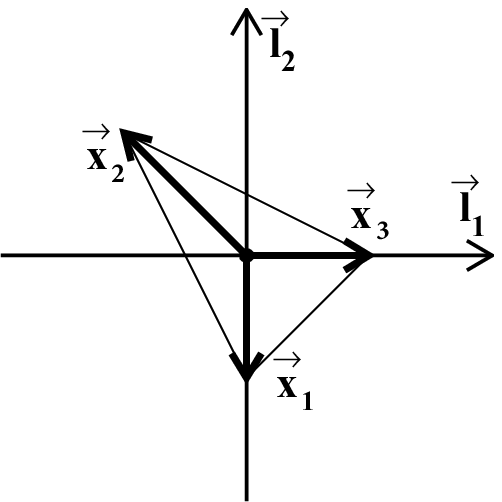,width=0.30\textwidth}}
$ \qquad\qquad\qquad
\lambda\ =\ \frac{2\pi}3, \quad a\ =\ \pi,
\qquad\qquad\qquad\qquad \lambda\ =\ \frac{5\pi}6, \quad a\ =\
\frac{7\pi}6,$
\vspace{-8mm}
\end{figure}
\[
\begin{array}{lcl}
x^{(2)}_1= -\frac\varrho{\sqrt3}, \;
x^{(2)}_2=\frac\varrho{2\sqrt3},  \;
x^{(2)}_3=\frac\varrho{2\sqrt3},
&\qquad&
x^{(2)}_1=-\frac\varrho{2\sqrt3}, \;
x^{(2)}_2=-\frac\varrho{2\sqrt3}, \;
x^{(2)}_3=\frac\varrho{\sqrt3}.
\end{array}
\]

\subsubsection{Three-body problem in celestial mechanics
(the Laplace case). Self-consistent field in classical mechanics}

Suppose that an attractive Newtonian potential is acting between three
particles. It can be shown that there exists a solution for which all
three particles stay in the vertices of an equilateral triangle while
each particle moves along an elliptic trajectory about the common
center-of-mass as if there was a central body the mass of which is
equal to the sum of masses of the three particles.

Assume that $\dot a\dot\theta=\dot\varphi_2$ is equal to zero. If
the particles form an equilateral triangle, we have $a=0$, and the
distance between the particles is $\varrho$. The potential energy in
this case is equal to $U=-3/\varrho$, so that the equations of
motion take the form
\begin{eqnarray}
\varrho-\varrho\left[\frac14 \dot\lambda^2 +\dot\varphi^2_1
+\dot\varphi_1\dot\lambda\right] = 0,
\qquad
\frac12\ddot\lambda+\dot\varphi_1+\frac{\dot\varrho}\varrho
\dot\lambda+\frac2\varrho\dot\varphi_1\dot\varrho\ =\ 0.
\label{1.54}
\end{eqnarray}
Introducing a new variable
\begin{equation}
\dot\psi\ =\ \frac12\,\dot\lambda+\dot\varphi_1\,,
\label{1.55}
\end{equation}
we obtain the Kepler equations
\begin{eqnarray}
\dot\varrho\ \varrho\dot\psi^2+\frac3{\varrho^2}\ =\ 0,
\qquad
   \psi+\frac2\varrho\dot\varrho\dot\psi\ =\ 0.
\label{1.56}
\end{eqnarray}
If we now express $\vec x_1,\vec x_2$ and $\vec x_3$ in terms of
$\varrho$, we get the equations of three ellipses. It is easy to
prove that in the case $a\neq0$ the equations will not lead to the
Kepler equations.  It should be noted that the type of the solution
is independent of the form of the potential; indeed, we use only the
fact that the forces acting on any of the three particles are
directed to the centre-of-mass of the triangle.


\section{Conclusion}

The present review paper is, in a way, an addition to the book
\cite{book-4}, considering the three-body problem from a different
point of view: we apply here the group theoretical method to its
investigation.

The Hamiltonian of the three-body problem consists of a kinematic part and a
part due to the interactions. Here we investigate mainly the kinematic part.
The group theoretical properties of low-lying states are defined by the
structure of the interaction while in the higher excitations the kinematic part
of the components is prevailing. The investigation of the highly excited states
is rather relevant in different actual problems. As an example we may consider
the three-quark baryon
states: the quark model predicts, obviously, more bound states than can be
observed experimentally. Another interesting problem is that
of the molecular systems with short range pair-interactions which lead to
three-particle systems with binding energies close to zero
\cite{mi-fa}. In such loose systems the details of the short range
interactions are, as a rule, not too important, while the quantum numbers of
pair interactions may turn out to be essential.

In the present paper we have described the quantum mechanical consideration of
the kinematical part of the Hamiltonian. The quantum mechanical investigation
can be easily generalized to the relativistic case. The reason is that in the
general case the kinematical parts have the same structure, namely: $ s_{12} +
s_{13} + s_{23} \longrightarrow \vec{k}^2_{12} + \vec{k}^2_{13} +
\vec{k}^2_{23} $ (let us remind here that the masses can be considered to be
equal).

The non-relativistic operator can be easily changed to a
relativistic one, {\it i.e.} written in a covariant form. The
covariant form of the operators, similar to the operators of angular
variables, were considered in detail for two-particle systems see
Ref.~\refcite{book-4}.

There are no principal difficulties in writing three-particle angular
momentum operators. In other words, the wave functions of the composite
systems investigated here can be successfully applied to relativistic
problems.

\section{Acknowledgments}

The authors would like to mention here the name of Ya.A.
Smorodinsky, who was one of the founders of the group-theoretical
description of the three-body problem, with special gratitude. We
thank our colleagues A. Frenkel, A. Luk\'acs, G. Stepanova
 and N.Ya. Smorodinskaya for their help.

\end{document}